\documentclass[12pt]{article}

\usepackage{a4wide}
\usepackage{hyperref}
\usepackage{graphicx}
\usepackage{amsmath}
\usepackage{pifont}
\usepackage{caption}
\usepackage{subcaption}
\usepackage{multirow}
\usepackage{slashed}
\usepackage{braket}
\usepackage{comment}
\usepackage{breqn}

\renewcommand{\Re}{\mathrm{Re}}

\newcommand{\calT}{\mathcal{T}}
\newcommand{\h}{h}
\newcommand{\calO}{{\cal O}}
\newcommand{\VmA}{$V$-$A$}
\newcommand{\VpA}{$V$+$A$}

\newcommand{\matel}[3]{\Braket{#1|#2|#3}}
\newcommand{\vev}[1]{\Braket{#1}}

\newcommand{\GeV}{\,{\rm GeV}}
\newcommand{\MeV}{\,{\rm MeV}}
\newcommand{\CBBL}{C}

\newcommand{\tick}{\ding{51}}
\newcommand{\cross}{\ding{55}}

\begin{document}

\begin{flushright}
Edinburgh/13/05 \\
CP3-Origins-2013-007 DNRF90
 \\
DIAS-2013-7
\end{flushright}

\begin{center}
   {\large \bf \boldmath Isospin asymmetries in $B\to (K^*,\rho) \gamma/ l^+ l^-$ 
   and $B \to K l^+ l^-$ in and beyond the Standard Model}
    \vskip1.3cm {\sc
 James Lyon$^{a,}\footnote{J.D.Lyon@sms.ed.ac.uk} $ \& Roman Zwicky$^{a,}$\footnote{Roman.Zwicky@ed.ac.uk}
  \vskip0.5cm
        {\em  $^a$ {\sl School of Physics and Astronomy, University of Edinburgh, 
    Edinburgh EH9 3JZ, Scotland}}}
\end{center}

\begin{abstract}
We compute  the isospin asymmetries in $B \to (K^*,\rho) \gamma$ and $B\to (K,K^*,\rho) l^+l^-$ for low lepton pair invariant mass $q^2$,
within the Standard Model (SM) and beyond the SM (BSM) in a generic dimension six operator basis.
Within the SM the CP-averaged isospin asymmetries for $B \to (K,K^*,\rho) ll$, between 
$1\GeV^2  \leq q^2   \leq 4m_c^2$, are predicted to be small (below 1.5\%) though with significant cancellation.  In the SM  the non-CP averaged asymmetries for  $B \to \rho ll$
deviate by $\approx \pm 5\%$ from the CP-averaged ones. 
We provide physical arguments, based on resonances, of why isospin asymmetries have to decrease for large $q^2$ (towards the endpoint). 
Two types of isospin violating effects are computed:
ultraviolet (UV) isospin violation due to differences between operators coupling to up and down quarks,
and infrared (IR) isospin violation where a photon is emitted from the spectator quark and is hence proportional to the difference between the up- and down-quark charges.
These isospin violating processes may be subdivided into weak annihilation (WA), quark loop spectator scattering (QLSS) and a chromomagnetic contribution. 
Furthermore we discuss generic selection rules based on parity and angular momentum for the $B \to Kll$ transition as well as specific selection rules valid for WA at leading order in the strong coupling constant.
 We clarify that the relation between the  $K$ and the longitudinal part of the $K^*$ 
only holds for leading twist and for left-handed currents.  
In general the $B \to \rho ll$ and $B \to K^*ll$ isospin asymmetries are structurally different 
yet the closeness of  $\alpha_{\rm CKM}$ to ninety degrees allows us to construct a (quasi) null test
for the SM out of the respective isospin symmetries. 
We provide and discuss an update on ${\cal B}(B^0 \to K^{*0}\gamma)/{\cal B}(B_s \to \phi \gamma)$ which is sensitive to WA.
\end{abstract}
\newpage

\setcounter{footnote}{0}
\renewcommand{\thefootnote}{\arabic{footnote}}

\tableofcontents

\section{Introduction}

The isospin asymmetry in $B \to K^* ll$ 
gave, in recent years, some indication of being of opposite sign to the Standard Model (SM) prediction 
\cite{Aubert:2008ps,Wei:2009zv}. This trend has not been confirmed by the new LHCb data in the year of 2012 \cite{Aaij:2012cq},
yet in $B \to K ll$ a negative deviation from zero has been measured at the level of $4\sigma$ taking into 
account the entire $q^2$-spectrum. The isospin-asymmetry in $B \to Kll$ is expected to be 
small in the SM and therefore it is important to assess this observable.
Isospin asymmetries of the $B \to K^* \gamma$ and $B \to \rho \gamma$  \cite{Amhis:2012bh} are found to agree and deviate by two standard deviations from the SM  giving rise to constraints and curiosity for future measurements respectively.

Independent of any signs of deviation, the isospin asymmetry contributes to the 
microscopic investigation of flavour changing neutral currents (FCNCs).
It is sensitive to a large number of four Fermi  operators of the $\Delta F = 1$-type and complements the constraints from non-leptonic decays. 

In this paper we consider two types of isospin violating effects: \emph{ultraviolet (UV) isospin violation}  due to 
unequal coupling of up and down quarks to Fermi operators as well 
as \emph{infrared (IR) isospin violation from spectator emission} of the intermediate photon which is therefore proportional to the charge difference of up and down quark.
The latter mainly result from processes for which the large energy release of the $b$-quark is transferred to the spectator quark which then emits an energetic photon as the external kinematics require. 
The processes  
are depicted in Fig.~\ref{fig:isoSM}
and are from left to right: weak annihilation (WA), quark loop spectator scattering (QLSS) and 
the contribution from the chromomagnetic operator which we shall simply denote 
by ${\cal O}_8$ hereafter.
Isospin effects in quantum chromodynamical (QCD) quantities such as masses and decay constants 
are known to be just below the sub-percent level, to be discussed later on, and therefore small in comparison to the precision accessible to near future experiments. 

In this paper we have computed WA in LCSR, extending earlier results for $q^2 = 0$ \cite{Ali:1995uy,Khodjamirian:1995uc}, QLSS within QCD factorisation (QCDF) and ${\cal O}_8$ we take from our recent computation \cite{DLZ12}.
Furthermore we include a complete set of dimension six operators relevant at leading 
order of the strong coupling constant $\alpha_s = g_s^2/4\pi$.  By mapping a specific model into an operator basis, 
such as the family model considered in \cite{Ishida:2013sya}, one can get the effects with our estimates.

Various aspects of the isospin asymmetry in $B\to K,K^* l^+l^-$ decay have been calculated previously.
The closely related decay $B\to V\gamma$ has been computed using QCD factorisation (QCDF) in
\cite{Kagan:2001zk} and using a mixture of QCDF and light-cone sum rules (LCSR) in \cite{Ball:2006eu}. A program computing the isospin asymmetry $B \to K^* \gamma$ 
in the minimal supersymmetric SM has been reported in \cite{Mahmoudi:2007vz}.
$B\to K,K^* l^+l^-$ was computed using QCDF in \cite{Feldmann:2002iw}, and a mixed approach
was recently employed for $B \to Kll$ in \cite{Khodjamirian:2012rm}. 
In this paper we improve on these works by including a complete basis of dimension six operators for WA and QLSS, our recent ${\cal O}_8$-computation \cite{DLZ12}, 
and the complete set of twist-3 terms for WA.

Essential results of the paper are that the isospin asymmetries of $B \to K,K^*,\rho ll$
are small in the SM; especially and systematically for high $q^2$. 
The isospin asymmetries of the $K^*$- and $\rho$-mesons  turn out to be very similar at small $q^2$ up 
to form factor ratios due to the, presumably, accidental closeness of the Cabibbo-Kobayashi-Maskawa (CKM) angle $\alpha_{\rm CKM}$ to ninety degrees.
This prompts us to define a quantity $\delta_{a_I}$ which serves as a (quasi) null test of the SM. 
The results of the generic dimension six operators are given in form of tables and complement constraints from non-leptonic decays
and $B \to (K^*,\rho) \gamma$ decays, which we discuss.

This work is written in the language of $B \to K/K^*$-transition. Adaptation to other light vector mesons is generally straightforward, with the exception of WA for
$B^0 \to \rho^0$- and $B_s \to \phi$-transitions which we
discuss in the corresponding section in some detail. The paper is organised as follows:
in section \ref{sec:rates} we present formulae for pseudoscalar and vector meson decay rates in terms of helicity amplitudes, 
the relation of the latter to form factors and quark loop functions and finally
formulae for the isospin asymmetries in a linear approximation.
In section \ref{sec:weak-annihilation} the complete operator basis of dimension six operators contributing
at ${\cal O}(\alpha_s^0)$ to the WA subprocess is given;
with results in subsection \ref{sec:WAresults} and detailed formulae in  appendix \ref{app:wa-formulae}.
In section \ref{sec:quark-loop-spectator} we present the complete dimension six operator basis 
contributing to QLSS at order ${\cal O}(\alpha_s)$ along with results computed in QCDF.
In section \ref{sec:iso} we present the $B \to K^*ll$ and $B \to K^*\gamma$ isospin asymmetries.
In subsection \ref{sec:selection} we discuss selection rules:
those applicable to any scalar $\to$ scalar $ll$ decay in \ref{sec:Ksel},
those particular to WA in the factorisation approximation \ref{sec:WAsel}
and we also discuss to what extent the $K$ distribution amplitude (DA) does or does not correspond to the $K^*_\parallel$ in \ref{sec:non-relation}.
In subsection \ref{sec:q2-dependence} we discuss $q^2$ dependence of the isospin asymmetries,
and in subsection \ref{sec:isospinSM}  we discuss the kaon isospin asymmetries in the SM respectively. 
 In section \ref{sec:rho-meson} we present the isospin asymmetry of $B \to \rho \gamma/ll$ as well as an extension of the operator basis for $\rho^0$-channel.
 In section  \ref{sec:isospinBSM}  we cover aspects of isospin BSM: in \ref{sec:constraints} we (briefly) discuss constraints and in \ref{sec:dai} we propose
 the ratio of the $K^* \gamma$- and $\rho \gamma$-asymmetries as a (quasi) null test of the SM.
 We end the paper with a summary and conclusions in section  \ref{sec:discussion}.
 
 We give an updated prediction for ${\cal B}(B^0 \to K^{*0}\gamma)/{\cal B}(B_s \to \phi \gamma)$  in appendix \ref{app:update}, which was recently measured by LHCb.
Various explicit results can be found in appendix \ref{app:results}, such 
as numbers for  the generic dimension six operator basis for all isospin asymmetries discussed 
 in this paper in \ref{app:tables}.
Aspects of gauge invariance  of the WA and QLSS computations, which turn out to be rather intricate, can be found in appendix  \ref{app:gauge-invariance}.

\section{Decay rate and form factors}
\label{sec:rates}

The effective Hamiltonian in the SM, to be extended in upcoming sections, 
is parametrised by  
\begin{equation}
\mathcal H_{\mathrm{eff}} = \frac{G_F}{\sqrt 2}\left( \sum_{i=1}^2 
(\lambda_u C_i \mathcal O_i^u + \lambda_c C_i \mathcal O_i^c )       -\lambda_t \sum_{i=3}^{10} C_i \mathcal O_i \right)  \;, \qquad 
\lambda_i \equiv V_{is}^*V_{ib} \;,
\label{eq:effective-ew-hamiltonian}
\end{equation}
where the $b \to s$ unitarity relation reads $\lambda_u + \lambda_c + \lambda_t = 0$.
The basis is given by \cite{Buchalla:1995vs}:
\begin{align}
\label{eq:SMbasis}
\mathcal O_1^q &= (\bar s_i q_j)_{V-A}(\bar q_j b_i)_{V-A} &
\mathcal O_2^q &= (\bar s_i q_i)_{V-A}(\bar q_j b_j)_{V-A} \nonumber \\
\mathcal O_3 &= (\bar s_i b_i)_{V-A} \sum_q (\bar q_j q_j)_{V-A} &
\mathcal O_4 &= (\bar s_i b_j)_{V-A} \sum_q (\bar q_j q_i)_{V-A} \nonumber \\
\mathcal O_5 &= (\bar s_i b_i)_{V-A} \sum_q (\bar q_j q_j)_{V+A} &
\mathcal O_6 &= (\bar s_i b_j)_{V-A} \sum_q (\bar q_j q_i)_{V+A} \nonumber \\
\mathcal O_7 &= -\frac{e m_b}{8\pi^2}\bar s \sigma\cdot F (1+\gamma_5)b &
\mathcal O_8 &= -\frac{g_s m_b}{8\pi^2}\bar s \sigma\cdot G (1+\gamma_5) b \nonumber \\
\mathcal O_9 &= \frac{\alpha}{2\pi}(\bar l\gamma^\mu l)(\bar s \gamma_\mu (1-\gamma_5) b) &
\mathcal O_{10} &= \frac{\alpha}{2\pi}(\bar l\gamma^\mu\gamma_5 l)(\bar s \gamma_\mu (1-\gamma_5) b) \;, & 
\end{align}
where $i,j$ are colour indices, $(\bar s b)_{V\pm A}=\bar s \gamma^\mu (1\pm\gamma_5) b$ and $\calO'_{7-10}$ will denote the operators with opposite chirality as usual,
and we have taken the opposite sign of $\mathcal O_{7,8}$\footnote{
This corresponds to a covariant derivative $D_\mu = \partial_\mu - i Q  e A_\mu - i g_s A_\mu$ and  interaction vertex $+i( Qe + g_s\frac{\lambda^a}{2}) \gamma^\mu$ 
in agreement with \cite{Beneke:2001at} but differing from \cite{Ali:1995uy,Khodjamirian:1995uc,Buchalla:1995vs,Chetyrkin:1996vx}.}.
These are the same conventions as in  \cite{Beneke:2001at}.
Details of the calculation of the Wilson coefficients $C_{1-10}$ are given in appendix \ref{app:wilson-coefficients}.
Furthermore, $e = \sqrt{4 \pi \alpha} > 0 $ where $\alpha$ is the fine structure constant and $G_F$ is the Fermi constant.



We parameterise the amplitude as follows:
\begin{equation}
\begin{split}
&_{\rm out}\left\langle M(p) l^+(l_1) l^-(l_2)\mid B(p+q)\right\rangle_{\rm in}  \\
&= \frac{G_F}{\sqrt 2}\lambda_t \frac{\alpha m_b}{q^2\pi}
\left(\bar u(l_1)\gamma_\mu v(l_2)\sum_i\mathcal \calT^V_i P^\mu_i + \bar u(l_1)\gamma_\mu\gamma_5 v(l_2)\sum_i\calT^A_i P^\mu_i\right)\delta^{(4)}(q-l_1-l_2) \;,
\label{eq:amplitude-definition}
\end{split}
\end{equation}
where  M stands, throughout this work, for  a light vector ($K^*$,$\rho$) 
or pseudoscalar ($K$) meson. 
The symbols  $u,v$  correspond to lepton polarisation spinors of mass dimension $1/2$  and  $q=l_1+l_2$ is the total momentum of the lepton pair. 
For lepton coupling we only allow $V$ and $A$-couplings as present in the SM. 
The isospin violating contributions all proceed through a photon and thus have a vectorial coupling. The axial coupling is included as it originates through $Z$-penguins 
and box diagrams which are significant in the SM.  In this work we do not include non-SM 
lepton couplings. 
The basis tensors $P_i^\mu$ \cite{DLZ12} are the standard choice for penguin form factors
\begin{alignat}{2}
& K^*: \quad  & P_1^\mu =& 2\epsilon^{\mu\nu\rho\sigma}\eta^*_\nu p_\sigma q_\rho  \;,\nonumber \\
&   &  P_2^\mu =& i\left[(m_B^2-m_{K^*}^2)\eta^{*\mu} - (\eta^* \cdot q)(2p+q)^\mu\right]  \;,\nonumber \\
&   &P_3^\mu =& i(\eta^* \cdot q)\left[q_\mu - \frac{q^2}{m_B^2-m_{K^*}^2}(2p+q)^\mu\right]  \;, \nonumber \\
& K:  &P_T^\mu =& \frac{1}{m_B+m_K}\left[(m_B^2-m_K^2)q^\mu - q^2 (2p+q)^\mu\right] \;,
\end{alignat}
in the sense that $\calT^V_i=C_7 T_i(q^2)+\text{corrections}$\footnote{The $\calT^V_i$ differ from 
the ones \cite{Beneke:2001at} in that we include the contributions of $C_9$ as well.}.
The basis for  pseudoscalar  and vector meson decays 
are  $P_{T}^\mu$  and $P_{1,2,3}^\mu$ respectively. Note we have implicitly assumed $m_l =0$ as otherwise there is additional direction proportional to $q^\mu$ which 
vanishes for the $V$ but is proportional to $m_l$ for $A$.
The four vector  $\eta$ denotes the vector meson polarisation.
We use the Bjorken \& Drell convention for the Levi-Civita tensor $\epsilon_{0123}=+1$.
In discussing physical quantities and problems it will prove advantageous 
to go over to the so-called \emph{helicity basis}:
\begin{equation}
\label{eq:Btrafo}
 \begin{pmatrix}
\h_0 \\
\h_+ \\
\h_-
\end{pmatrix} =
\underbrace{
 \begin{pmatrix}
 0 & a & b  \\
1/\sqrt{2} &-c/\sqrt{2}  & 0 \\ 
1/\sqrt{2} & +c/\sqrt{2} & 0 
\end{pmatrix} }_{\equiv B}
 \begin{pmatrix}
\calT_1 \\
\calT_2 \\
\calT_3
\end{pmatrix}   \;.
\end{equation}
which corresponds to $0,\pm$\footnote{The direction $0$ and $\pm$ are also known as the longitudinal and transversal polarisation directions.} polarisation of the vector meson.
Basis tensors corresponding to $h_{0,\pm}$ are given in appendix \ref{app:decay-width-tensors}.
The variables $a,b,c$ in the basis transformation matrix are given by:
\begin{equation}
\label{eq:abc}
 (a,b) \equiv \sqrt{\frac{\hat q^2}{8}} \frac{1}{\hat m_V} \left( \frac{1+3 \hat m_V^2- \hat q^2}{\sqrt{\lambda_V}},\frac{ -\sqrt{\lambda_V}}{1-\hat m_V^2}\right) \;, \quad c \equiv  \frac{1-\hat m_V^2}{\sqrt{\lambda_V}} = 1 +{\cal O}(\hat q^2)  \nonumber \;,
\end{equation}
where here and below hatted quantities are normalised with respect to the $B$-meson mass, $\hat q^2 \equiv q^2/m_B^2$,  $\hat m_V^2 \equiv m_V^2/m_B^2$ and  $ \lambda_V$ is the K\"all\'{e}n-function with normalised entries:
\begin{equation}
\label{eq:Kallen}
\lambda_{V} \equiv \lambda_V(1, \hat m_V^2, \hat q^2) = ((1 + \hat m_V)^2 - \hat q^2)((1 - \hat m_V)^2 - \hat q^2)\;.
\end{equation}
For the $K$-meson there is no polarisation and no freedom in choosing a basis. 

The decay rates are given by\footnote{The IR sensitive $1/q^2$ factor in the $B \to V ll$ rate for $m_l\to 0$ is
compensated by a virtual lepton loop in the limit $q^2 \to 0$ as the collinear lepton pair
is indistinguishable from a photon. This corresponds to the famous Bloch-Nordsieck cancellation mechanism.
Furthermore we note that $|h_0|^2 \sim q^2$ by virtue of 
\eqref{eq:abc} and \eqref{eq:Btrafo} and corresponds to the well-known decoupling of 
the zero helicity mode towards $q^2 \to 0$. In the differential rate into the pseudoscalar \eqref{eq:scalar-decay-rate} the $q^2$ has been factored out from $|h_T|^2$ to cancel the explicit pole.}
\begin{align}
\frac{d\Gamma}{dq^2}[B\to K^*l^+l^-] &=  \left[\frac{\lambda_V^{3/2}}{ q^2}\right] \left(\frac{\alpha}{4\pi}\right)^2 \,  c_F c_L \sum_{i=V,A} \left[\left|\h^i_+\right|^2 + \left|\h^i_-\right|^2 + |h^i_0|^2\right]  \;,
\label{eq:vector-decay-rate}  \\[0.1cm]
\frac{d\Gamma}{dq^2}[B\to K l^+l^-] &=   \left[ \frac{\lambda_{P}^{3/2}}{2 (m_B+m_K)^2 } \right] \left(\frac{\alpha}{4\pi}\right)^2 \,  c_F c_L \sum_{i=V,A} |\h^i_T|^2  \;,
\label{eq:scalar-decay-rate}
 \\[0.1cm]
\Gamma[B\to K^*\gamma] &=   \left[ \frac{3}{4}  \lambda_V^{3/2} \right] \left(\frac{\alpha}{4\pi}\right) \,  c_F
\left[|\h^V_+|^2 + |\h^V_-|^2 \right]|_{q^2 = 0} \;,
\label{eq:photon_rate}
\end{align}
where $c_F \equiv  (G_F^2 |\lambda_t|^2 m_b^2 m_B^3/12\pi^3)$, $\h_T^i \equiv \calT_T^i$
and $c_L=(1+2m_l^2/q^2)\sqrt{1-4m_l^2/q^2}$ accounts for nonzero lepton mass.
 An important observation is that 
for $m_V \to 0$ the rate remains bounded\footnote{Note we do not want to invoke the $m_V \to 0$ limit per se as it is well known that massless and massive representations differ in a discontinuous fashion.}  provided that 
 \begin{equation}
\label{eq:X2X3general}
\h^i_0 = {\cal O}(m_V^0) \quad \Rightarrow \quad \calT^i_2 =    \frac{\lambda_V}{(1- \hat m_V^2)(1+3\hat m_V^2-\hat q^2)} \calT^i_3 + {\cal O}(m_V) \;.
\end{equation}
This expression reduces to the form we have given in our previous work \cite{DLZ12} in the appendix in the $m_V \to 0$ limit. In essence the relation between ${\calT}_2$ and ${\calT}_3$ cancels the explicit $1/m_V$ in $h_0$ which appears through \eqref{eq:Btrafo} and \eqref{eq:abc}.
Note in the SM $h_+ \ll h_-$ by virtue of the \VmA-interactions. 
In \cite{Beneke:2001at}, which operates in the heavy quark limit, $h_+ \to 0$. 
In our work $h_+$ is vital as we allow for right-handed structures that violate isospin.

The axial lepton amplitudes $\calT^A_i$ arise only from the $\mathcal O_{10}$ operator and are given in terms of standard form factors by:
\begin{align}
\calT^A_1 &=  \frac{C_{10} q^2 V(q^2) }{2m_b(m_B+m_{K^*})}  \;,
&\calT^A_3 &= -C_{10} \frac{m_{K^*}}{m_b} A_3(q^2)  \;,  \nonumber \\
 \calT^A_2 &=  \frac{C_{10} q^2 A_1(q^2)}{2m_b(m_B-m_{K^*})}  \;,
& \calT^A_T &= C_{10} \frac{m_B+m_K}{2m_b} f_+(q^2) \;.
\label{eq:lepton-axial-ff}
\end{align}
We will split the vector lepton amplitudes into isospin sensitive and insensitive parts denoted
by $\calT^q$ and $\calT^{0}$ respectively with $q$ being the light flavour of 
the $B$-meson:
\begin{align}
\label{eq:Tdec}
\calT^V_i &= \calT^{V,0}_i + \calT^{V,q}_i  \;, 
& \calT^{V,q}_i &= C_8^{\mathrm{eff}} G^q_i(q^2) + W^q_i(q^2) + S^q_i(q^2) \;.
\end{align}
Note, we have absorbed the Wilson coefficient (WC) for WA and QLSS into the functions $W^q_i(q^2)$ and $S^q_i(q^2)$ respectively as there are quite a few of them.
The WC $C^{\rm eff}_{7,8,9}$ correspond to scheme and basis independent WCs which include quark loop contributions and will be defined further below.
The symmetric part is approximated throughout this work by the $C^{\rm eff}_{7,9}$ contributions, which in terms of standard form factors is given by:
\begin{align}
\calT^{V,0}_1(q^2) &= \frac{C_9^{\mathrm{eff}}(q^2) q^2 V(q^2)}{2m_b(m_B+m_{K^*})} + C_7^{\mathrm{eff}} T_1(q^2) \;, &
\calT^{V,0}_3(q^2) &= -C_9^{\mathrm{eff}}(q^2) \frac{m_{K^*}}{m_b} A_3(q^2) + C_7^{\mathrm{eff}} T_3(q^2) \nonumber \\
\calT^{V,0}_2(q^2) &= \frac{C_9^{\mathrm{eff}}(q^2) q^2 A_1(q^2) }{2m_b(m_B-m_{K^*})} + C_7^{\mathrm{eff}} T_2(q^2) \;, &
\calT^{V,0}_T(q^2) &= C_9^{\mathrm{eff}}(q^2)\frac{m_B+m_K}{2m_b} f_+(q^2) + C_7^{\mathrm{eff}} f_T(q^2)
\label{eq:lepton-vector-sym-ff}
\end{align}
\begin{figure}
\includegraphics[width=\textwidth]{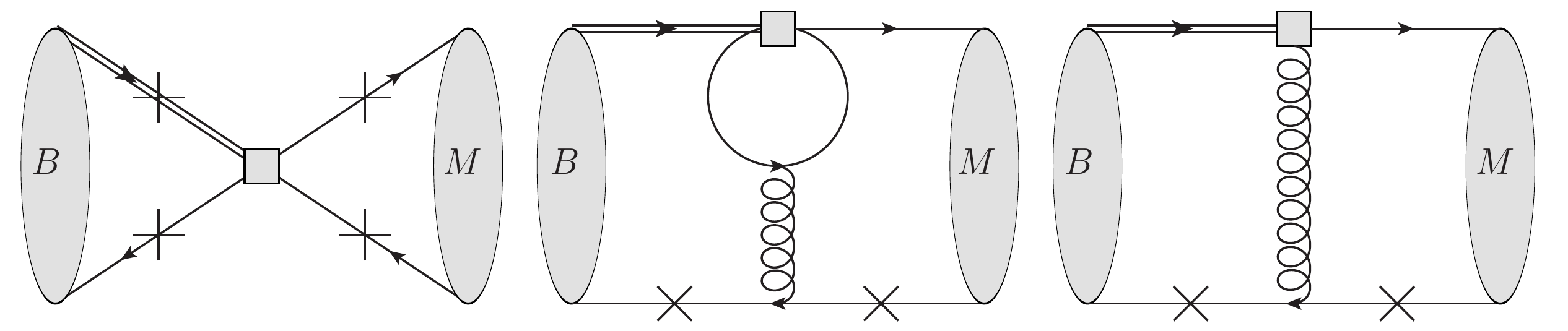}
\caption{\small Isospin violating processes included in our calculation.
Crosses indicate possible photon emission points. Throughout this paper double lines stand for the $b$-quark flavour.
(left) Weak annihilation (WA) (middle) Quark loops with spectator scattering (QLSS) (right) Chromomagnetic operator ${\cal O}_8$. 
Note that  in WA we have indicated photon emission from quarks with $Q_b$-charge as well 
as WA is sensitive to UV isospin violation where the four Fermi operator 
with spectator quark $u$ and $d$ appear in unequal proportion.}
\label{fig:isoSM}
\end{figure}
The isospin sensitive diagrams are shown in Fig.~\ref{fig:isoSM}.
The weak annihilation amplitude, denoted $W^q_i(q^2)$, (Fig.~\ref{fig:isoSM},left)
originates from  $\mathcal O_{1-6}$ and is computed using LCSR in section \ref{sec:weak-annihilation}.
Spectator scattering with a quark loop, denoted $S^q_i(q^2)$, (Fig.~\ref{fig:isoSM},middle) arises from $\mathcal O_{1-6}$ 
as well and is computed using QCD factorisation in section \ref{sec:quark-loop-spectator}.
The spectator contributions due to $\mathcal O_8$ (Fig.~\ref{fig:isoSM},right) are denoted by  
$G_i^q(q^2)$ and are taken from our recent work \cite{DLZ12}.  
For the short distance  form factors in \eqref{eq:lepton-axial-ff} and \eqref{eq:lepton-vector-sym-ff}
we use the fits in \cite{Ball:2004rg,Ball:2004ye}, recomputed with updated hadronic inputs
as in  \cite{DLZ12}.
Quark loop contributions, other than the ones with a gluon connecting to the spectator, 
are absorbed into the effective  WCs. The structures proportional to the $m_b$-mass 
are independent of $q^2$ and described by:
\begin{align}
C_7^{\mathrm{eff}} &= C_7 - \frac{4}{9}\CBBL_3 - \frac{4}{3}\CBBL_4 + \frac{1}{9}\CBBL_5 + \frac{1}{3}\CBBL_6 \;, &
C_8^{\mathrm{eff}} &= C_8 + \frac{4}{3}\CBBL_3 - \frac{1}{3}\CBBL_5 \;.
\end{align}
The other contributions are taken care off by $ C_9^{\mathrm{eff}}(q^2)$ \cite{Grinstein:1988me}, 
 \begin{equation}
 C_9^{\mathrm{eff}}(q^2) = C_9 + Y(q^2)  \;,
 \end{equation}
 where 
\begin{equation}
\begin{split}
Y(q^2) =& h(q^2,m_c)\left(-\frac{\lambda_c}{\lambda_t}\left(3\CBBL_1 + \CBBL_2\right) + 3\CBBL_3 + \CBBL_4 + 3\CBBL_5 + \CBBL_6\right) \\
& - \frac{h(q^2,m_b)}{2}\left(4\CBBL_3 + 4\CBBL_4 + 3\CBBL_5 + \CBBL_6\right) \\
& - h(q^2,0)\left(\frac{\lambda_u}{\lambda_t}\left(3\CBBL_1+\CBBL_2\right) + \frac{1}{2}\left(\CBBL_3 + 3\CBBL_4\right)\right) + \frac{4}{27}\left(\CBBL_3 + 3\CBBL_4 + 8\CBBL_5\right)  \;,
\end{split}
\end{equation}
with $h(s,m_q)$ being photon vacuum polarisation which we quote in section \ref{sec:QLSS},
and we restored the factors $\lambda_u$ and $\lambda_c$ factors explicitly especially in view of the fact that for the $b \to d$ transition the hierarchies differ from the $b \to s$ transitions.
 
\subsection{Definition of isospin asymmetries} 
 
The experimentally accessible isospin asymmetry $a_I(q^2)$ and its CP-average $\bar{a}_I$\footnote{For the $K^*$ this merely doubles the statistics and reduces experimental uncertainties 
in the production.  For the pseudoscalar $K$ this is essential as the $K_S^0$ is detected in experiment 
which is a linear superposition of  $|s \bar d \rangle$ and $|d \bar s \rangle$ eigenstates which implies averaging.}, which are functions
of the lepton pair invariant mass $q^2$, are defined as follows:
\begin{eqnarray}
a_I^{\bar 0 -}(q^2) &\equiv&  \frac{d A_I^{\bar 0 -}}{dq^2} \equiv \frac{c_M^2 d\Gamma[\overline B^0\to \overline M^0 l^+l^-]/dq^2 - d\Gamma[B^-\to M^- l^+ l^-]/dq^2}{c_M^2 d\Gamma[\overline B^0\to \overline M^0 l^+l^-]/dq^2 + d\Gamma[ B^-\to M^- l^+ l^-]/dq^2} \;, \nonumber \\[0.1cm]
\bar{a}_I(q^2)  &\equiv&  \frac{1}{2}\left( a_I^{\bar 0 -}(q^2) + a_I^{0 +}(q^2) \right) \;,
\label{eq:dai-def}
\end{eqnarray}
where $a_I^{0 +}$ corresponds to the CP-conjugated process of $a_I^{\bar 0 -}$.
The constant $c_M$ is given by $c_K=c_{K^*}=1$ and $c_\rho=\sqrt{2}\,$\footnote{This accounts for $\rho^0\sim(\bar uu-\bar dd)/\sqrt{2}$ since the leading decay amplitude only couples to the $\bar dd$ component of the $\rho^0$.}.
A similar definition without differential applies for the $B \to K^*/ \rho \gamma$ transition.
The definition above makes clear the meaning of isospin in this context:
it is understood throughout this paper to mean a rotation between $u$ and $d$ quarks,
with no corresponding rotation between $b$ and $t$ quarks,
as is understood in the case of the electroweak precision parameter $\rho$ for instance.

Assuming that the decay rate is dominated by $C_{7,9,10}$, only taking linear terms into account 
as in \cite{Kagan:2001zk,Feldmann:2002iw}, we arrive at:
\begin{align}
\label{eq:dai}
 \frac{d A_I^{\bar 0 -}}{dq^2} [B\to K^* l^+l^-] &= \frac{{\displaystyle \sum_{i=\{0,\pm\}}}  \Re\left[\h^{V,0}_i(q^2) \Delta^{V,d-u}_i(q^2)  \right]}{
{\displaystyle \sum_{i=\{0,\pm\}}} \left[|\h^{V,0}_i(q^2)|^2 + \left|\h^A_i(q^2)\right|^2\right]} + {\cal O}([\Delta^{V,d-u}_i(q^2)]^2)   \;, \nonumber \\
 \frac{d A_I^{\bar 0 -}}{dq^2}[B\to Kl^+l^-] &= \frac{\Re\left[\h^{V,0}_T(q^2)  \Delta^{V,d-u}_T(q^2) \right]}{|\h^{V,0}_T(q^2)|^2 + \left|\h^A_T(q^2)\right|^2} + {\cal O}([\Delta^{V,d-u}_T(q^2)]^2)  \;, \nonumber \\
 a_I^{\bar 0 -}  [B\to K^* \gamma]&=  \frac{{\displaystyle \sum_{i=\pm}} \, \Re\left[\h^{V,0}_i(0) \Delta^{V,d-u}_i(0)  \right]}{
{\displaystyle \sum_{i=\pm}} \left[|\h^{V,0}_i(0)|^2 + \left|\h^A_i(0)\right|^2\right]} + {\cal O}([\Delta^{V,d-u}_i(0)]^2)  \;,
\end{align}
where $ \Delta^{V,d-u}_\iota(q^2) \equiv  \left(\h^{V,d}_\iota(q^2) - \h^{V,u}_\iota(q^2)\right)$.
It is worth mentioning that in the above formulae we have explicitly and implicitly neglected 
effects from  phase space differences, isospin corrections to 
QCD quantities such as decay constants, and the light quark masses $m_q$. 
The latter are negligibly small and QCD corrections to isospin are known to be small:
for example the pseudoscalar decay constant differs by
roughly $0.5\%$  between the neutral and charged case; see e.g. \cite{isoOLD} for a computation in chiral perturbation theory. 
For the $B$-mesons isospin effects are even smaller as $m_{B^0} -m_{B^\pm} = 0.32(6) \MeV$\footnote{The mass difference between the two neutral kaons is about one percent and relatively large and rather exceptional as a result of the Gell-Mann--Oakes--Renner relation $m_K^2 = -2 (m_q+m_s) \vev{\bar q q}/f_K^2 + ..$.} \cite{Beringer:1900zz} , which is minuscule in comparison with the heavy quark scale
$\bar \Lambda = m_B - m_b \simeq 600 \MeV$.
Thus in summary it is expected that isospin violation arising from the form factors
will not reach the one percent level and we shall therefore not discuss them any further.

\section{Weak annihilation (WA) contribution}
\label{sec:weak-annihilation}

The WA process $B^-\to W^-\to K^{(*)-}$ 
is described by the ``tree-level'' operators $\mathcal O_{1,2}$ in a process 
as shown in Fig. \ref{fig:isoSM}(left).
By extension, the same name is also given to diagrams with the same arrangement of quark lines involving $\mathcal O_{3-6}$, 
though they arise from renormalisation running and short distance penguins.

The WA contribution to $B\to K^*\gamma/l^+l^-$ 
has previously been computed at ${\cal O}(\alpha_s)$ using LCSR at $q^2 = 0$ in \cite{Ali:1995uy,Khodjamirian:1995uc}
and in QCD factorisation at $q^2 = 0$ \cite{Bosch:2001gv} and for $q^2 > 0$ \cite{Beneke:2001at,Feldmann:2002iw}.  
We extend the LCSR computation to higher $q^2$ including  twist-3 corrections from the $h_\parallel$ DA relevant to the longitudinal $K^*$ component which were neglected 
in \cite{Beneke:2001at,Feldmann:2002iw}.

\subsection{Complete WA-basis of dimension $6$ operators at $\calO(\alpha_s^0)$}
We include all four quark operators $\bar q \Gamma_1 b\, \bar s \Gamma_2 q$
which potentially  contribute at ${\cal O}(\alpha_s^0)$\footnote{As we shall see shortly there are further selection rules e.g. parity at ${\cal O}(\alpha_s)$ and Lorentz covariance to all orders for the $K$.}:
\begin{align}
 O^{{\rm WA}}_1 &\equiv \bar q b \, \bar sq &
 O^{{\rm WA}}_2 &\equiv \bar q\gamma_5 b\,\bar sq &
 O^{{\rm WA}}_3 &\equiv \bar qb\,\bar s\gamma_5 q &
 O^{{\rm WA}}_4 &\equiv \bar q\gamma_5 b\,\bar s\gamma_5 q \nonumber \\
 O^{{\rm WA}}_5 &\equiv \bar q\gamma_\mu b\,\bar s\gamma^\mu q &
 O^{{\rm WA}}_6 &\equiv \bar q\gamma_\mu\gamma_5 b\,\bar s\gamma^\mu q &
 O^{{\rm WA}}_7 &\equiv \bar q\gamma_\mu b\,\bar s\gamma^\mu\gamma_5 q &
 O^{{\rm WA}}_8 &\equiv \bar q\gamma_\mu\gamma_5 b\,\bar s\gamma^\mu\gamma_5 q \label{eq:OWA} \\
&&  O^{{\rm WA}}_9 &\equiv \bar q\sigma_{\mu\nu} b\,\bar s\sigma^{\mu\nu} q &
 O^{{\rm WA}}_{10} &\equiv \bar q\sigma_{\mu\nu}\gamma_5 b\,\bar s\sigma^{\mu\nu} q \;,\nonumber
\end{align}
parametrised by the effective Hamiltonian: 
\begin{equation}
{\cal H}^{{\rm WA,q}} = -\frac{G_F}{\sqrt{2}} \lambda_t  \sum^{10}_{i=1} a_i^q  O_i^{{\rm WA}}   \;,
\label{eq:h-effective-wa}
\end{equation}
where we suppress the $q$ superscript on the operators $O_i^{\rm WA}$ throughout this work.
Note that at ${\cal O}(\alpha_s)$, as well as for the $\rho^0$ for ${\cal O}(\alpha_s^0)$ to be discussed in section \ref{sec:Heffrho},  the size of the basis doubles as octet combinations 
of the type $ O_1^{{\rm WA},8}  \equiv (1/4) \bar q \lambda^a b \, \bar s \lambda^a q$ contribute as well. 

\subsubsection{Projection on SM-basis (WA in SM)}
\label{sec:WASM}

In the SM the operators \eqref{eq:OWA} obey minimal flavour symmetry (MFV) 
\cite{MFV1,MFV2,MFV3,MFVmicro1,MFVmicro2}
and may be expressed in the form $\bar q \Gamma P_L b \, \bar s P_R\Gamma q$ \eqref{eq:SMbasis}.
Since WA fixes the quark flavours and couples to only a single colour structure, two independent combinations of SM WCs ($\Gamma   \in \{ \mathbf{1},\gamma_\mu  \}$) 
appear in  each $B\rightarrow Mll$ process.
For a $bq \to sq$ process, with $q = u,d$, the couplings are given by
\begin{align}
\label{eq:aSM}
\text{SM:   }  \text{scalars:  }  a^q_1 &= -a^q_2 = a^q_3 = -a^q_4 = -2\left(\frac{\CBBL_5}{N_c} + \CBBL_6\right) \nonumber \\
 \text{vectors:  } a^q_5 &= -a^q_6 = -a^q_7 = a^q_8 = \left(\frac{\CBBL_3}{N_c} + \CBBL_4\right) - \delta_{qu}\frac{\lambda_u}{\lambda_t}\left(\frac{\CBBL_1}{N_c} + \CBBL_2\right) \nonumber \\
 \text{tensors:  } a^q_9 &= a^q_{10} = 0 \;,
\end{align}
where $a_{5-8}$ are the only ones which are non-degenerate in $q$ and $N_c=3$ denotes 
the number of colours.
The role of $\calO_{1,2}^u$ in the SM is exceptional as there is no $\mathcal O_{1,2}^d$ counterpart. It corresponds to what we called UV isospin violation. 
In particular radiation from all quarks in Fig.~\ref{fig:isoSM}(left) contributes to the isospin asymmetry.
Therefore the isospin asymmetry in the SM is sensitive to three independent combinations of the four quark WCs $\CBBL_{1-6}$.
Note, in the SM 
the effect is CKM suppressed in $b \to s$ contrary to $b \to d$. In the latter case the closeness of $\alpha_{\rm CKM}$ to ninety degrees leads to a suppression of the effect; to be discussed and exploited in further sections.

\begin{table}[h]
\center
\begin{tabular}{l | l | c | cccc |  cccc |  cc}
\multicolumn{2}{c|}{} & Twist & \multicolumn{4}{|c}{} & \multicolumn{4}{c}{Operator $O^{\rm WA}_n$} & & \\
\multicolumn{2}{c|}{} & & 1 & 2 & 3 & 4 & 5 & 6 & 7 & 8 & 9& 10 \\
\hline
\multirow{4}{*}{$B\to K$}
&    cov. ($\alpha_s^0$)&  & \cross & \underline{\cross} & \underline{\cross} & & \cross & \underline{\cross} & \underline{\cross} & & \cross & \underline{\cross} \\
& $\chi$-even ($\phi_K$) & 2 & & & & & & & & I,F$_{\rm c}$ & & \\
& $\chi$-odd ($\phi_{P,\sigma}$) & 3  & & & & I,F & & & & & & \\
\cline{2-13}
&  cov.   ($\alpha_s^n$, $n >0$) & & \tick & \underline{\cross}  & \underline{\cross} & \tick & \tick &  \underline{\cross} &   \underline{\cross} &  \tick & \tick &  \underline{\cross} \\
\hline \hline
\multirow{5}{*}{$B\to K^*$}
& cov. ($\alpha_s^0$) & &  \cross & & \cross & & & & \cross & \cross \\
& $\chi$-even  ($\phi_\parallel$) & 2 & & & & & I & I,F$_{\rm c}$ & & & & \\
& $\chi$-even  ($g_\perp^{(v)},g_\perp^{(a)} $) & 3 & & & & & I & I,F$_{\rm c}$ & & & & \\
& $\chi$-odd ($\phi_\perp$) & 2 & & F & & F & & & & & I & I \\
& $\chi$-odd ($h_\parallel^{(t)},h_\parallel^{(s)} $) & 3 & & F & & & & & & & & I \\
\cline{2-13}
& cov. ($\alpha_s^n$, $n >0$) &   & \tick &  \tick  & \tick & \tick & \tick &  \tick &   \tick &  \tick & \tick &   \tick
\end{tabular}
\caption{\small Operators contributing to  WA.  
The acronyms t. and  cov. stand for twist and  for covariance respectively and  $\chi$-odd/even for odd/even chirality.
Furthermore: a)  (I,F) radiation from inital (I) and or final (F) state; (F$_{\rm c}$) corresponds solely to  
a (local) contact term contribution  from final state radiation. The latter are then necessarily to all orders in the twist expansion.   b)
(\tick) contribution expected in initial and final state c)  (\cross) no contribution due to parity invariance of strong interactions in the factorisation approximation ${\cal O}(\alpha_s^0)$
d)  (\underline{\cross}) no contribution, in any order of $\alpha_s$ and twist, as chirality necessitates 
a  Levi-Civita tensor structure  for which there are not enough independent vectors for contraction (relevant for pseudscalar final state).
We should also note that  $g_\perp^{(v,a)}$ and $h_\parallel^{(s,t)}$ are related to $\phi_\perp$ an $\phi_\parallel$ by
Wandzura-Wilczek type relations \cite{Ball:1998sk}. At our level of approximation, $m_q = 0$ and no 3-particle DA, this corresponds to Eqs.(4.15/16) and (3.21/22)\cite{Ball:1998sk}  respectively. See also appendix \ref{app:light-DA} for further comments.
This means that when $g_\perp^{(v,a)}$ and $h_\parallel^{(s,t)}$ are computed 
$\phi_\parallel$ and $\phi_\perp$ are needed to render the computation gauge invariant at
the relevant ${\cal O}(m_V)$-level. 
Thus for (I,F$_{\rm c}$), contrary to (F) itself,  it is not possible to properly distinguish between twist $2$ and $3$ 
which is reflected in the table. 
}
\label{tab:OWA}
\end{table}

\subsection{WA at leading order ${\cal O}(\alpha_s^0)$}
\label{sec:wa-factorisation}

\begin{figure}
\centering
\begin{subfigure}[b]{0.45\textwidth}
\includegraphics[width=\textwidth]{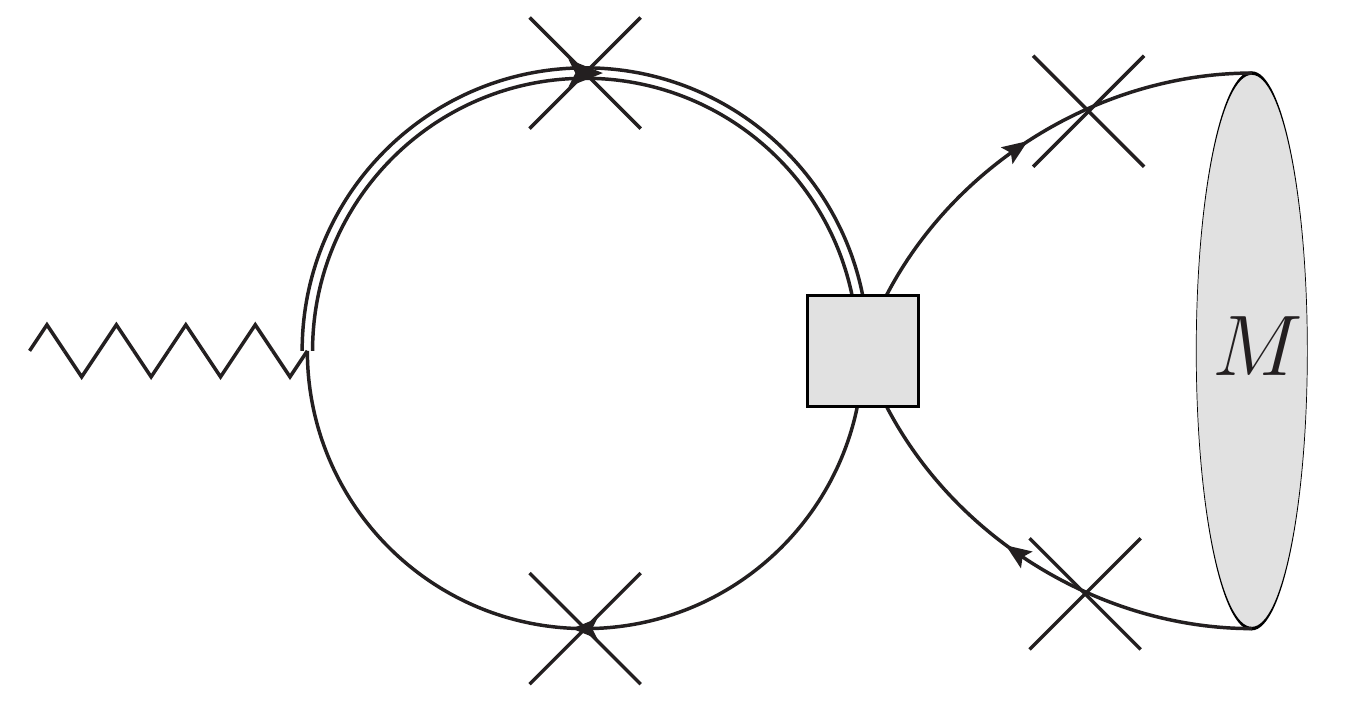}
\caption{\small Perturbative photon contributions.}
\label{fig:wa-diagram-pert}
\end{subfigure}
\begin{subfigure}[b]{0.45\textwidth}
\includegraphics[width=\textwidth]{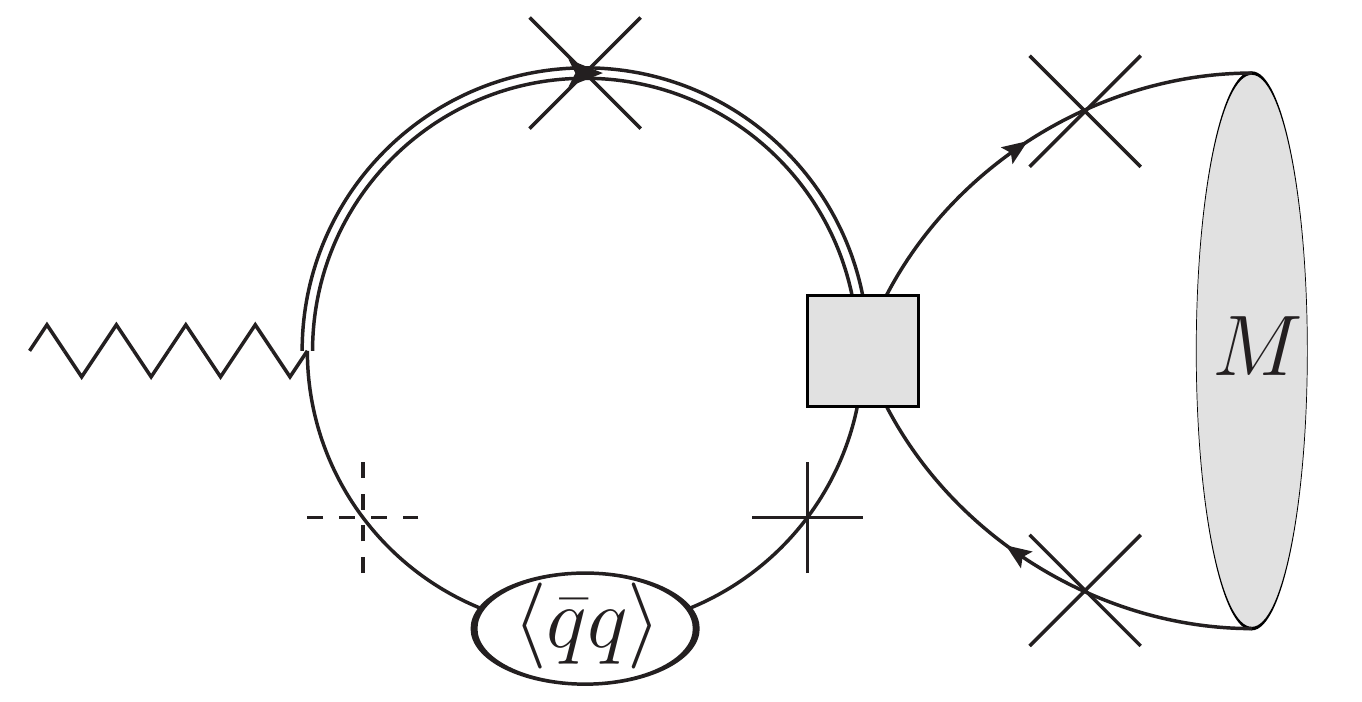}
\caption{\small OPE quark condensate contribution.}
\label{fig:wa-diagram-ope}
\end{subfigure}
\caption{
\small Weak annihilation Feynman diagrams for $B\to M l^+l^-$.
The zigzag line is the $B$-meson current insertion.
Crosses mark possible photon insertions,
although the contribution from the insertion at the dashed cross is zero.
}
\label{fig:wa-diagrams}
\end{figure}

\begin{figure}
\center
\includegraphics[width=0.45\textwidth]{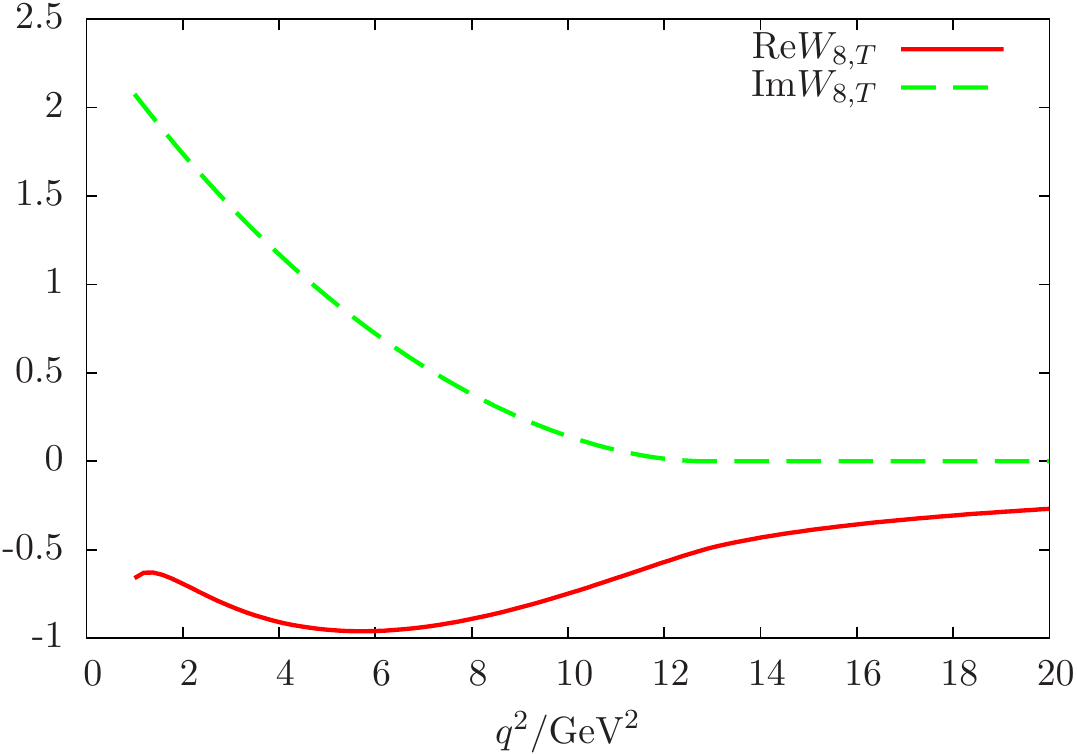}
\caption{
\small WA contribution to $B\to Kll$ for from the $O^{WA}_8$ operator \eqref{eq:OWA} 
as defined implicitly in \eqref{eq:Tdec}.
The imaginary part originates from the emission of $\rho,\omega,..$-meson and conversion into the photon,
which is analogous to the $\calO_8$ contribution c.f. Fig 4(right)\cite{DLZ12}.
}
\label{fig:wak8}
\end{figure}

The WA matrix element with uncontracted photon polarisation tensor $\epsilon(q)_\rho$ reads
\begin{eqnarray}
{\cal A}^{*\rho}|_{\text{WA}}  &=& \langle X\gamma^*(\rho)| \bar q \Gamma_1 b\,\bar s \Gamma_2 q | B \rangle|_{\text{WA}}  \nonumber \\[0.1cm]
&=&
\underbrace{\langle X|\bar s \Gamma_2 q|0\rangle \langle \gamma^*(\rho)|\bar q \Gamma_1 b | B \rangle}_{\text{initial state radiation (ISR)}}
+ \underbrace{\langle X \gamma^*(\rho) |\bar s \Gamma_2 q|0\rangle \langle 0|\bar q \Gamma_1 b | B \rangle}_{\text{final state radiation (FSR)}} +{\cal O}(\alpha_s) \;.
\label{eq:wa-factorisation}
\end{eqnarray}
We shall call the first and second term initial (ISR) and final state radiation (FSR) 
respectively.
The computation of these two contributions is performed, as previously stated, using LCSR and further details are deferred to section \ref{sec:lcsr}.
The computation is valid as long as $q^2$ is away from partonic and hadronic thresholds.  This means that the $\rho,\omega$-resonance region has to be treated with care and that the computation is valid say $1$-$2\GeV^2$ below the $J/\Psi$-resonance region. 
For $B \to V \gamma$ (partial) effects of the $\rho,\omega$-mesons are included
into the photon DA \cite{Khodjamirian:1995uc,Ali:1995uy,Ball:2002ps}. For $q^2 > 1 \,\GeV^2$ the corresponding leading twist effects effects are included in the quark condensate contributions 
and appear as  $\vev{\bar q q}/q^2$ in the results.  We refrain from using our computations between $0$ and $1\GeV^2$.
A salient feature  due to the resonance region is the appearance of an imaginary part.
In the partonic computation this results from the photon emitted from 
the light valence quark of the $B$-meson corresponding to the cross in the lower left of Fig.\ref{fig:wa-diagram-pert}. In the hadronic picture this corresponds 
to the emission of $\rho,\omega,..$-mesons and conversion into the photon;
the analogous $\calO_8$-case can be found in reference 
\cite{DLZ12} figure 4(left).
In Fig.~\ref{fig:wak8} we plot the $W(q^2)_T$ \eqref{eq:Tdec} matrix element
for $a_8$-contribution (with $a_8 = a_8^u = a_8^d$), which illustrates the point made above.

We restrict ourselves to leading twist-2 and twist-3 DAs c.f. appendix \ref{app:light-DA}.
We neglect 3-particle DAs and quark mass corrections and thus the twist-3 2-particle DAs for the $K^*$ may be written in terms of
the twist-2 DAs via the so-called Wandzura-Wilczek relations \cite{Ball:1998sk}.
We include the first two moments in the Gegenbauer expansion and thus have a total of four input parameters to the $K^*$ in addition to the decay constants.
The $K$ DAs $\phi_P$ and $\phi_\sigma$ are also related \cite{Ball:2004ye}, but here we use the asymptotic forms for reasons discussed in appendix \ref{app:DA}.

In Tab.~\ref{tab:OWA} the selection rules for the operators are depicted. It is apparent 
that selection rules are at work. We refer the reader to section \ref{sec:WAsel} where 
the WA selection rules in the factorisation, c.f. Eq.~\eqref{eq:wa-factorisation} approximation are discussed in some detail.

At this point, we wish to briefly discuss the issue of gauge invariance (GI) in the factorisation approximation \eqref{eq:wa-factorisation}.
At the level of the amplitude \eqref{eq:wa-factorisation} electromagnetic GI,
that is invariance under  $\epsilon^*(q)_\mu \to \epsilon^*(q)_\mu + q_\mu$, implies:
\begin{equation} 
\label{eq:WIem}
q_\rho {\cal A}^{*\rho}|_{\text{WA}} = 0 \;.
\end{equation}  
When the mesons are neutral this equation is satisfied for  
ISR and FSR separately.  
When the mesons are charged  the two terms cancel each other when they are treated at the same level of approximation.  
From Tab.~\ref{tab:OWA} we infer that for the $K$ and $K^*$ this is the case 
for ${\cal O}_{4,8}$ and ${\cal O}_{5,6}$ respectively.  
For the current-current operators $ {\cal O}_{5-8}$ the final state radiation is, by virtue of 
the (axial)vector WI, equal to a contact term. This was implicitly used 
in \cite{Khodjamirian:1995uc,Ali:1995uy} and analysed in more clarity and detail in 
\cite{Khodjamirian:2001ga}.  In the case ${\cal O}_{4}$ for the $K$ there is ISR and FSR 
and no Ward identity (WI) at hand which complicates the issue.  More precisely this  necessitates 
the same approximation scheme be used for both ISR and FSR
as discussed and illustrated at length in  appendix \ref{app:wa-gauge-invariance}.

\subsection{Light cone sum rules}
\label{sec:lcsr}

We calculate initial state terms using the technique known as \textit{light cone sum rules}
\cite{Balitsky:1989ry,Colangelo:2000dp} which  originates from QCD sum rules
\cite{Shifman:1978bx,Shifman:1998rb} and the light-cone operator product expansion (LC-OPE).
We extract the matrix elements required in \eqref{eq:wa-factorisation} from the correlation function
\begin{equation}
\Pi(q^2,p_B^2) = i\int d^4x e^{-ip_B\cdot x} \langle \gamma^*(q) M(p)|T\{J_B(x) \mathcal O(0)\}|0\rangle \;,
\label{eq:wa-matrix-element}
\end{equation}
where 
\begin{equation}
J_B = im_b\bar b\gamma_5 q, \qquad \langle\bar B|J_B(0)|0\rangle = m_B^2 f_B  \;,
\end{equation}
is the interpolating current for the $B$-meson.
By application of Cauchy's theorem we can express this matrix element as
\begin{equation}
\Pi(q^2,p_B^2) = \frac{m_B^2 f_B}{m_B^2 - p_B^2}\langle \gamma^*(q) M(p)|\mathcal O(0)|B(p_B)\rangle
+ \frac{1}{2\pi i} \oint_{\Gamma_{\mathrm{NP}}} \frac{ds}{s-p_B^2} \Pi(q^2,s)  \;,
\label{eq:disp-hadron}
\end{equation}
where the integration contour $\Gamma_{\mathrm{NP}}$ separates the pole at $p_B^2=m_B^2$ of the desired matrix element
$\langle \gamma^*(q) M(p)|\mathcal O(0)|B(p_B)\rangle$ from all other poles and branch cuts.
The matrix element in \eqref{eq:wa-matrix-element} may also be calculated within the LC-OPE, and after
applying Cauchy's theorem to the LC-OPE result we get that, at large spacelike $p_B^2$:
\begin{equation}
\Pi(q^2,p_B^2) = \frac{1}{2\pi i} \oint_\Gamma \frac{ds}{s-p_B^2} \Pi^{\rm LC-OPE}(q^2,s)  \;,
\label{eq:disp-pert}
\end{equation}
where the contour $\Gamma$ encloses
all poles and branch cuts of this function.
The sum rule is obtained by equating the two representations \eqref{eq:disp-hadron} and \eqref{eq:disp-pert} and making the approximation, known as semi-global quark-hadron duality,
 $\Pi(q^2,s) = \Pi^{\rm LC-OPE}(q^2,s)$ under the integral in \eqref{eq:disp-hadron}.
A Borel transformation is also applied to reduce the sensitivity to the duality threshold,
which in this case only requires the relation
\begin{equation}
B_{p_B^2\to M^2}\left[\frac{1}{x-p_B^2}\right] = \frac{e^{-x/M^2}}{M^2} \;,
\end{equation}
yielding the final form of the sum rule
\begin{equation}
\label{eq:final_SR}
\begin{split}
\langle \gamma^*(q) M(p)|\mathcal O(0)|B(p_B)\rangle &= \frac{1}{f_B m_B^2}\frac{1}{2\pi i}\int_{\Gamma\backslash\Gamma_{\mathrm{NP}}}ds \exp\left(\frac{m_B^2-s}{M^2}\right) \Pi_P(q^2,s) \\
&\equiv \frac{1}{f_B m_B^2}\int_{\text{cut}}^{s_0} ds \exp\left(\frac{m_B^2-s}{M^2}\right) \rho(q^2,s) \;,
\end{split}
\end{equation}
where $\Gamma\backslash\Gamma_{\mathrm{NP}}$ is the difference between the integration contours in
\eqref{eq:disp-hadron} and \eqref{eq:disp-pert}.
The resulting contour will lie along either side of the real line and thus the final sum rule may be expressed
in terms of an integral over the density function $\rho$ on the real line from the lowest perturbative
state mass ($m_b^2$ here) to the duality threshold $s_0$.
In the full theory the lowest lying multi-particle state coupling to the current $J_B$
occurs at $(m_B+2m_\pi)^2$, and $s_0$ is an effective parameter which is adjusted so that the continuum contribution matches that of QCD.
In practice this means that one expects 
$\sim(m_B+2m_\pi)^2 \simeq 30.9 \GeV^2  < s_0   < (m_B+m_\rho)^2 \simeq 36.6 \GeV^2 $ with $s_0$ somewhat closer to the upper boundary as the other case is $\alpha_s$ suppressed.

\subsection{WA results}
\label{sec:WAresults}

Due to our choice of basis for the four quark operators,
it is convenient to present our results for the $K^*$ in the following linear combinations, 
which is basically the helicity basis, 
\begin{align}
W^q_V(q^2) &= W^q_1(q^2) = \frac{1}{\sqrt{2}}(W_-(q^2)+W_+(q^2))  \;, \nonumber \\
W^q_A(q^2) &= c W^q_2(q^2) = \frac{1}{\sqrt{2}}(W_-(q^2)-W_+(q^2))  \;, \nonumber \\
W^q_0(q^2) &= a W^q_2(q^2) + b  W^q_3(q^2)   \;,
\end{align}
where $V$ and $A$ are the PC and PV transverse decay modes,
and the constants $a,b,c$ are defined in (\ref{eq:Btrafo},\ref{eq:abc}).
The matrix elements  $W_\iota^q(q^2)$, with $\iota \in \{T,V,A,0\}$, are decomposed as follows
\begin{equation}
W^q_\iota(q^2) = \sum_{j=1}^{10} a^q_j\left[
F^q_{j,\iota}(q^2) +  I^q_{j,\iota}(q^2) \right]  \;.
\end{equation}
The functions $I$ and $F$ stand for ISR and FSR and are further parametrised as
\begin{align}
I^q_{j,\iota}(q^2) =&  \frac{1}{f_B m_B^2}\left(\vev{\bar qq}\exp\left(\frac{m_B^2-m_b^2}{M^2_{\rm WA}}\right)V^q_{j,\iota}(q^2) + \int_{m_b^2}^{s_0} ds\, \exp\left(\frac{m_B^2-s}{M^2_{\rm WA}}\right)\rho^q_{j,\iota}(q^2,s) \right)   \;, \nonumber \\[0.1cm]
F^q_{j,i}(q^2) =& f_{K^*}^\perp f_B \left(\frac{m_B}{m_b}\right)^2\int_0^1 f^q_{j,i}(q^2,u) \; du   \;,
\end{align}
for the $K^*$-meson with $i\in\{V,A,0\}$ and 
\begin{align}
F^q_{j,T}(q^2) =& \mu_K^2 f_B^{\mathrm{wti}} \left(\frac{m_B}{m_b}\right)^2 \int_0^1 f^{q}_{j,T}(q^2,u) \; du \;,
\label{eq:wa-if-split}
\end{align}
for the $K$-meson.
We take the Borel parameter $M^2_{\rm WA}=9(2)\GeV$ to be the same for all WA processes, although this is not strictly necessary since
in principle the results should be independent of it within a reasonable range, and a calculation involving higher twist
and/or $\alpha_s$ corrections would usually extremise the result w.r.t. the Borel parameter.
We take the duality threshold as $s_0=35(1)\GeV^2$.
The quoted uncertainty in the Borel parameter and the duality threshold are the ranges over which we vary them to provide an estimate of the error of the LCSR method.
The use of $f_B^{\mathrm{wti}}$ in $F^q_{j,T}(q^2)$ arises because this is the only case where both physical initial and final state radiation contribute,
and thus we must choose $f_B^{\mathrm{wti}}$ as the sum rule approximation of $f_B$ which corresponds to our approximation of the initial state radiation contribution in order to fulfil the Ward identity:
\begin{equation}
f_B^{\mathrm{wti}} = \frac{m_b^2}{f_B m_B^4}\left[\frac{3}{8\pi^2}\int_{m_b^2}^{s_0} \exp\left(\frac{m_B^2-s}{M^2_{\rm WA}}\right) \frac{(s-m_b^2)^2}{s} ds - m_b\vev{\bar qq}\exp\left(\frac{m_B^2-m_b^2}{M^2_{\rm WA}}\right)\right] \;.
\end{equation}
This procedure is discussed further in appendix \ref{app:wa-gauge-invariance}.
The occurrence of $f_B$ in $I^q_{j,\iota}$ is evaluated using the leading order sum rule including the $\vev{\bar qq}$ and $\vev{\bar qGq}$ condensates \cite{Aliev:1983ra}
\begin{dmath}
(m_B^2 f_B)^2 = m_b^2 \exp\left(\frac{m_B^2-m_b^2}{M^2_{f_B}}\right)
\left(\frac{3}{8 \pi^2}\int_{m_b^2}^{s_0}  \exp\left(\frac{m_b^2-s}{M^2_{f_B}}\right)\frac{(s-m_b^2)^2}{s} ds
-m_b\vev{\bar qq}_\mu - \frac{m_b}{2 M^2_{f_B}} \left(1 - \frac{m_b^2}{2 M^2_{f_B}}\right) \vev{\bar qGq}_\mu\right)  \;,
\end{dmath}
where $M^2_{f_B}=5.0(5)\GeV$ is used.
Quark condensates are taken at $\mu=1\GeV$ to be $\vev{\bar qq}=(-0.24(1)\GeV)^3$ and $\vev{\bar qGq}=(0.8(1) \GeV)^2 \vev{\bar qq}$ which are the same values as in \cite{DLZ12}.
The occurrence of $f_B$ in $F^q_{j,i}$ is taken from lattice data as $f_B=191(5)\MeV$ \cite{Bazavov:2011aa,Na:2012kp}.
Formulae for DAs of the external light mesons are given in appendix \ref{app:light-DA}.
Formulae for all functions appearing on the RHS of \eqref{eq:wa-if-split} are given in appendix \ref{app:wa-formulae}.

\subsection{WA at $q^2 = 0$ --- photon DA replaces some $\vev{\bar qq}$-contributions}
\label{sec:WAq2eq0}

\begin{figure}
\includegraphics[width=\textwidth]{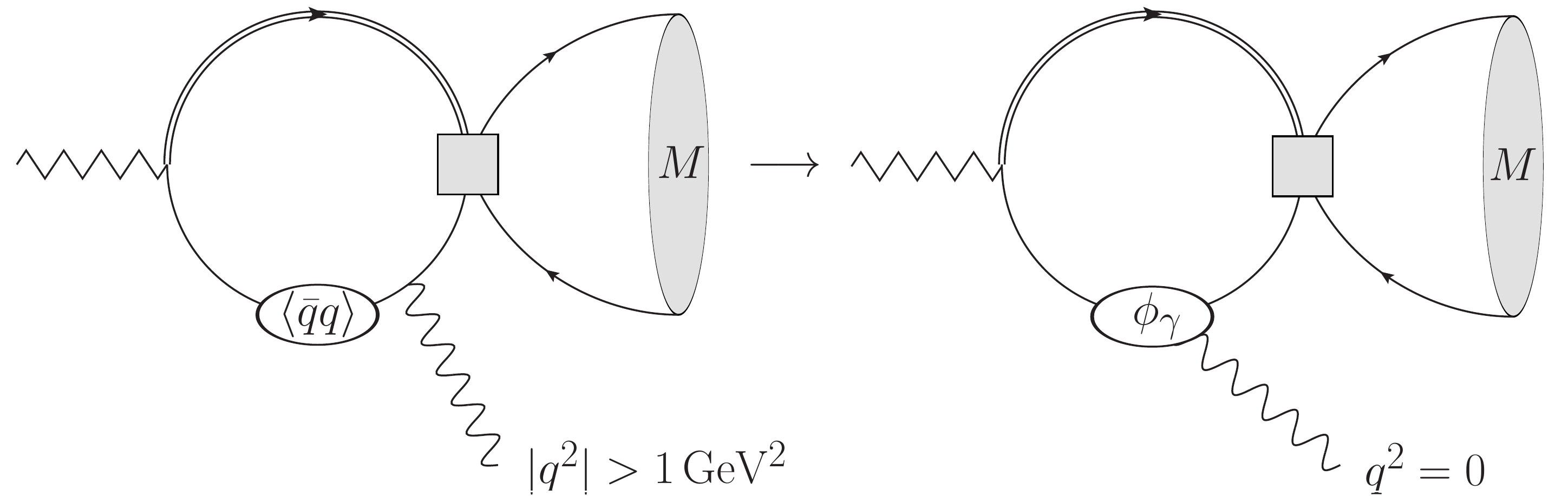}
\caption{\small Quark condensate contribution to be replaced by the  photon DA contribution 
for $q^2 =0$ case i.e. $B \to V \gamma$. The important point to realise is that both diagrams are gauge variant and produce, together with the other diagram in Fig.~\ref{fig:wa-diagram-ope}(right) a fully gauge invariant result.}
\label{fig:wa-diagram-lcope}
\end{figure}

The local OPE for $q^2 \neq 0$, used in diagrams like Fig.~\ref{fig:wa-diagram-lcope}(left) 
for the light quark propagator originating from the $J_B$-current, results in terms
like $Q_q\vev{\bar qq}/q^2$ c.f. \eqref{eq:wa-if-split} which cannot be a good description at 
$q^2 =0$. 
The resolution to this apparent paradox is to replace the this term by the photon DA\footnote{One might also pose the problem the other way around, starting from the photon DA at $q^2=0$ and asking how the latter is to be modified when $q^2 > 0$. The primary effect can be covered by the quark condensate contribution.} as sketched in Fig.~\ref{fig:wa-diagram-lcope}. 

This type of computation has been completed in \cite{Khodjamirian:1995uc,Ali:1995uy} 
for the vector-current operators ${\cal O}^{{\rm WA}}_{5-8}$. 
Our calculation, essentially, extends this to the complete four quark operator basis \eqref{eq:OWA}.  Aspects of GI and contact terms as well as 
a difference in the $(Q_b-Q_q) \vev{\bar qq}$ contribution with reference \cite{Khodjamirian:1995uc} are discussed in 
appendix \ref{app:qq-photonDA}. We note that the latter are small and have not been 
included in \cite{Ali:1995uy}.

Our results are given, such that $I^q_{j,i}(q^2) \to I^q_{j,i}|_\gamma$ in \eqref{eq:wa-if-split},
\begin{equation}
I^q_{j,i}|_\gamma = \frac{1}{f_B m_B^2}\left(\vev{\bar qq}\exp\left(\frac{m_B^2-m_b^2}{M^2_{\rm WA}}\right)V^{q,\gamma}_{j,i} + \int_{m_b^2}^{s_0} ds\, \exp\left(\frac{m_B^2-s}{M^2_{\rm WA}}\right)\rho^{q,\gamma}_{j,i}(s) \right)  \;, \\[0.1cm]
\label{eq:wa-if-split-0}
\end{equation}
where we re-use our result from $q^2\neq 0$ for the density via
\begin{equation}
\rho^{q,\gamma}_{j,i}(s) = \rho^{q}_{j,i}(0,s) + \vev{\bar qq} \widetilde{\rho}^{q,\gamma}_{j,i}(s)  \;,
\end{equation}
and $V^{q,\gamma}_{j,i}$ and $\widetilde{\rho}^{q,\gamma}_{j,i}(s)$ are given in appendix \ref{app:q2eq0}.

\section{Quark loop spectator scattering (QLSS)}
\label{sec:quark-loop-spectator}

\begin{figure}
\center
\includegraphics[width=0.3\textwidth]{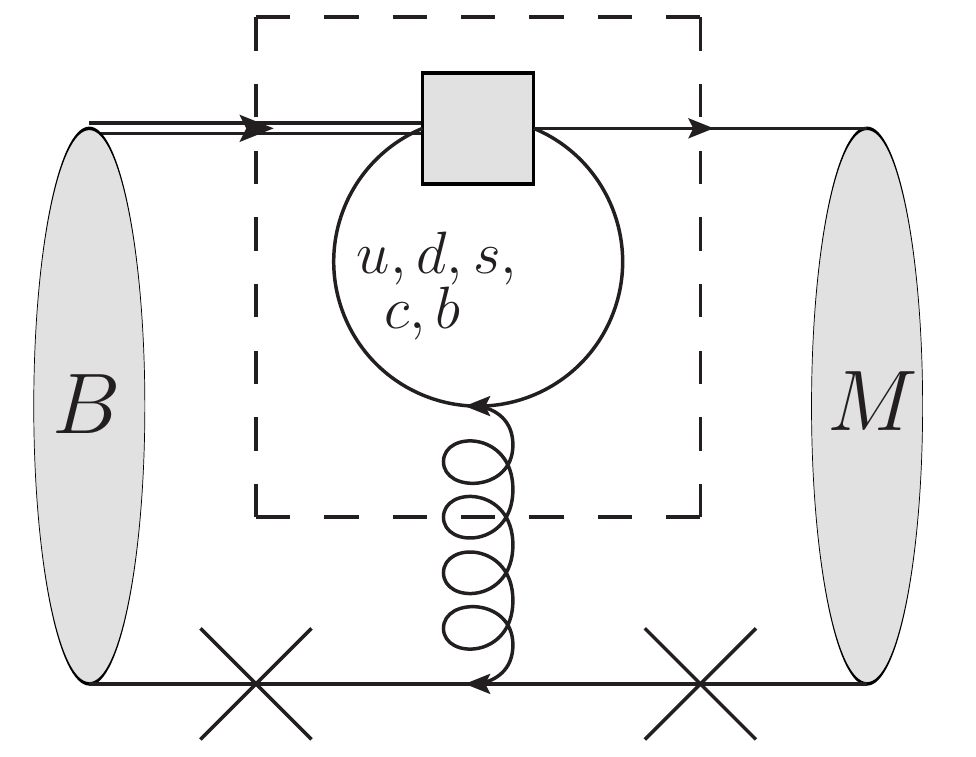}
\caption{\small Hard gluon scattering through a charm loop.
Isospin symmetry violating photon insertions are indicated by crosses.
The segment in the dashed box is computed first without the DAs attached to simplify the calculation.}
\label{fig:quark-loop-spectator}
\end{figure}

The QLSS topology parallels the ${\cal O}_8$-contribution
as can be seen from Fig.~\ref{fig:isoSM}.  
We would expect an LCSR computation of this contribution to include some long distance (LD) contributions, in analogy 
with the ${\cal O}_8$ contribution (intermediate multiparticle states with quantum numbers $(\bar s q)_{J^P = 0^\pm}$ c.f. Fig.~ 4(left) in \cite{DLZ12}. 
On a computational level though spectator scattering differs from the analogous ${\cal O}_8$-computation by the additional nested quark loop which makes the computation rather involved; especially in view of the fact that we further expect a non-trivial analytic structure including anomalous thresholds \cite{DLZ12}.
Thus the evaluation of this contribution with LCSR is beyond the scope of this paper.
We therefore resort to QCDF where it would seem that these LD contributions are, at least 
at leading order in ${\cal O}(\alpha_s)$, absent. 
In QCDF QLSS has been computed previously for the $B\to K^*l^+l^-$ \cite{Feldmann:2002iw} and for the $B\to K^*\gamma$ case in \cite{Kagan:2001zk}. 
We extend these computations by including a complete basis of four quark operators.
Elements of GI are discussed in appendix \ref{app:QL-GI}. 

\subsection{Complete QLSS-basis of dimension $6$ operators at $\calO(\alpha_s)$} 

We now turn to the discussion of the relevant operators contributing to 
QLSS.  The vectorial coupling of the gluon in Fig.\ref{fig:quark-loop-spectator} 
imposes efficient selection rules such that only four out of the ten operator types, 
as listed in Eq.~\eqref{eq:OWA}, can contribute:
\begin{alignat}{2}
\label{eq:OQLSS}
Q^{4f}_{1 L(R) } \equiv \frac{1}{4}\bar  f  \lambda^a \gamma_\mu f\,\bar s_{L(R)}  \lambda^a \gamma^\mu  b  \;, \quad 
Q^{4f}_{2 L(R) } \equiv \frac{1}{4}\bar  f  \lambda^a \sigma_{\mu\nu} f\,\bar s_{L(R)}  \lambda^a \sigma_{\mu\nu}  b \;.
\end{alignat}
Quark  flavours and colour matrices are arranged, differently  from  Eq.~\eqref{eq:OWA},
in a way that is convenient for the QLSS computation. 
 Since we shall set the light quark masses 
to zero the light flavour $u,d,s$ are effectively degenerate and it proves economic to 
introduce the $SU(3)_F$ singlet operator 
\begin{equation}
Q^{4SU(3)_F}_{xL(R)} \equiv  (Q^{4u}_{xL(R)}+ Q^{4d}_{xL(R)} + Q^{4s}_{xL(R)}) \;, \quad x = 1,2 \;.
\end{equation}
Finally, the relevant effective Hamiltionian for QLSS becomes,
\begin{equation}
{\cal H}^{\rm QLSS} = -\frac{G_F}{\sqrt{2}} \lambda_t  \sum_{x,\chi,f} s_{x\chi}^f  Q^{4f}_{x\chi}    \;,  \quad x=1,2\;, \;\; \chi = L,R \;, \;\; f=SU(3)_F,c,b  \;,
\label{eq:qlss-hamiltonian}
\end{equation}
with $s^f_{x\chi}$ being  WCs. 
The somewhat baroque
sum over the indices $x,\chi,f$, which amounts to twelve operators at this stage, will prove economic in the end.
We would like to stress that this basis 
in linearly independent, though not orthogonal, to the WA basis \eqref{eq:OWA}.  This is the case as in WA 
the flavour $f$ is, modulo a few exceptions, fixed by the spectator flavour and since in QLSS $f$ is either $c,b$ or the sum of light flavours linear independence follows.

\subsection{QLSS results}
\label{sec:QLSS}

The computation of QLSS may be broken up into two stages:
first, the intermediate process shown in the dashed box of Fig.~\ref{fig:quark-loop-spectator} is computed,
and second, the results of this computation are combined with the remainder of the diagram Fig.~\ref{fig:quark-loop-spectator}.

To start, the intermediate process can be written, by virtue of Lorentz-covariance, as
\begin{equation}
{\emph a}^\mu = \matel{s g(r,\mu)}{H_{\rm eff}}{b} = \sum_{i=L,R} \left[K_{1,i}^\mu F_{1,i}(r^2) + K_{2,i}^\mu F_{2,i}(r^2) \right]  \;,
\end{equation}
where $r$ is the gluon momentum and $\mu$ is the gluon polarisation index.
The two tensor structures
\begin{align}
K_{1,(L,R)}^\mu &= \frac{r^\mu \slashed r - r^2\gamma^\mu}{r^2} P_{L,R}  \;, &
K_{2,(L,R)}^\mu &= \frac{r^\mu - \slashed r\gamma^\mu}{r^2} P_{L,R} \;,
\end{align}
are the only ones allowed by gauge invariance $r \cdot {\emph a} = 0$. 

In our parametrisation \eqref{eq:qlss-hamiltonian}, the functions $F_{x,\chi}$ are given by
\begin{equation}
F_{x,\chi} = s_{x,\chi}^{SU(3)} H_x(s,0)  + s_{x,\chi}^{c} H_x(s,m_c)  + s_{x,\chi}^{b} H_x(s,b) \;,
\end{equation}
and it is clear that $F_2 \sim m_f$, where $f$ is the flavour of the quark running in the loop, by virtue of dimensional analysis.
This means that $s_{2L(R)}^{SU(3)}$ is heavily suppressed and not present in our approximation where we set the light quark masses to zero
and so only ten of the twelve operators in \eqref{eq:qlss-hamiltonian} effectively contribute.
The functions $H_x$ result from loop integrals for the vector and tensor currents and are given by
\begin{align}
H_1(s,m) &= -\frac{1}{96\pi^2}\left[9h(s,m) + 4\right]   \;,  \nonumber \\
H_2(s,m) &= -\frac{m}{4\pi^2} B_0(s,m^2,m^2)  \;,
\end{align}
where the function $h(s,m_q)$ is the vacuum polarisation (this form from \cite{Feldmann:2002iw}), with  $z\equiv\frac{4m_q^2}{s}$,
\begin{align}
h(s,m_q) &= -\frac{4}{9}\left(\log\frac{m_q^2}{\mu^2} - \frac{2}{3} - z\right) - \frac{4}{9}(2+z)\sqrt{|z-1|}
\begin{cases}
\displaystyle\arctan\frac{1}{\sqrt{z-1}} & z > 1 \\[1em]
\displaystyle\log\frac{1+\sqrt{1-z}}{\sqrt{z}} - \frac{i\pi}{2} & z \leq 1 
\end{cases} \nonumber \\
B_0(s,m_q^2,m_q^2) &= 2 - \log\frac{m_q^2}{\mu^2} + 2\frac{\frac{9}{4}h(s,m_q) + \log\frac{m_q^2}{\mu^2} - \frac{2}{3} - z}{2+z} \;,
\end{align}
which can be found in any standard textbook on quantum field theory.

Before quoting the full result let us detail a few points of the computation.
The $B$-meson DA used is \cite{Beneke:2000wa,Beneke:2000ry}
\begin{equation}
\matel{0}{\bar q_\alpha(x) [x,0] b_\beta(0)}{B} = \frac{-i f_B m_B}{4}\int_0^\infty dl_+ e^{-il_+ x_-/2}
\left[\frac{1+\slashed v}{2}\left\{\phi_+(l_+)\slashed n_+ + \phi_-(l_+)\slashed n_-\right\}\gamma_5\right]_{\beta\alpha}
\label{eq:b-meson-da}
\end{equation}
Definitions of the vectors and various terms involved along with a more complete version of this formula are given in appendix \ref{app:B-DA},
however this version of the formula contains all terms involved in our computation of the diagram in Fig.~\ref{fig:quark-loop-spectator}.

To this end let us mention that 
corrections of ${\cal O}(q^2/m_B^2)$ which come from neglecting the $l_-$-direction of the light quark in the $B$-meson DA and aspects of the $B$-meson DA including 
the transverse derivative 
are discussed in appendix \ref{app:O(q2/mB2)}.
Since we have restricted ourselves to $q^2<4m_c^2$,
using the $B$-meson light-cone DA \eqref{eq:b-meson-da} seems reasonable. 
A more interesting question is to what extent the shape of the $q^2$-distribution 
is accurate or trustworthy. 

We write the results directly in the helicity basis, $S_{\pm,0}$, whose transformation 
to the $\calT_{1,2,3}$ basis is given in \eqref{eq:Btrafo}, for the $K^*$-meson:
\begin{align}
\begin{split}
S_-^q(q^2) =& \sqrt{2} Q_q \frac{C_F}{N_c} \frac{16\pi^3\alpha_s f_B m_B}{m_b}
\int_0^1 du\, \left(F_{1,L}(\bar u m_B^2 + uq^2) - \frac{1}{m_B} F_{2,R}(\bar u m_B^2 + uq^2)\right) \\
& \times\left[ \frac{f_{K^*}^\perp\phi_\perp(u)}{\bar u m_B^2 + uq^2}
- \frac{f_{K^*} m_{K^*}}{2\lambda_+(q^2)(m_B^2-q^2)}\left(g_\perp^{(v)}(u) - \frac{g_\perp^{(a)'}(u)}{4}\right)\right] \\
& - \frac{F_{2,R}(\bar u m_B^2 + uq^2)}{m_B}\Bigg[
\frac{f_{K^*}^\perp\phi_\perp(u) u (m_B^2-q^2)}{2(\bar um_B^2 + uq^2)^2} - \frac{f_{K^*} m_{K^*}}{2\lambda_+(q^2)(m_B^2-q^2)}\frac{g^{(a)}_\perp(u)}{4\bar u}
\Bigg] \;,
\label{eq:charm-loop-perp}
\end{split}    \\
S_+^q(q^2) =& \; (L \leftrightarrow R)  \;,  \\
\begin{split}
\label{eq:S0}
d \cdot  S^{q,V}_0(q^2) &= - \, Q_q \frac{C_F}{N_c} \frac{32\pi^3\alpha_s f_B m_B}{m_b}
\frac{f_{K^*} m_{K^*}}{\lambda_-(q^2)(m_B^2-q^2)}
\int_0^1 du\,\phi_\parallel(u) \\
& \times \left[F_{1,A}(\bar u m_B^2 + uq^2) + \frac{m_B}{\bar u(m_B^2-q^2)}F_{2,A}(\bar u m_B^2 + uq^2)\right]  \;,
\end{split} 
\end{align}
with $d  \equiv -\frac{\sqrt{2} m_B m_V}{\sqrt{q^2} E}$. 
For the $K$-meson we get
\begin{align}
\begin{split}
\label{eq:ST}
S^q_T(q^2) =& -Q_q\frac{C_F}{N_c} \frac{(m_B+m_K) 16 \pi^3 \alpha_s f_B f_K}{m_B m_b \lambda_-(q^2)}
\int_0^1 du\,\phi_K(u) \\
& \times \left[F_{1,V}(\bar u m_B^2 + uq^2) - \frac{m_B}{\bar u(m_B^2-q^2)}F_{2,V}(\bar u m_B^2 + uq^2)\right] \;,
\end{split}
\end{align}
where we have used $F_{1,V(A)}(s) \equiv F_{1,R}(s) \pm F_{1,L}(s)$ for the sake of compact notation.
At this point we would like to specify some details of the computation. 
There are two types of terms that appear:
\begin{eqnarray}
X_1  &=& \int_0^\infty  dl_+ \phi_\pm(l_+) H_1(l_+)  \;, \\[0.1cm] 
X_2  &=& \int_0^\infty  dl_+ \phi_\pm(l_+) \frac{H_2(l_+)}{l_+ - q^2/m_B  - i\epsilon}  \;,
\end{eqnarray}
where the kernels $H(l_+)$ are smooth and the denominator in the second term
originates from a propagator which in turn cancels for the type one term.
The following recipe is applied:
\begin{eqnarray}
X_1  &\to&  \left( \int_0^\infty  dl_+ \phi_\pm(l_+) \right) H_1(0) = \text{constant} \cdot H_1(0)   \;,
\\[0.1cm] 
X_2  &\to&  \left( \int_0^\infty  dl_+ \frac{\phi_\pm(l_+)}{l_+ - q^2/m_B  - i\epsilon}  \right)  H_2(\bar l_+) = 
\frac{H_2(\bar l_+)}{\lambda_\pm(q^2)} \;, 
\end{eqnarray}
where $\bar l_+ \equiv q^2/m_B$, $1/\lambda_\pm(q^2)$ is further detailed in appendix 
\ref{app:B-DA}.
The term in \eqref{eq:charm-loop-perp} proportional to $\phi_\perp(u)$  is of the first type and all others are of the second type as can be seen in Eqs.~ (\ref{eq:charm-loop-perp}-\ref{eq:ST}).
The equation \eqref{eq:charm-loop-perp} disagrees with \cite[eq. (24)]{Feldmann:2002iw} by a factor of 2 in the $g^{(v,a)}_\perp$
term but agrees with the result in \cite{Kagan:2001zk} in the limit $q^2 \to 0$.
The definitions of the $B$-meson moment functions $\lambda_\pm(q^2)$
and the $K$ and $K^*$ DA functions are given in appendix \ref{app:light-DA}.
An important remark is that the Wandzura-Wilczek type equations of motions  ($m_s=0$) \eqref{eq:WW} for the $K^*$ have been used.

We note that we have included the $K^*$ DAs $\phi_\parallel$, $\phi_\perp$ and $g^{(v,a)}_\perp$.
In light of Tab.~~\ref{tab:OWA} it might seem that we should have also included $h^{(s,t)}_\parallel$,
however here we are considering the leading $1/m_B$ term,
so it is the $g^{(v,a)}_\perp$ term which requires justification.
It is included because the $\phi_\perp$ coefficient does not contain a $1/\lambda_\pm$ factor as it might, and therefore
occurs at the same power of $1/m_B$ as $g^{(v,a)}_\perp$.
Since the $\phi_\parallel$ term comes with a $1/\lambda_\pm$ factor, an $h^{(s,t)}_\parallel$ term
would be $\calO(1/m_B)$ suppressed so we neglect it.

It should be noted that there is an endpoint (infrared) divergence proportional to $F_{2,(R,L)}(0)$ in $S_\pm(q^2\to 0)$ arising from the $\bar u\to 0$ integration region which is of the same type as ${\cal O}_8$. 
There are three ways to deal with it: a) only use it for $q^2 > 0$ in which case an IR sensitivity remains; b) employ the cut-off model \cite{Kagan:2001zk,Feldmann:2002iw},
or c) perform a local subtraction to render the QCDF result finite and then use the IR finite result from LCSR \cite{DLZ12} to compensate.
We choose option (c). To get an infrared finite result we write
\begin{equation}
F_{2,i}(r^2) = \left[ F_{2,i}(0) \right]_{\rm LCSR} + \left[ F_{2,i}(r^2) - F_{2,i}(0) \right]_{\rm QCDF}  \;,
\end{equation}
where the contribution of each term in square brackets to $B\to K^{(*)}ll$ is computed as indicated in the subscript.
The LCSR term is therefore a local operator and the $\calO_8$ result \cite{DLZ12} applies,
and the QCDF term vanishes for $r^2\rightarrow 0$ restoring IR finiteness of \eqref{eq:charm-loop-perp} at $q^2=0$.
The slightly inconsistent feature is that the LCSR computation contains LD contributions which are not present at this level of approximation in QCDF as explained at the very beginning of this section.

At last we wish to add some remarks about imaginary parts. 
In the partonic picture the charm quark can go on-shell, whose importance has been emphasised  
in \cite{Kim:2008rz},  as is visible from the formulae. 
Fortunately the momentum that enters the charm loop depends on the momentum fraction 
of the light meson as $\bar u m_B^2 + u q^2$ and is sufficiently smeared out that a partonic 
description seems tolerable.
Conversely the interpretation of such effects in the real hadronic world would be a $D \bar D$-thresholds for 
which is, compared to a single resonance, sufficiently tame to be described by partons 
within our quoted uncertainties. 

\subsection{Projection on SM-basis (QLSS in the SM)}
\label{sec:QLSSinSM}

In the SM using na\"ive dimensional regularisation we have 
\cite{Feldmann:2002iw}
\begin{eqnarray}
\text{SM:} \quad F_{1,L}(s) &=& \frac{3}{32\pi^2}\Bigg[h(s,m_c)\left(-\frac{\lambda_c}{\lambda_t}\CBBL_2 + \CBBL_4 + \CBBL_6\right) + h(s,m_b)(\CBBL_3 + \CBBL_4 + \CBBL_6) \nonumber  \;,  \\
&+& h(s,0)\left(-\frac{\lambda_u}{\lambda_t}\CBBL_2 + \CBBL_3 + 3\CBBL_4 + 3\CBBL_6\right) - \frac{8}{27}(\CBBL_3 - \CBBL_5 - 15\CBBL_6)\Bigg]  \;, \nonumber  \\
F_{1,R}(s) &=& 0   \;, \\
F_{2,R}(s) &=& \frac{m_b}{8\pi^2}(C_8^{\rm eff} - C_8)  \;, \nonumber
 \\
F_{2,L}(s) &=& \frac{m_s}{8\pi^2}(C_8^{\rm eff} - C_8) \to 0 + {\cal O}(m_s)   \;.
\end{eqnarray}
The MFV-symmetry of the SM implies that $F_{1,R} =0$. 
The operators $Q^{4f}_{2\chi}$ are not present in the SM in $d=4$ but give contributions
in the spirit of evanescent operators in naive dimensional regularisation. 
They render the effective WC $C_8^{\rm eff}$ scheme independent \cite{Misiak:1991dj,Ciuchini:1993ks}.
It is worth pointing out that in the SM the charm loop dominates as 
it originates  from $\calO_2^c$ which is proportional to a large WC $C_2\approx 1$ and is not CKM suppressed.
All other WCs are small as can be seen in Tab.~\ref{tbl:wilson-coefficient-example}.

\section{Isospin asymmetries $B \to K^{(*)} \gamma/ ll$}
\label{sec:iso}

We shall first make a few generic remarks on selection rules and related issues in 
subsection \ref{sec:selection}, reflect on the $q^2$-behaviour from various viewpoints 
in subsection \ref{sec:q2-dependence}
and then discuss the isospin asymmetries of $B \to K^{(*)}$ in the SM in section \ref{sec:isospinSM}.
Discussion of isospin asymmetry beyond the SM is deferred to section \ref{sec:isospinBSM}.

\subsection{Generic remarks on selection rules}
\label{sec:selection}

In total there are $32$ operators potentially contributing to the isospin asymmetry at the level
of  ${\cal O}(\alpha_s)$-correction we are considering. Schematically they decompose as follows:
\begin{equation}
\label{eq:32}
32 = 2_{{\cal O}_8^{(')} \eqref{eq:SMbasis}} + 20_{WA \eqref{eq:OWA}} + 10_{\rm QLSS \eqref{eq:OQLSS}}\;,
\end{equation}
where the prime denotes \VpA{} chirality as previously defined. We note that ${\cal O}^{SU(3)}_{2L(R)}$ gives no contribution in the limit of all light quark masses set to zero since 
it is proportional to $m_f$ as pointed out in section \ref{sec:QLSS}.
The number $32$ will be reduced further for the $K$ and $K^*$ below.

We shall discuss below general selection rules for the $K$ in subsection \ref{sec:Ksel} before discussing 
more specific selection rules for WA in the factorisation approximation in subsection 
\ref{sec:WAsel}
and then comment on the (non)-relation between the $K$ and $K^*_\parallel$-amplitude
in subsection \ref{sec:non-relation}.

\subsubsection{Parity \& angular momentum selection rule for $K$}
\label{sec:Ksel}

For the $K$ there is a parity selection rule. We note that
\begin{equation}
B[0^-] \to K[0^-] (\gamma^*[1^-] \to l^+l^-[1^-])  \quad  \Rightarrow \quad  \text{ p-wave; i.e.} \;\; 
l =1 \;, \quad 
\end{equation}
where $[J^P]$ denotes total angular momentum and parity respectively and $l$ is the spatial angular momentum of the decay product. Thus the (left hand side) LHS and (right hand side) RHS of the decay 
have odd parity and the decay is therefore induced 
by parity conserving (PC) operators. This means that operators of the type  
$\bar s\dotsm\gamma_5 q\,\bar q \dotsm b$,  $\bar s\dotsm q\bar q \dotsm \gamma_5  b$,
where $\dotsm$ stand for strings of $\gamma$-matrices not including $\gamma_5$, do not contribute.
This reduces the number of operators for the $K$ (by a factor of two) 
down to $1_{{\cal O}_8}$ and $5_{\rm QLSS}$ using the notation of Eq.~\eqref{eq:32}. 
It seems worthwhile to emphasize that the selection rules are generic. 
In the next subsection we are going to discuss 
WA  in the factorisation approximation \eqref{eq:wa-factorisation}  for which there are additional selection rules.

\subsubsection{WA selection rules in the factorisation approximation}
\label{sec:WAsel}

In the factorisation approximation, automatic at ${\cal O}(\alpha_s^0)$ we are considering for WA, there are more stringent selection rules. 
They come from the fact that the  Dirac traces of the $B$-meson and the $K^*$-meson close separately,
and so $\gamma_5$-matrices cannot be commuted from one end to the other.
Selection rules arising due this effect derive from the matrix element which does not emit the photon,
i.e. the initial state in a FSR diagram and vice-versa.
We call the matrix element imposing the selection rule the local matrix element (LME) since it is the matrix element of an external state coupling to a local operator.

The $O^{\rm WA}_{9,10}$ operators must be considered separately:
in the case of the $K$ they do not contribute in the factorisation approximation since the LME
will be a pseudoscalar coupling to an antisymmetric tensor operator and no such tensor can be formed from a single vector.
On the other hand in the case of the $K^*$ both operators contribute since an antisymmetric tensor $p^{[\mu}\eta^{\nu]}$
is available and by $\epsilon^{\mu\nu\rho\sigma}\sigma_{\rho\sigma}=2i\sigma^{\mu\nu}\gamma_5$
the two different parities are trivially related.
For $O^{\rm WA}_{1\dots8}$, the LME will impose a selection rule since the external state will only couple to a local scalar or vector operator of the correct parity.
This reduces the number of applicable operators by a factor of $2$.
If this is further combined with the global parity constraint for the $K$ case so for $B\to Kll$ only 2 of the 8 operators remain.

Combining these rules we expect $8/2 + 2 = 6$ and $8/2/2+0 = 2$ operators to contribute to WA in 
the factorisation approximation for the $K^*$ and the $K$-respectively.
This is indeed the case as the reader may verify from Tab.~\ref{tab:OWA} or Tab.~\ref{tab:breakdown}. 
With respect to the latter table note that we have not taken into account the 
degeneracy in $q = u,d$ in the previous wording and this is why the numbers are $12$ and $4$ rather than $6$ and $2$.

\begin{table}
\center
\begin{tabular}{l | c |  c l  | c l |  c   }
          & $C^{(')}_8$ & WA & Eq.\eqref{eq:OWA} &  QLSS &  Eq.\eqref{eq:OWA} & 
          total  \\[0.1cm]
          \hline
$K^*$ &  2[1] & 12[3] & $a^q_{2,4,5,6,9,10}$  & 10[3] & all no i=2,f=SU(3)  &  24[7] \\
$K$    &  1[1] & 4[3] & $a^q_{4,8}$                  & 5[3]  &  idem no $\chi=A$ &   10[7]
\end{tabular}
\caption{\small Operators contributing to isospin in $B \to K^{(*)} ll$. In square brackets 
we denote the number of operators that are present in the SM for the respective channel. 
In this counting we neglect $C_8'$ as $C_8'/C_8 \simeq m_s/m_b$. 
}
\label{tab:breakdown}
\end{table}

\subsubsection{On the (non)-relation between $K^*_{\parallel}$ and $K$}
\label{sec:non-relation}

There is some conventional wisdom, throughout the literature, 
that the longitudinal polarisation of the $K^*$ corresponds to that of the $K$.
We shall argue that this is true in the SM at leading twist and falls apart thereafter.

The main points follow from the fact the the longitudinal polarisation can be decomposed 
as follows:
\begin{equation}
\label{eq:eta0_dec}
\eta(0)^\mu  \equiv \eta^\mu_\parallel = 
\frac{p^\mu}{m_{K^*}} + (q^\mu \calO(m_{K^*}) + p^{\mu}  \calO(m_{K^*})) \;.
\end{equation}
Thus at leading twist, $m_{K^*}^2 =0$, we see that $\eta^\mu_\parallel$ and $p^\mu$ play the same r\^ole. Since the former is a pseudo-vector and the latter is a vector we also see 
that they couple to opposite parity: 
 $K$ only couples to PC operators, as pointed out in subsection \ref{sec:Ksel}, and $K^*_\parallel$ only couples
to PV operators. 
Due to the \VmA{}-interactions in the SM there is a link between the corresponding 
WCs and this makes it clear that the statement at the beginning of this subsection is true.

On the other hand it is then also clear that the presence of right-handed currents, i.e. \VpA{}-interactions, invalidates the statement. Furthermore the $ \calO(m_{K^*})$-corrections, by virtue of 
\eqref{eq:eta0_dec},  are going to bring in new structures as well and we can therefore 
not expect the correspondence to hold at twist $4$.  
Examples:
\begin{itemize}
\item \emph{Isospin asymmetry plots  $K$ versus $K^*_{\parallel}$  (Fig.~\ref{fig:1}):} 
We notice that the  $K$ and $K^*_{\parallel}$ are correlated  since the SM, due to the $V$-$A$-interactions, satisfy 
the conditions discussed above. 
The differences, which is essentially a shift of the shape, are due to sizeable next-leading twist effects.
\item \emph{Working example at leading twist:}  From the  formulae in appendix \ref{app:wa-formulae}  one infers that
the parity related $O^{\rm WA}_6$ and $O^{\rm WA}_8$ contributions are (indeed)
proportional to each other ($a_6(K^*_\parallel) \sim a_8(K)$).
\item \emph{Non-working example at non-leading twist:} 
From Tab.~\ref{tab:OWA} we  infer that  $O_4^{\rm WA}$ couples to ISR for the $K$ but the corresponding
PV operator $O_2^{\rm WA}$ does not for the $K^*$.
 \end{itemize}
The latter point deserves some further explanation. Should there be an extension of the 
correspondence from $\phi_{p,\sigma}$ to $K^*$-amplitudes then it would be through the same chirality DA and necessarily involve the chiral-odd DA $\phi_\perp$ and $h_\parallel^{(s,t)}$ in Tab.\ref{tab:OWA}. That this cannot hold can also be seen as follows:
the chiral-odd $K^*$ 2-particle DAs included in our calculation
have independent coefficients to the chiral-even ones.
In the case of the $K$ this is not so: the chiral-odd DAs are fully constrained by equations of motion and arise from
finite quark masses, 3-particle and higher twist DAs and chiral symmetry breaking, the last of which is the only effect we include.

\subsubsection{Implications of  selection rules on twist-expansion hierarchy}
\label{sec:comment}

In practice selection rules such as the ones depicted in Tab.~\ref{tab:OWA} enforce a rethinking  of the matters of the twist expansion.
More precisely we mean that if a large WC  does not contribute at leading twist 
but say only at next leading twist then it does not seem wise to truncate at leading twist. 
Thus in practice this implies that one should  
expand to the twist order such that all sizeable WC, allowed by the fundamental selection rules such as the ones quoted in section \ref{sec:Ksel}, contribute to the amplitude.

The $K$ shall serve as an explicit example. For the latter we see that at leading and next leading twist $a_8^q$ and $a_4^q$ contribute which correspond to 
 $(C_3/N_c+ C_4)$ and $(C_5/N_c+ C_6)$ in the SM \eqref{eq:aSM}.
From the size of the WC \ref {tbl:wilson-coefficient-example}, we infer that the latter could be of importance especially in view of the fact that the next leading twist DA $\phi_{p,\sigma}$ is known to be chirally enhanced which can be inferred from its normalisation \eqref{eq:KDA}. The reader is referred
to Figs.~\ref{fig:1},\ref{fig:2} to convince him or herself of this fact.

\subsection{Generic remarks on $q^2$-dependence}
\label{sec:q2-dependence}
Below we add a few generic remarks on the $q^2$-dependence ranging from 
the validity the LCSR  up to pointing out the dominant contributions. 
The latter are particularly useful 
for understanding the isospin asymmetries within and beyond the SM.

\begin{itemize}
\item \emph{Physical spectrum and approximation ranges:}
the physical spectrum of the decays ranges from  $4m_l^2 < q^2 < (m_B-m_K^{(*)})^2 = 22.9(19.3)\GeV^2$.
In this work we compute the isospin asymmetries at low $q^2$ (large recoil) 
where the LCSR (WA and ${\cal O}_8$) and QCDF (QLSS) results are naturally trustworthy. 
We restrict ourselves to the interval of $[1,8]\, \GeV^2$ whose boundary is limited 
by the nearby  $\omega,\rho,\rho'$ resonances from below and the charmonium resonances, 
commencing at $q^2 = m_{J/\Psi}^2 \simeq (3.1GeV)^2  \simeq 9.6 \GeV^2$, from above. 
While it is plausible to assess effects of the latter close to these regions we consider
it too difficult to asses them locally and thus refrain from doing so. 
\item \emph{Isospin asymmetry in $B\rightarrow K^*ll$ decreases for high $q^2$}:
to understand the possible size of isospin asymmetry for a given WC as a function 
of $q^2$ it is helpful to look at the WCs $C_{9,10}$ and note that: 
\begin{itemize}
\item[a)] They are large as  compared to the other WC (c.f.Tab.\ref{tbl:wilson-coefficient-example}),
partially as a result of a $1/\sin \theta_W^2 \simeq 4$ enhancement, where  $\theta_W$ is the Glashow-Weinberg angle.
\item[b)] We may write the leading terms in the $B\rightarrow Mll$ decay rate as:
\begin{eqnarray}
\label{eq:hasq2}
h_T &\sim& \phantom{\sqrt{q^2}}  [ C_{9,10}^{\rm eff} {\cal O}(1) + C_7^{\rm eff} {\cal O}(1)]  \;, \nonumber \\[0.1cm]  
h_0 &\sim&\sqrt{q^2}[ C_{9,10}^{\rm eff} {\cal O}(1) + C_7^{\rm eff} {\cal O}(1)]  \;,\nonumber \\[0.1cm]
h_\pm &\sim& \phantom{\sqrt{q^2}} [ C_{9,10}^{\rm eff} {\cal O}(q^2/m_B^2) + C_7^{\rm eff} {\cal O}(1)]   \;,
\end{eqnarray}
This behaviour can be inferred from Eqs.~(\ref{eq:Btrafo},\ref{eq:lepton-vector-sym-ff},\ref{eq:lepton-axial-ff}).
Another way to look at it is to realise that $\CBBL_{9,10}$ should never be sensitive to $1/q^2$ in front
of the rate as they are \emph{not} generated by an intermediate photon but through intermediate 
$Z$-bosons and box diagrams.
\end{itemize}
We therefore see that at low $q^2$ in $B\rightarrow K^*ll$ isospin violating terms only compete against $C_7^{\rm eff}$,
but at high $q^2$ they must compete with the much larger $C_{9,10}$ and hence the asymmetry
decreases for large $q^2$.
In $B\rightarrow Kll$ no such argument applies as in $h_T$ $C_7^{\rm eff}$ and $C_{9,10}$ are on equal footing.

\item \emph{High $q^2 \leq (m_B-m_K^{(*)})^2$ region:} In this paper we have not assessed the isospin asymmetry at high $q^2$, that is low recoil, per se. We shall discuss it from two viewpoints which fortunately lead to the same conclusion, namely that the isospin violating effects get smaller for large $q^2$; that is to say the short-distance  form factor contribution with 
small isospin effects are dominant.

\begin{itemize}
\item
\emph{Form factor} contributions (FFCs) in the high-$q^2$ region.  In that region 
the  $C_{7,9,10}$-FFC  are expected to be enhanced
by the presence of the nearby resonance at $q^2 = m_{B_s^*}^2$, which can for be seen from the plots and form factor parametrisations in \cite{Ball:2004rg} whereas no such 
enhancement is present in the isospin violating 
 (IVQE) terms\footnote{This statement is only corrected by UV isospin violating effects in WA at ${\cal O}(\alpha_s)$.}.
Let us add that  IVQE terms, arising from intermediate off-shell photons,
will be enhanced by light resonances at $q^2 = m_\rho^2, m_\omega^2,\dots$,
as can be seen for example in Fig. \ref{fig:wak8}, and also by a heavy resonance $\Upsilon(\bar bb)$. 
Thus in summary the isospin asymmetry is expected to be suppressed both by small WCs
and competing with a resonant isospin-symmetric term, and thus should be very small at high $q^2$.

\item \emph{Low recoil-OPE:}
Some time ago an OPE in $q^2$ and $m_b^2$  
was proposed \cite{Grinstein:2004vb} for the low recoil region, which was implemented 
into phenomenology \cite{Bobeth:2010wg} and reinvestigated from a theoretical viewpoint 
in \cite{Beylich:2011aq}. In this language the FFC come as dimension three matrix elements and IVQE originate from higher dimensional operators (dimension 6 for WA and
dimension 5 for QLSS and $\calO_8$) and are therefore naturally small.
\end{itemize}
\end{itemize}

\subsection{Isospin asymmetries $B \to K^{(*)} \gamma/ ll$ in the SM}
\label{sec:isospinSM}

\begin{figure}
\center
\includegraphics[width=0.40\textwidth]{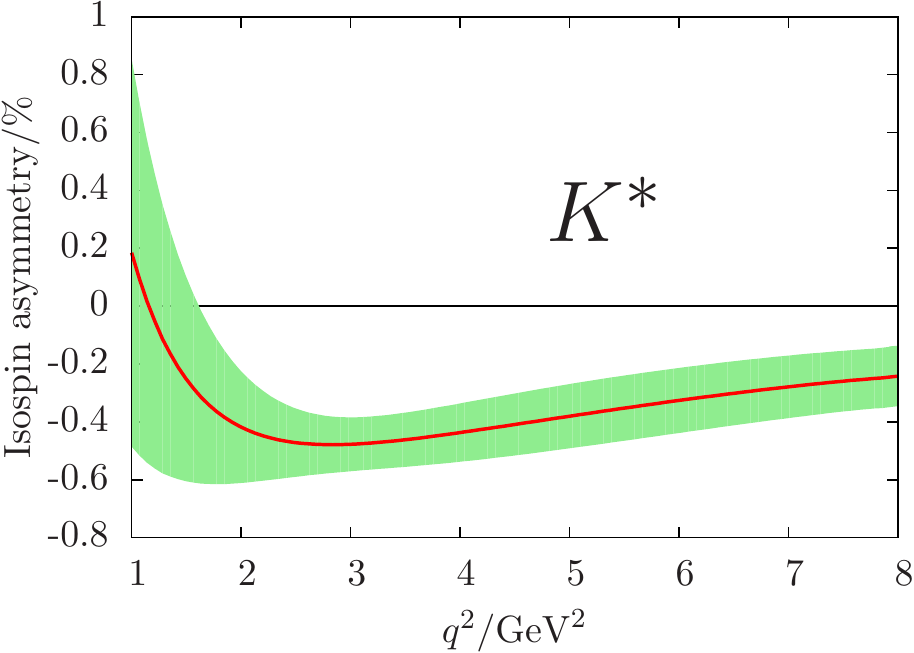}
\includegraphics[width=0.40\textwidth]{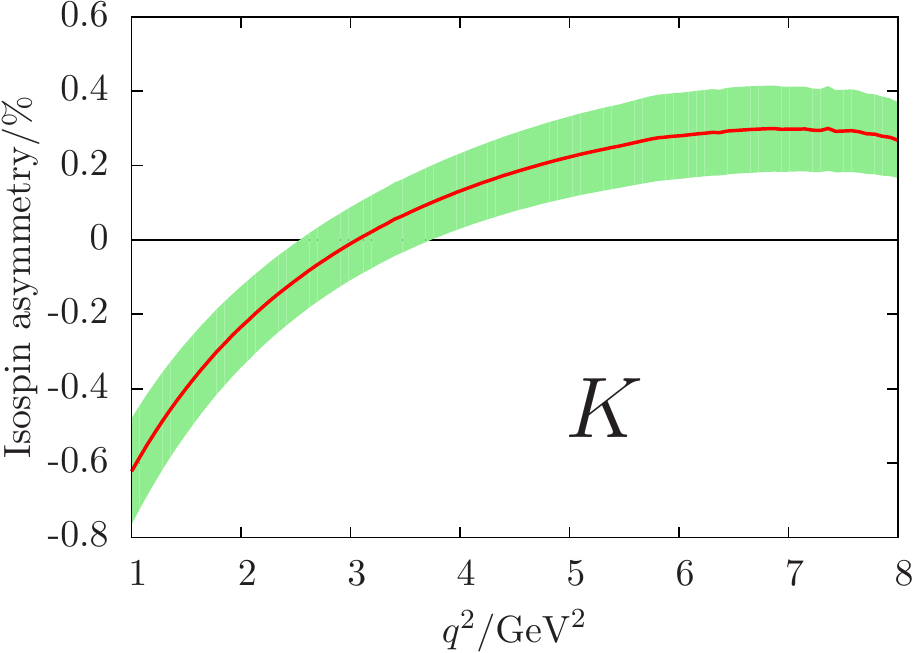}
\includegraphics[width=0.40\textwidth]{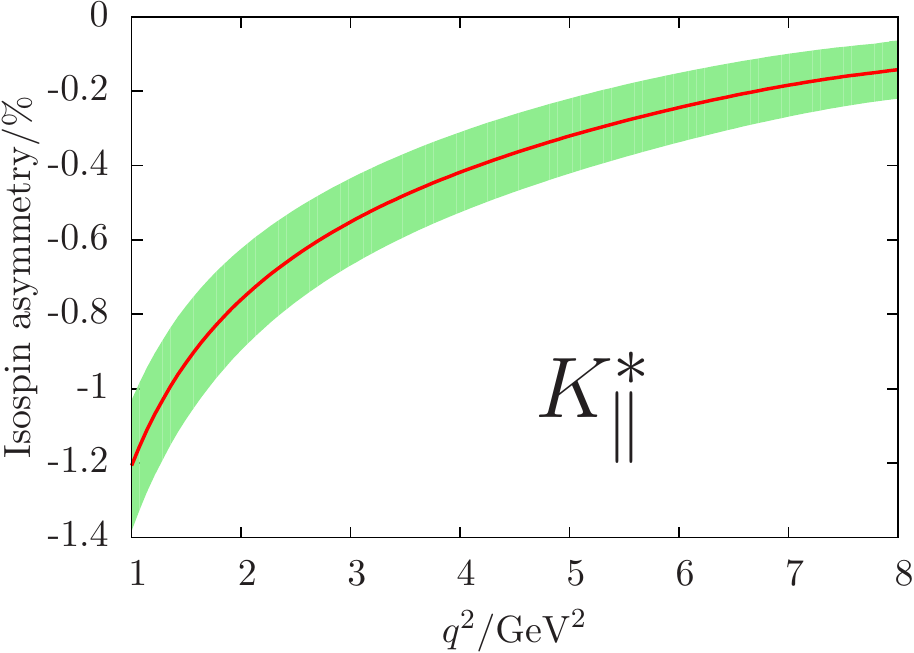} 
\includegraphics[width=0.40\textwidth]{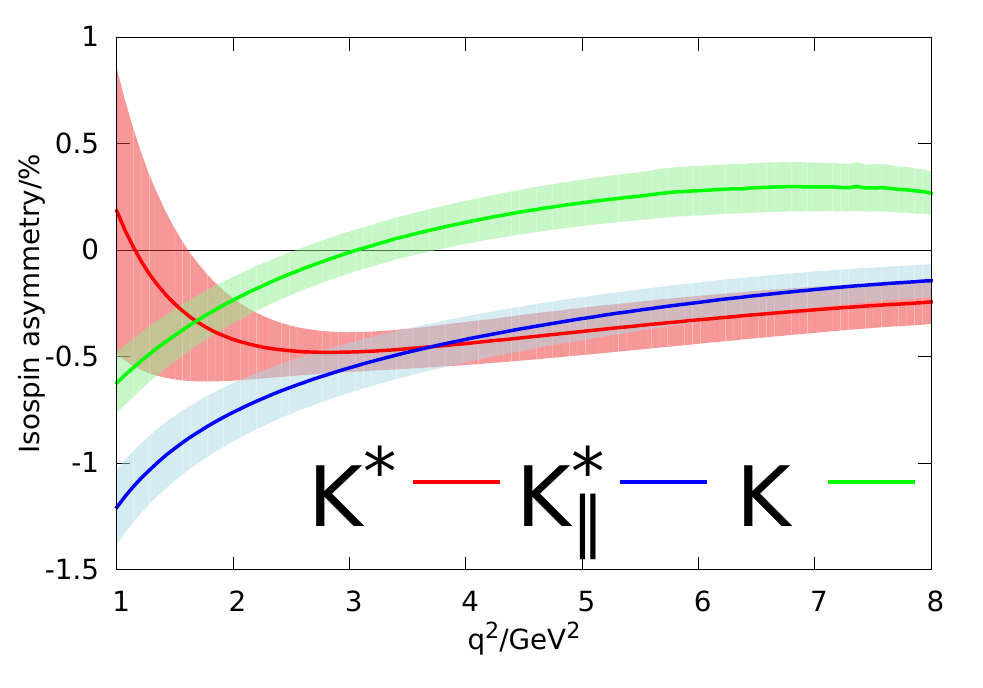}
\includegraphics[width=0.35\textwidth]{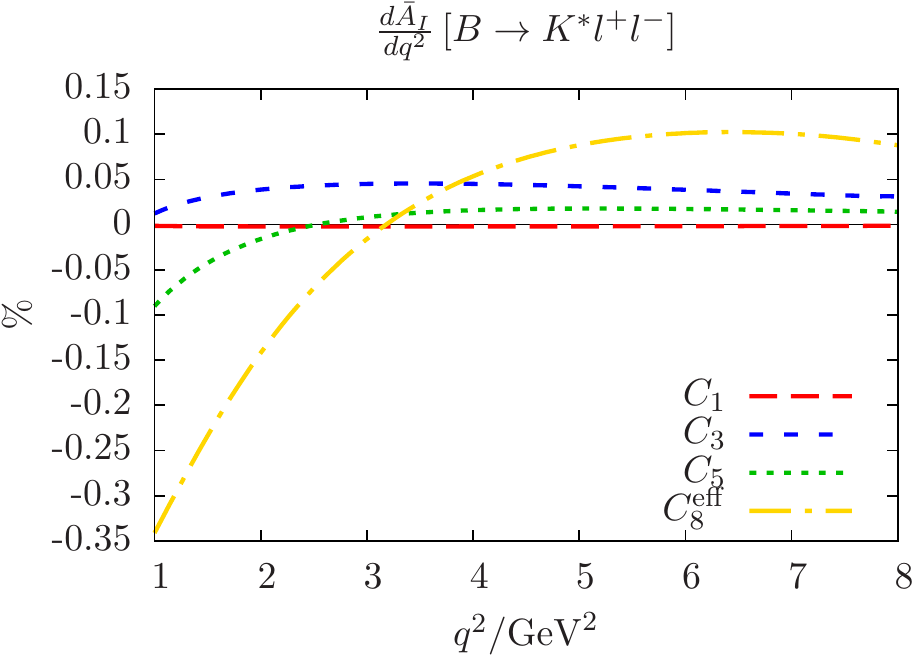}
\includegraphics[width=0.35\textwidth]{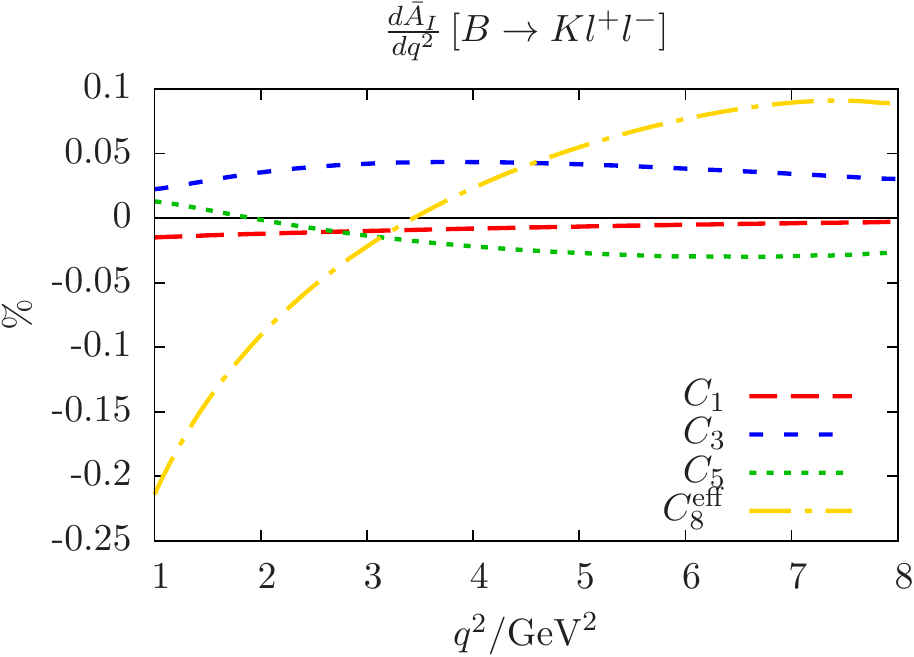}
\includegraphics[width=0.35\textwidth]{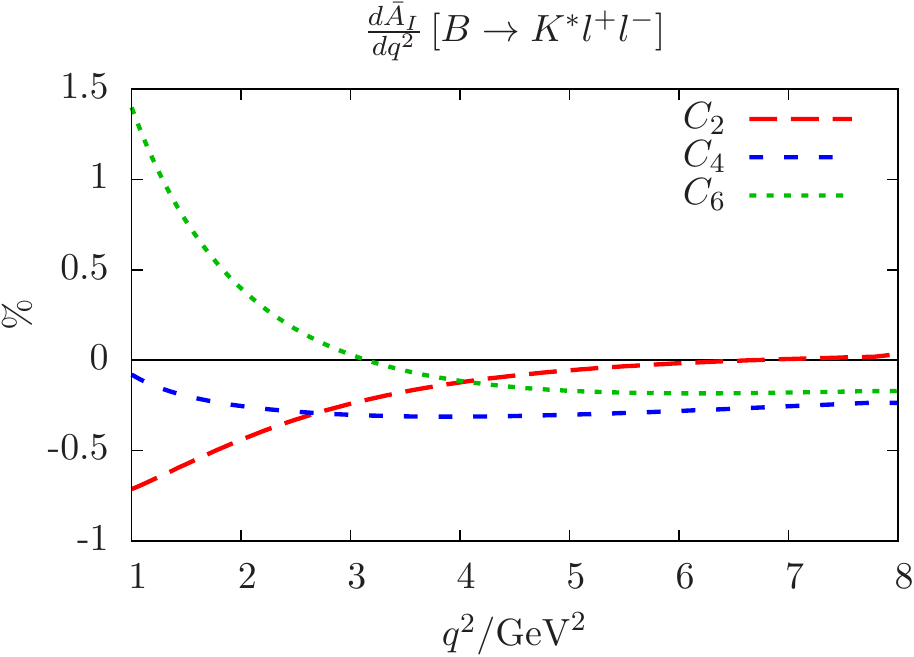}
\includegraphics[width=0.35\textwidth]{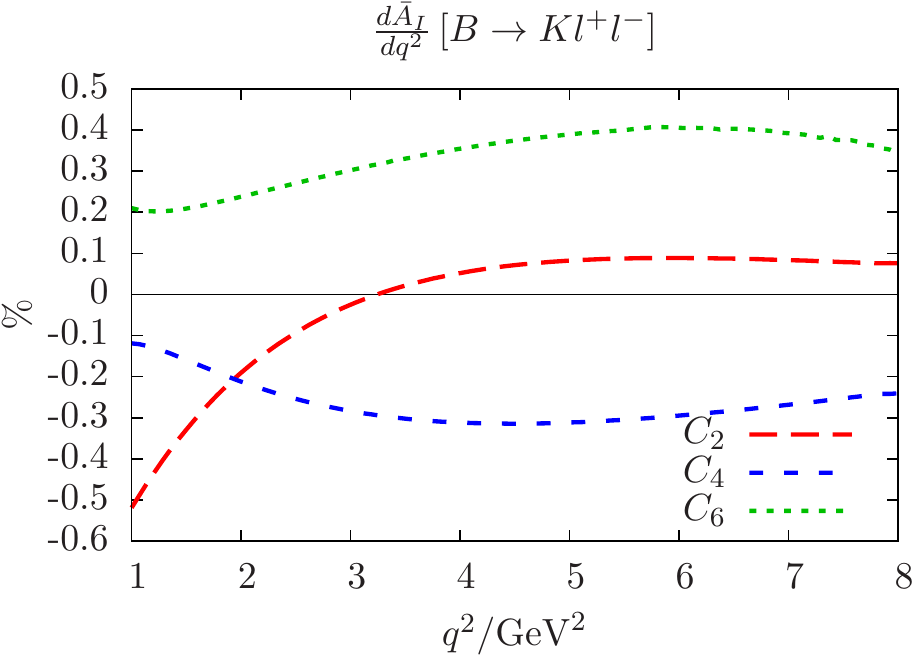}
\caption{\small (left(right),first row) Isospin asymmetry for $B \to K^{(*)} ll$ with green error bands. 
(left(right),second row) Isospin asymmetry for $B \to K^{*}_{\parallel} ll$ for
the longitudinal ($0$-helicity $L$ and $\parallel$ mean the same in this context!) part of $K^*$-meson denoted by subscript $L$ as well 
as the three previous graphs shown on one plot. Comments are deferred to the text.
(left(right),third and fourth row) Contribution of different SM operators to the isospin asymmetry in $B \to K^{(*)} ll$. The bottom ones are the sizeable contributions. The dominance of $[\CBBL_6 +(\CBBL_5/3)]$ has been found previously.
Its decrease is due to the relevant operators $O^{\rm WA}_{1-4}$ entering at subleading twist in the longitudinal part which is dominant at high $q^2$.
See appendix \ref{app:uncertainty} for details of the calculation of the green error bands.
}
\label{fig:1}
\end{figure}

The plots of the $B \to K^{(*)}ll$ isospin asymmetries are given in Fig.~\ref{fig:1}, including 
a plot of the longitudinal part (zero helicity) of the $K^*$ DA (c.f. section \ref{sec:non-relation} for comments), and the values are tabulated in Tab.~\ref{tab:1} (appendix \ref{app:tables}).
Important aspects on which operators or WC contribute were discussed in the previous subsection.  The feature that is obvious is that the isospin asymmetry is small 
on the scale of $-100\%$ to $100\%$ for $B \to K^{(*)}ll$ for $q^2 > 1 \GeV^2$; specifically below the $2\%$ level.
Qualitatively they agree with previous determination e.g. $B \to K^*ll$ \cite{Feldmann:2002iw} 
and $B \to K ll$ \cite{Khodjamirian:2012rm}\footnote{We differ from these references in that we compute WA and ${\cal O}_8$ in LCSR 
which includes LD contributions as argued at the beginning of section \ref{sec:QLSS}. 
Moreover we include twist-3 contributions for WA for the reasons mentioned in 
section \ref{sec:comment}.  The QLSS contributions are effectively treated in the same way. Whereas their result is small it differs from ours quantitatively which is explained by 
the differences mentioned above. 
}. A few qualitative remarks on the size of the contributions
can be found in the caption. Generically the asymmetries are dominated by WA which can also be inferred indirectly from Fig.~\ref{fig:2} in the next subsection.
Let us quote here the world average of the $B \to K^* \gamma$ isospin asymmetry from 
the Heavy Flavour Averaging Group (HFAG) \cite{Amhis:2012bh}
\begin{equation}
\label{eq:HFAG}
\bar a_I(K^* \gamma)_{\rm HFAG} = 5.2(2.6)\%     \;, \qquad \bar a_I(K^* \gamma)_{\rm LZ} = 4.9(2.6)\%  \;.
\end{equation}
which compared with our value turns out to be really close. Our value is also close 
to values previously found by \cite{Kagan:2001zk,Feldmann:2002iw,Ball:2006eu}.
The calculation of the theoretical uncertainty is detailed in appendix \ref{app:uncertainty}.
To what extent this constrains the dimension six operators and therefore $B \to K^{(*)}ll$ is 
discussed in section \ref{sec:constraints}.

Let us briefly discuss the three contributions in Fig.~\ref{fig:isoSM}  considered in this paper. 
\begin{itemize}
\item \emph{WA:}  The SM contributions $[a_i^q]^{SM}$ \eqref{eq:aSM} are given in term of the $C_i$ in Eq.~\eqref{eq:aSM}.
For WA one has to distinguish between
$\bar q  b  \bar s q $ operators (omitting the Lorentz indices)  
as  generated from tree and penguin processes. 
When  originating from penguin processes ${\cal O}_{3-6}$,  an equal amount of 
$q = u,d$ is generated and the process is dominated by the top quark penguin which 
results in $\lambda_t \sim \lambda^2$ CKM prefactor. The $q = u$ case also 
has a tree contribution $\calO^u_{1,2}$ which is then proportional to $\lambda_u \sim \lambda^4$. 
Thus a priori it seems difficult to judge whether the  loop  suppression or the CKM-suppression is more effective\footnote{For $D \to V \gamma$ and decays such as $D_s^+ \to \rho^+ \gamma$ there is no CKM suppression at all and since all other subprocesses are small, WA dominates these decays as we have argued in \cite{LZ12} in appendix A.}. Inspecting Fig.~\ref{fig:1} and taking into account that WA is the leading effect we see that the answer depends on $q^2$ and the Dirac structure:
$\CBBL_6$ dominates the isospin asymmetry for $B\to K^*ll$ but
for $B\to Kll$ the $\CBBL_{2,4}$ contributions are of similar magnitude to $C_6$.

\item \emph{QLSS:} QLSS is dominated by the charm loop as the latter originates from 
the tree operators ${\cal O}^c_{1,2}$. Whereas this contribution is not CKM suppressed $\lambda_c \sim \lambda^2$ it is of course loop suppressed. 

\item \emph{${\cal O}_8$:} The chromomagnetic ${\cal O}_8$-contribution   has been discussed in a separate paper \cite{DLZ12}. For the $B \to K^{(*)}$-transition the  matrix element 
is found it to be  rather small; as compared to the QCD penguin form factors 
$T_1(0)$. The total and isospin violating parts were found to be in the $6\%$ and $2\%$-range, as compared to $T_1(0)$, respectively.
An interesting aspect is the large strong phase attributed to LD contributions. The phase is
of importance for CP-violation in new physics searches in 
$D \to V\gamma$ \cite{LZ12},  but not for CP-averaged isospin as the latter is only sensitive to
the real part of strong phases; at least in the linear approximation c.f. \eqref{eq:dai}. 
Furthermore we should point out that we neglect the $\calO_8'$ contribution in 
the SM, as $C_8'/C_8 \simeq m_s/m_b$ by virtue of the MFV-symmetry.
\end{itemize}

Some of the points discussed above and in the previous subsection are summarised 
in Tab.~\ref{tab:Ooverview}.

\begin{table}
\center
\begin{tabular}{l c | l c | l c | l c }
$b \to s(d)$  & \multicolumn{2}{c }{}  WA & \multicolumn{2}{c}{} QLSS & \multicolumn{2}{c}{} $\calO_8$  \\ \hline
$\calO_X$ & WC & CKM & M.E. & CKM &  M.E. & CKM &  M.E. \\ \hline
${\cal O}_{1,2}$ & tree & $\lambda_u \sim \lambda^4(\lambda^3)$ & tree & $\lambda_c \sim \lambda^2(\lambda^3)$ & loop & - & - \\
${\cal O}_{3-6}$ & penguin & $\lambda_t \sim \lambda^2(\lambda^3)$ & tree & $\lambda_t \sim \lambda^2(\lambda^3)$ & loop & - & - \\
${\cal O}_8$      & penguin & - & - & - & - & $\lambda_t \sim$ 
$\lambda^2(\lambda^3)$ & loop  \\
\hline \hline
${\cal O}_7$  & penguin & \multicolumn{6}{c}{ not isospin sensitive \&  dominates low $q^2$} \\
${\cal O}_{9,10}$ &  penguin/box &   \multicolumn{6}{c}{ not isospin sensitive \&  dominates high $q^2$} 
\end{tabular}
\caption{\small SM operators contributing to the isospin asymmetry $\calO_{1-6,8}$ and operators not
contributing to the asymmetry $\calO_{7,9,10}$. This  table summarises the discussion in section \ref{sec:isospinSM}. \emph{WC} denotes whether the operator is generated 
by a tree or penguin process. \emph{CKM} denotes the CKM-suppression and $\lambda \simeq 0.22$ stands for the Wolfenstein parameter. In anticipation of 
$B \to \rho ll$ we have indicated the CKM hierarchy for $b \to d$ in parenthesis. \emph{M.E.} denotes whether the matrix element is a tree or loop level process.}
\label{tab:Ooverview}
\end{table}

\section{Isospin asymmetries in $B\to \rho \gamma/ll$}
\label{sec:rho-meson}

$B\to \rho$ ($ b \to d$) decays\footnote{In this work we refrain from including the 
isospin asymmetry $B \to \pi ll$.
$B^+ \to \pi^+ ll$, but not the neutral mode, has only been observed recently by the LHCb collaboration \cite{LHCb:2012de}.
Another logical extension would be to consider a $\rho$-$\omega$ asymmetry as in \cite{Ball:2006eu}. 
We refrain from doing so mainly because the latter suffers from a large theoretical uncertainty in the actual difference,
not to be confused with the separate values, of the
$\rho$ and $\omega$ form factors. This situation could be improved considerably 
through a dedicated study of the respective ratio of decay constants; both transversal  
and longitudinal.}
   differ from $B\rightarrow K^*$ ($ b \to s$) decays in two important respects:
1) WCs of the operators $\mathcal O_{1,2}$ are not CKM suppressed (c.f. Tab.~\ref{tab:Ooverview}) and 2)
$B^0\to \rho^0$, by virtue of $\rho^0 \sim (\bar u u - \bar d d)$,  contains additional diagrams with different arrangements of the four quark
operators,
and as a consequence 
also couples to colour octet operators (octet w.r.t. the basis \eqref{eq:OWA}).
We shall see shortly that the first point is effectively irrelevant as 
the relevant CKM angle $\alpha_{\rm CKM}=89(4)^\circ$ \cite{Beringer:1900zz} chooses to be close to Pythagorian perfection.

More precisely the  $\mathcal O_{1,2}$-contribution  in the SM 
comes with CKM-prefactor
\begin{equation}
\label{eq:alpha}
\frac{\lambda^{bd}_u}{\lambda^{bd}_t} = -\left|\frac{\lambda^{bd}_u}{\lambda^{bd}_t}\right|e^{-i\alpha_{\rm CKM}}  \;,
\end{equation}
where $\lambda^{bd}_i \equiv V_{id}^*V_{ib}$, in close analogy to  
$\lambda \equiv \lambda^{bs}_i \equiv V_{id}^*V_{ib}$ used previously.
Since the CP-averaged isospin asymmetry is sensitive to the real part, giving $\cos\alpha_{\rm CKM}=0.02(7)$,
the   relatively large magnitude of $\lambda^{bd}_u/\lambda^{bd}_t$ is overruled 
and  thus the overall contribution from $\mathcal O_{1,2}$ is very small.
Essentially there is then no interference of the $\mathcal O_{1,2}|_{\rm WA}$ with 
the leading contributions. This is why the non-CP averaged isospin asymmetry leads 
to rather different results (already in the SM c.f. section \ref{sec:CP-averaging}).

\subsection{Extending the effective Hamiltonian for $B^0\to\rho^0\gamma/ll$}
\label{sec:Heffrho}

For the $B^0\to\rho^0\gamma/ll$ decay ($\rho^0 \sim \bar u u - \bar d d$),
the basis  \eqref{eq:OWA} has to be extended to include
\begin{equation}
O^{{\rm WA}}_i = \bar q\Gamma_1b \, \bar d\Gamma_2 q
\implies O^{{\rm WA,8}}_i = \frac{1}{4} \bar q \lambda^a\Gamma_1b \, \bar d \lambda^a\Gamma_2 q\;,
\label{eq:OWA8}
\end{equation}
so that for example $O^{{\rm WA,8}}_1=\frac{1}{4}\bar q\lambda^ab\,\bar d\lambda^a q$,
and we modify the effective Hamiltonian \eqref{eq:h-effective-wa} to:
\begin{equation}
{\cal H}^{{\rm WA},q} = -\frac{G_F}{\sqrt 2}\lambda_t\sum_{i=i}^{10}\left[a^q_i O_i^{\rm WA} + a^{8q}_i O_i^{\rm WA,8}\right] \;.
\label{eq:h-effective-wa-8}
\end{equation}
In spite of all these operators being present, the basic situation presented in Tab.~\ref{tab:OWA} has not
changed: our calculation only picks up 6 linearly independent combinations of WCs in the $B^0\to\rho^0\gamma/ll$ case.
We therefore choose to present the isospin asymmetry for the $\rho$-meson in the following schematic way:
\begin{equation}
\label{eq:atildeandrho0}
\rho^\pm \leftrightarrow a_i^u \;, \qquad \rho^0  \leftrightarrow \tilde a_i^d = c_i^d a_i^d + c_i^{8d} a_i^{8d}  + 
c_i^{8u} a_i^{8u}
\end{equation}
with $c_i^x$ given in  appendix \ref{app:rho0-effective-coefficients}.

\subsection{Isospin asymmetries in $B\to \rho \gamma/ll$ in the SM}

The SM values of the new colour octet coefficients are:
\begin{align}
\label{eq:a8SM}
a_1^{8q} &= -a_2^{8q} = a_3^{8q} = -a_4^{8q} = -4 C_5 \;. \nonumber \\
a_5^{8q} &= -a_6^{8q} = -a_7^{8q} = a_8^{8q} = 2C_3 - 2\delta_{qu}\frac{\lambda_u}{\lambda_t} C_1 \;,  \\
a_9^{8q} &= a_{10}^{8q} = 0  \;.\nonumber
\end{align}
The ones for the colour singlet operators are the same as for the $K^*$ \eqref{eq:aSM}.
The formulae for $\tilde a$, in relation to $\rho^0$ \eqref{eq:atildeandrho0}, ar given 
in appendix \ref{app:rho0-effective-coefficientsSM} for the SM.

\begin{figure}
\center
\includegraphics[width=0.48\textwidth]{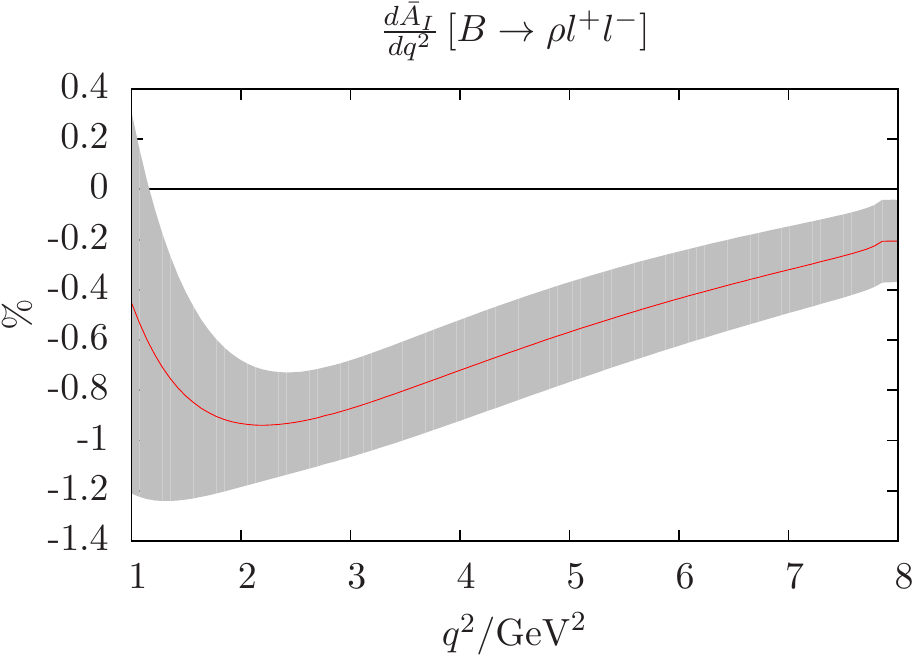} \\
\includegraphics[width=0.49\textwidth]{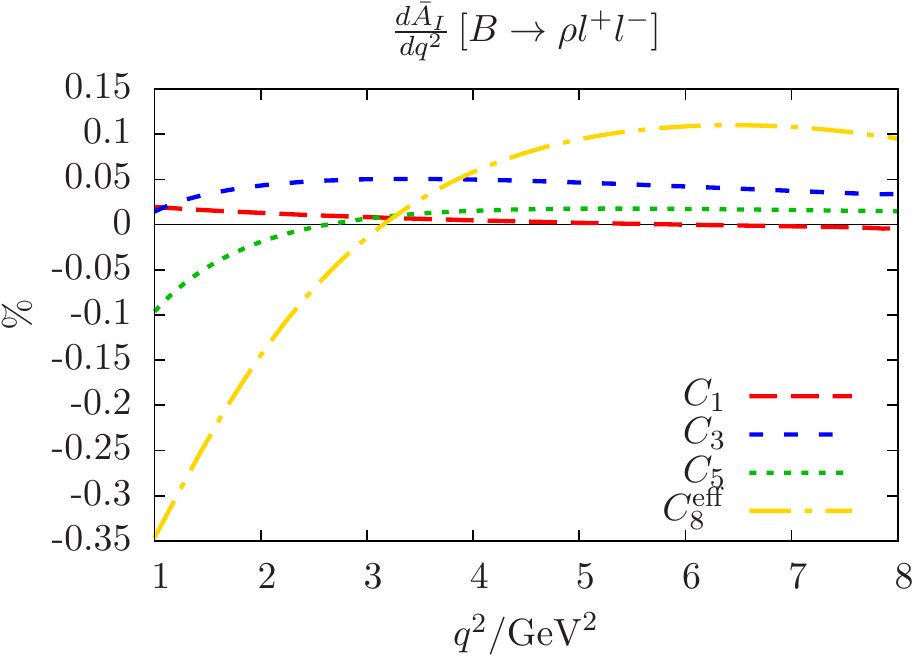}
\includegraphics[width=0.49\textwidth]{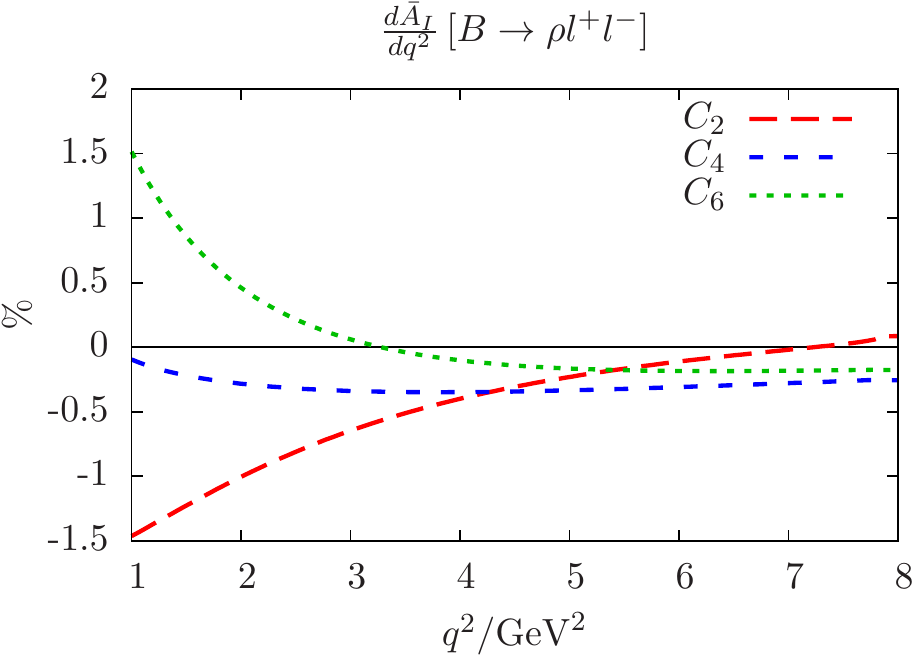} \\
\includegraphics[width=0.48\textwidth]{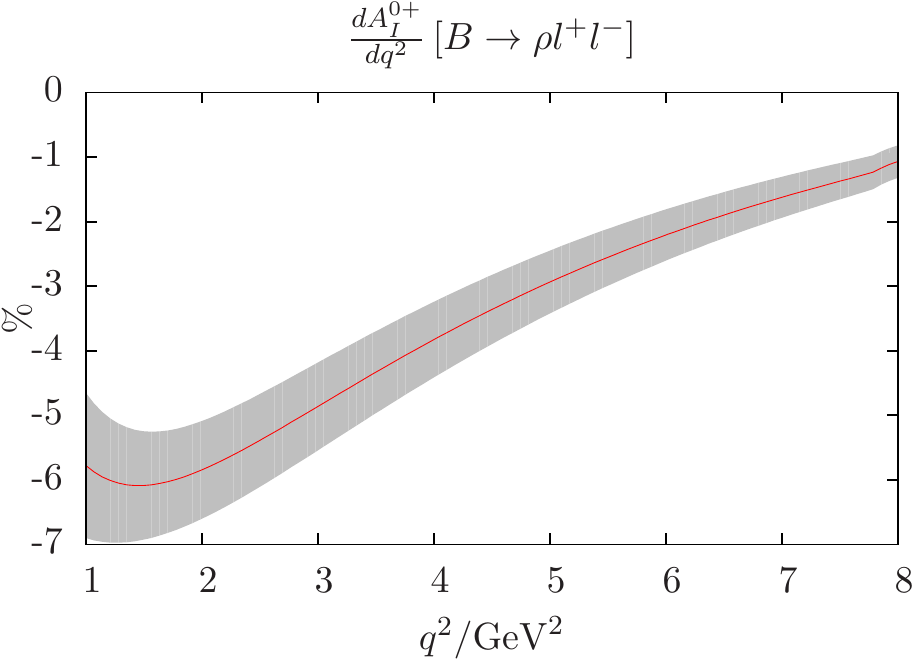}
\includegraphics[width=0.48\textwidth]{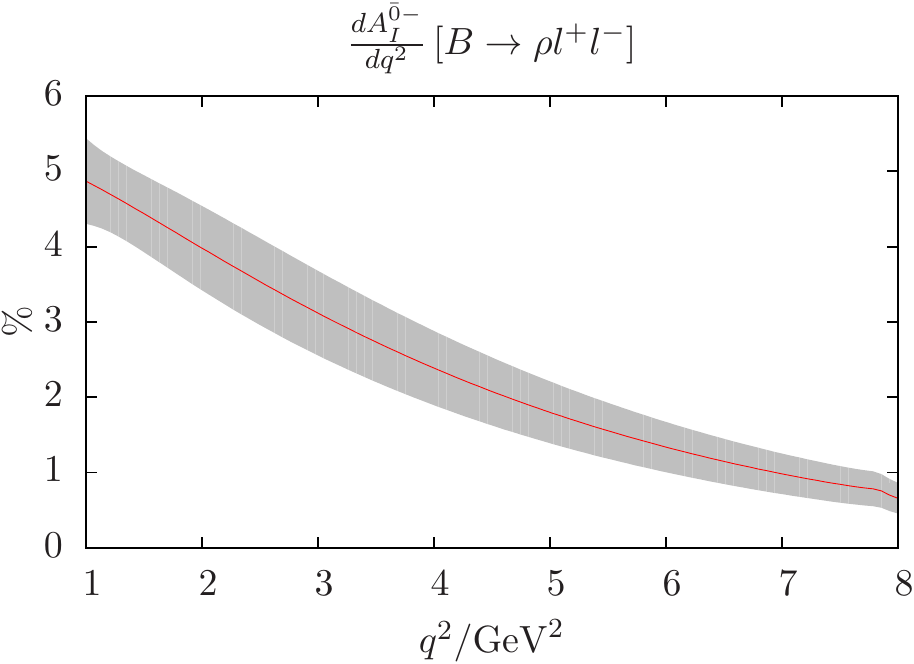}
\caption{\small (top) Isospin asymmetry for $B \to \rho ll$ with grey error bands. (left(right),middle) Contribution of different SM operators to the isospin asymmetry in $B \to \rho ll$.
The right hand graph shows sizeable contributions.
Note that unlike at $q^2=0$ the $\CBBL_2$ contribution is comparable to the $\CBBL_6$ contribution here;
this is due to a small weak phase arising from $C_9^{\rm eff}$ alleviating the $\cos\alpha_{\rm CKM}$ suppression a little.
See appendix \ref{app:uncertainty} for details of the calculation of the grey error band.
(left(right),bottom) Isospin asymmetry for $B \to \rho ll$  not CP-averaged $b \to d$ and 
$\bar b \to \bar d$-type. They do add up to the CP-average (top) but do deviate significantly from the latter as a result of strong and weak phases as discussed in section \ref{sec:CP-averaging}. No such effect is observed for the $K^*$ as explained in some detail in that section. 
}
\label{fig:1rho}
\end{figure}

Our results for $B\to \rho ll$, including breakdowns of operator dependence,
are shown in Fig.~\ref{fig:1rho} and Fig.~\ref{fig:2rho}, and
tabulated data is given in appendix \ref{app:tables} in Tab.~\ref{tab:1rho}.
The experimental measurement of the isospin asymmetry is defined differently to the $K^*$ case, as \cite{Taniguchi:2008ty}
\begin{equation}
\label{eq:Delta-aI-releation}
\Delta(\rho\gamma) = \frac{\tau_{B^0}}{2\tau_{B^+}}\frac{{\cal B}(B^+\to\rho^+\gamma)}{{\cal B}(B^0\to\rho^0\gamma)} - 1  = \frac{ -2\bar a_I(\rho\gamma) }{1 +  \bar a_I(\rho\gamma)}  \stackrel{a_I(\rho\gamma) \ll 1}{\approx} -2\bar a_I(\rho\gamma) \;,
\end{equation}
($  a_I(\rho\gamma)  = - \Delta(\rho\gamma) /(2 + \Delta(\rho\gamma) )$)
where a CP-averaged branching fraction is used.
In this normalisation, our result compares with the experimental result as \cite{Amhis:2012bh}
\begin{align}
\label{eq:Delta_rho}
\Delta(\rho\gamma)_{\rm HFAG} &= -46(17)\% \;, &
\Delta(\rho\gamma)_{\rm LZ} &= -10(6)\%  \;.
\end{align}
We shall quote $\Delta$ in percentage even though, contrary to $-1 \leq a_I \leq 1$, $\Delta$ is not bounded when $a_I \to -1$.
For completeness we further quote our result for the CP-averaged isospin asymmetry in $B\rightarrow\rho\gamma$ in the SM as
\begin{align}
\label{eq:aIrho}
\bar a_I(\rho\gamma)_{\rm HFAG} &= 30(^{-13}_{+16})\% \;, &  \bar a_I(\rho\gamma)_{\rm LZ} = 5.2(2.8)\%  \;,
\end{align}
where we have used Eqs.~(\ref{eq:Delta_rho},\ref{eq:Delta-aI-releation}) for computing 
what we call the HFAG value above.
Our result is comparable to that obtained in \cite{Ball:2006eu}\footnote{Note that \cite{Ball:2006eu} uses the opposite sign convention for $\Delta(\rho\gamma)$.}
and somewhat larger than that in \cite{Beneke:2004dp}, principally due to a different choice of $\alpha_{\rm CKM}$.
Our SM result is marginally consistent with the current experimental value,
that is to say they are exactly two standard deviations apart.
There is another way one can reflect on the experimental value 
\eqref{eq:Delta_rho}, namely one can extract  $|V_{\rm td}/V_{\rm ts}|$ from the ratio 
of branching fractions $R_{\rho/K^*} = \mathcal B(B \to\rho \gamma)/\mathcal B(B \to K^{*}\gamma) $ which can be applied for charged and neutral case separately. 
In view of the fact, to be discussed in the next section, that the isospin splitting 
of the ratios of $\rho$ and $K^*$-channel is accidentally small (i.e. SM: $R_{\rho^0/K^{*0}} \approx R_{\rho^+/K^{*+}}$ for current CKM angles),
we may infer from table 12 in Ref. \cite{Ball:2006eu} that 
\begin{equation}
\left| \frac{V_{\rm td}}{V_{\rm ts}}  \right|_{R_{\rho^0/K^{*0}}} = 0.229(25\%) \;, \quad 
 \left| \frac{V_{\rm td}}{V_{\rm ts}}  \right|_{R_{\rho^+/K^{*+}}} = 0.165(25\%) \;, \quad 
 \left|\frac{V_{\rm td}}{V_{\rm ts}}\right|_{\rm PDG\cite{Beringer:1900zz}} = 0.211(7) \;,
\end{equation}
and we have quoted the current value from Particle Data Group (PDG) for comparison.
We have given a rough estimate of the error which is mainly due to the 
$B \to \rho \gamma$ 
branching fraction (c.f. appendix \ref{app:delta-ai-calc}). Thus we infer that the discrepancy in $\Delta(\rho \gamma)$ \eqref{eq:Delta_rho} is presumably due to the $\rho^+$- rather than the $\rho^0$-channel.

\subsection{On subtleties of CP-averaging the isospin asymmetries}
\label{sec:CP-averaging}

In this paper we have computed CP averaged quantities which results in the linear approximation 
to taking the real part of the strong and weak phase separately, whereas no CP average implies taking real part of the product of the strong and weak phase.  Schematically:
\begin{equation}
\text{CP-average:} \quad {\rm Re}[e^{i \delta_{\rm strong}}]  {\rm Re}[ e^{i \phi_{\rm weak}}   ]   
\;, \qquad \text{no CP-average:} \quad {\rm Re}[e^{i \delta_{\rm strong}}e^{i \phi_{\rm weak}}   ]    \;.
\end{equation}
To be more precise the weak and the strong phase is the difference between 
the isospin-violating  and the isospin-conserving amplitude.
Thus in general there can be significant differences if both  $\delta_{\rm strong}$ 
and $\phi_{\rm weak}$ are sizeable. 

Are there sizeable strong phase differences?
The isospin conserving amplitude has a very small strong phase in the region we are considering and thus the question is whether 
there is a sizeable strong phase in the isospin violating amplitude.
The answer to this is no for $q^2 = 0$,  as only  $\calO_8{(')}$ contributes  with a strong phase at leading order in $\alpha_s$ (which is at least small  in the SM).
For $1\GeV^2  \leq q^2   \leq 4m_c^2$ however the answer is yes: the photon emitted from a light quark converts via an intermediate $\rho, \omega$-meson and gives raise to a tail in the imaginary part.
This is the case for all IR isospin violation and we refer the reader to figure 4(left) in \cite{DLZ12} for an illustration.
 
Are there large sizeable weak phase differences? In the SM this is the case for $B \to \rho ( b \to d)$ 
as can be seen from \eqref{eq:alpha} with  $\alpha_{\rm CKM}=89(4)$ but not for $B \to K,K^*(b \to s)$. In BSM scenarios this question is open modulo constraints, in particular  CP-observables.

We summarise the conclusions to be drawn from the discussion above in  Tab.~\ref{tab:CP-averaging}.
The plots for the non CP-averaged isospin asymmetries in  $B \to \rho ll$  are shown in Fig.~\ref{fig:2rho}.  We  see that the asymmetries raise up to $\pm 5\%$ in the  $1\GeV^2  \leq q^2   \leq 4m_c^2$-region.

\begin{table}[h]
\center
\begin{tabular}{l | c | c| c}
$\overline{\rm CP}$ effect & $B\to (K^*,\rho)\gamma$ &  $B\to K^{(*)} ll$  & $B\to \rho ll$ \\
\hline
SM & \cross & \cross & \tick \\
BSM ($\phi_{\rm weak}^{\rm BSM}$) &  \cross & \tick & \tick \end{tabular}\center
\caption{\small $\overline{\rm CP}$ effect stands for CP-averaging effect on  the isospin asymmetries and  \cross(\tick) mean  
(in)significant. Overview of the conclusions to be drawn from the analysis of section \ref{sec:CP-averaging}. Note that a enhancement of $\calO_8$ or sizeable radiative corrections to WA or QLSS would raise $\delta_{\rm strong}$ and could shift the situation slightly.}
\label{tab:CP-averaging}
\end{table}
In general it might therefore be interesting to measure non CP-averaged isospin asymmetries in future experimental determinations. This is certainly possible for 
the $K^*$ and the $\rho$ but not for the $K$, as it is observed through $K_S^0$ which is a superposition of the strangeness eigenstates $K^0$ and $\bar K^0$.

\section{Isospin asymmetries beyond the SM}
\label{sec:isospinBSM}

\begin{figure}
\center
\includegraphics[width=0.49\textwidth]{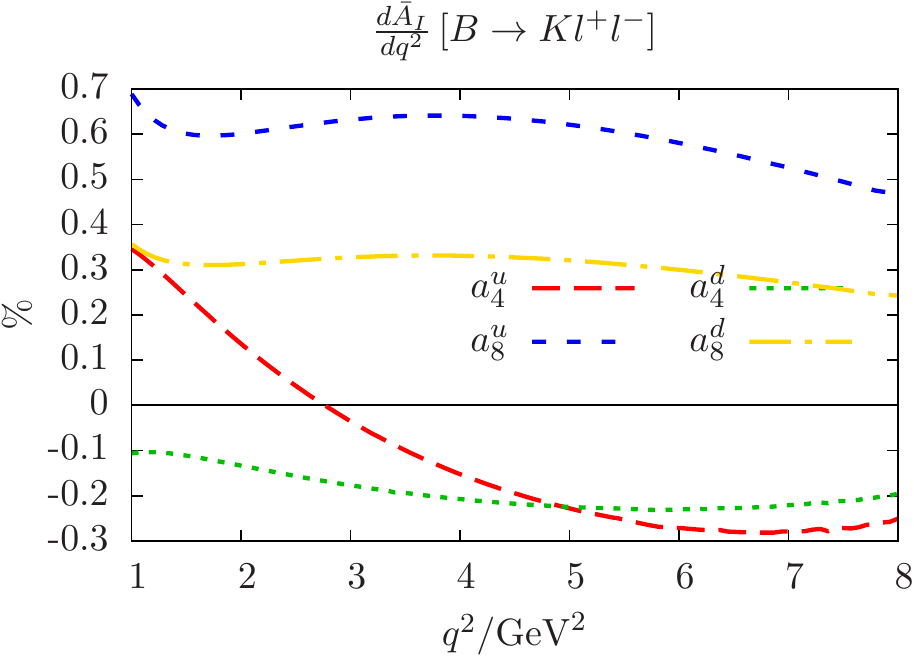}
\includegraphics[width=0.49\textwidth]{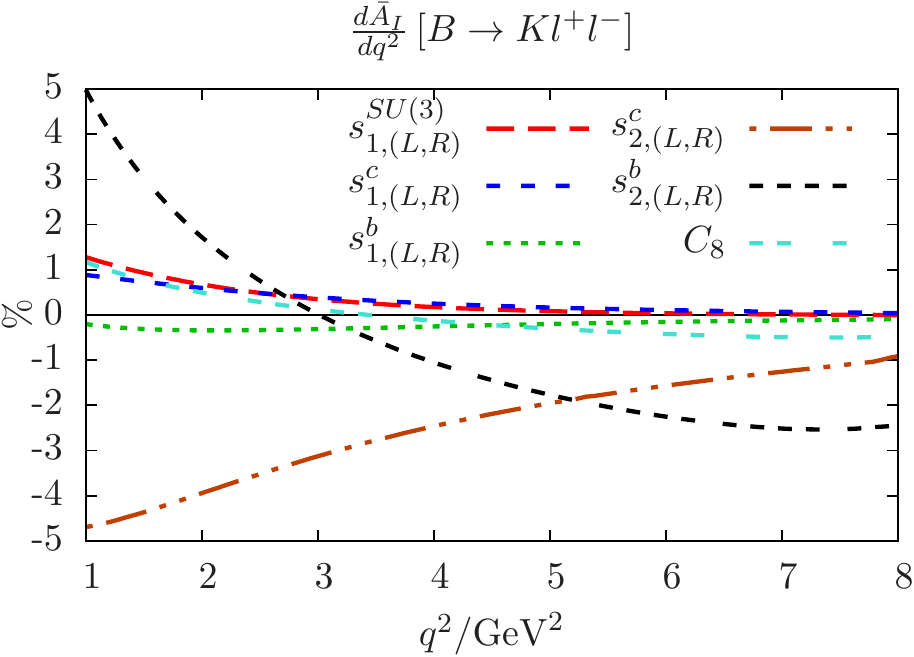}
\includegraphics[width=0.49\textwidth]{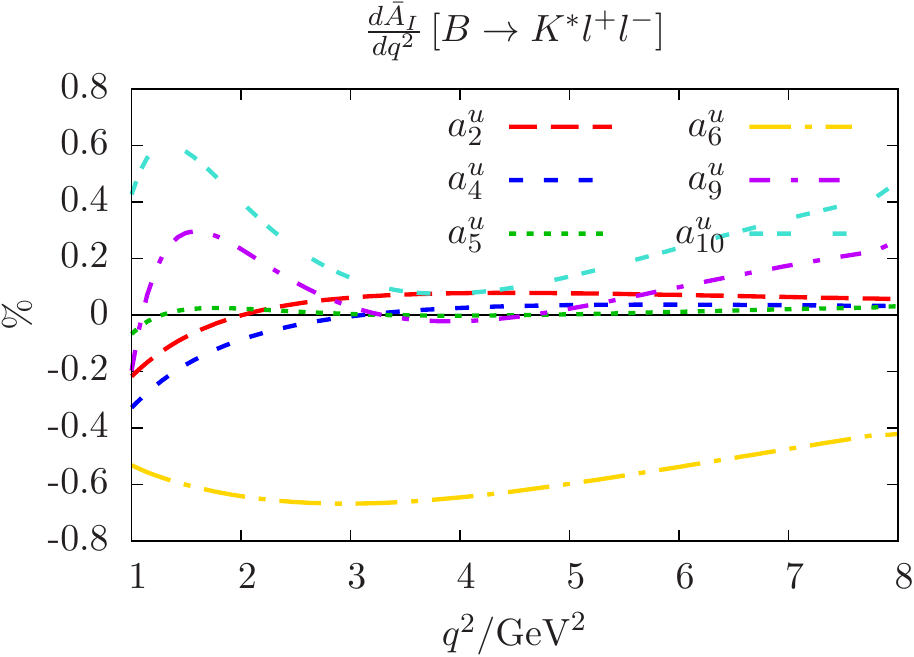}
\includegraphics[width=0.49\textwidth]{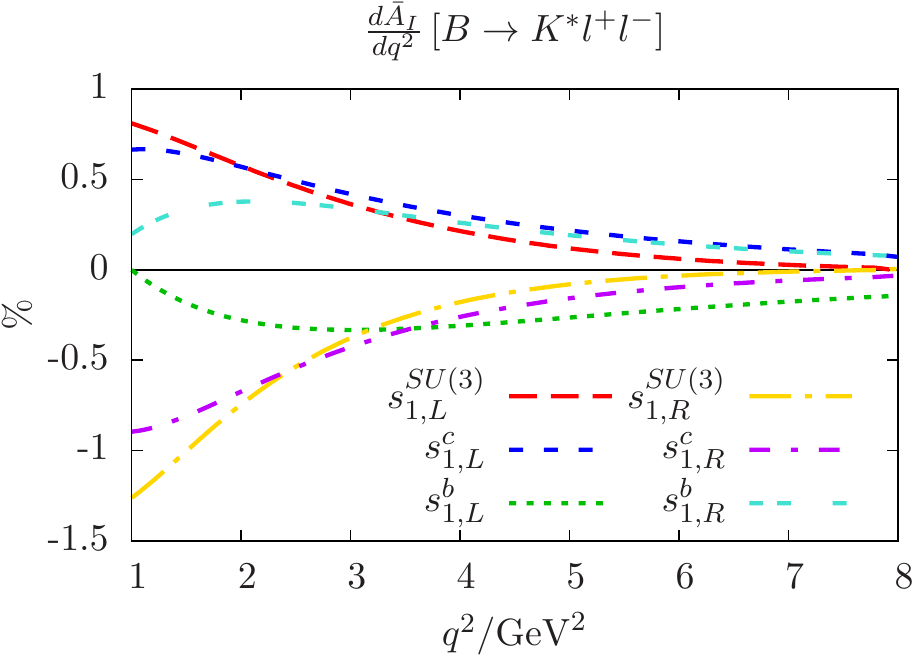}
\includegraphics[width=0.49\textwidth]{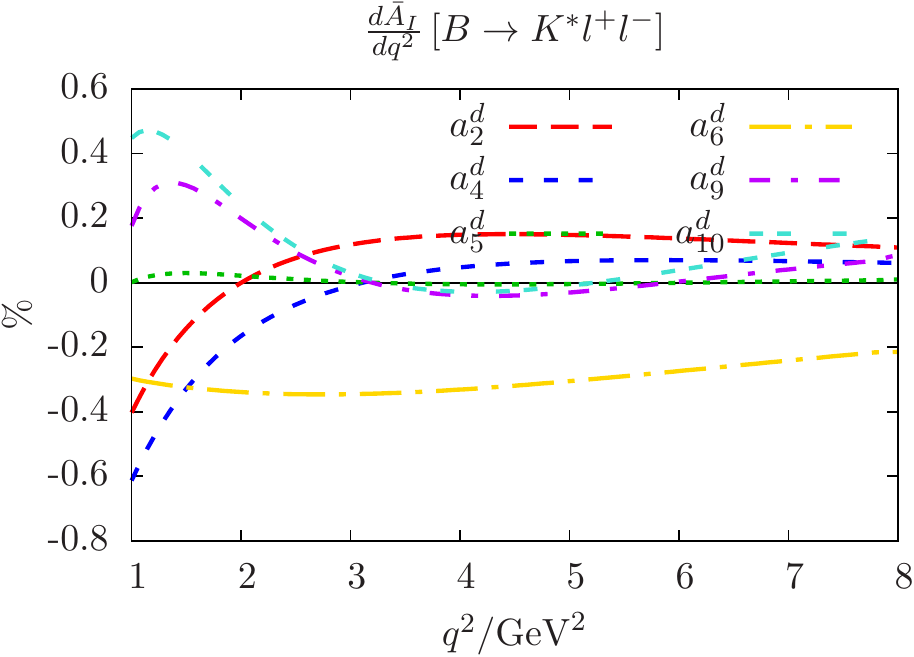}
\includegraphics[width=0.49\textwidth]{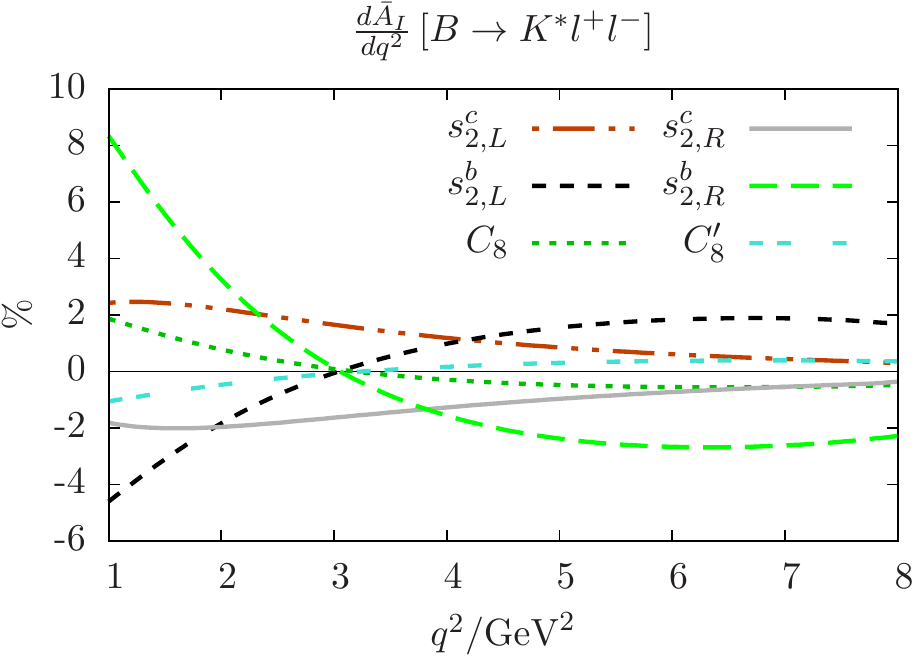}
\caption{\small  Breakdown of contributions of 
WA ($a_i$), QLSS ($s^f_{x,\chi}$) and ${\cal O}_8^{(')}$  to the isospin asymmetry 
$B \to K^{(*)}ll$  in the linear approximation \eqref{eq:dai}. We have split the contributions as  detailed in Tab.~\ref{tab:breakdown} into different graphs in order to make them more readable.
Note that we use $a^q_i=0.1$ and $s^f_{x,\chi}=1$ to produce these figures, as in the tables.}
\label{fig:2}
\end{figure}

\begin{figure}
\center
\includegraphics[width=0.49\textwidth]{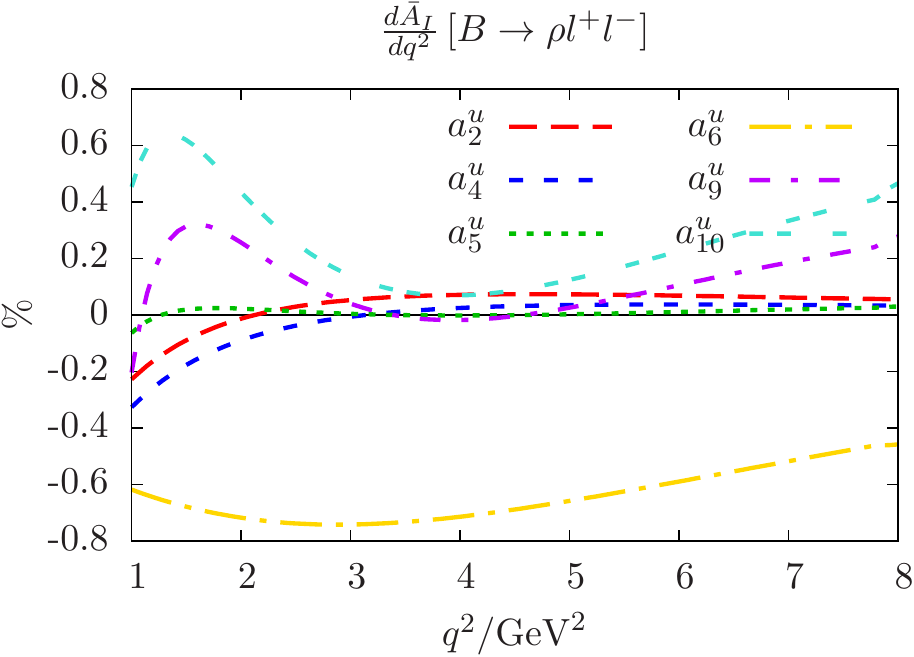}
\includegraphics[width=0.49\textwidth]{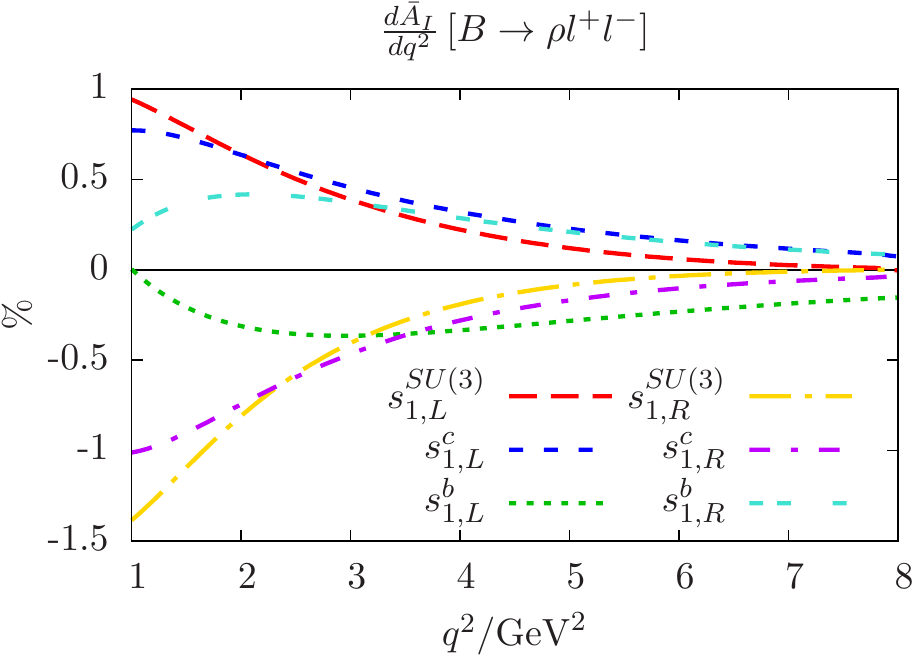}
\includegraphics[width=0.49\textwidth]{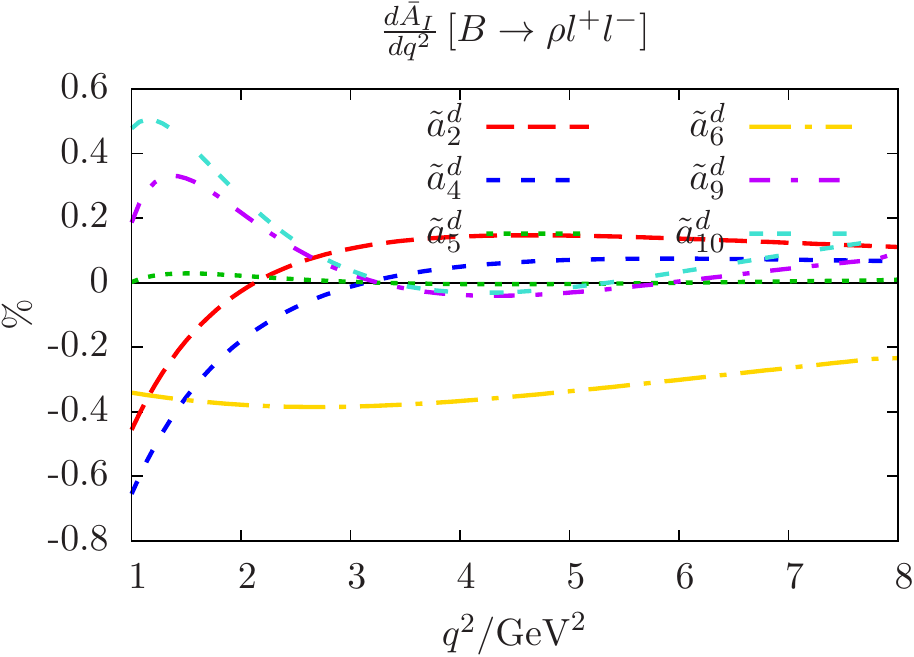}
\includegraphics[width=0.49\textwidth]{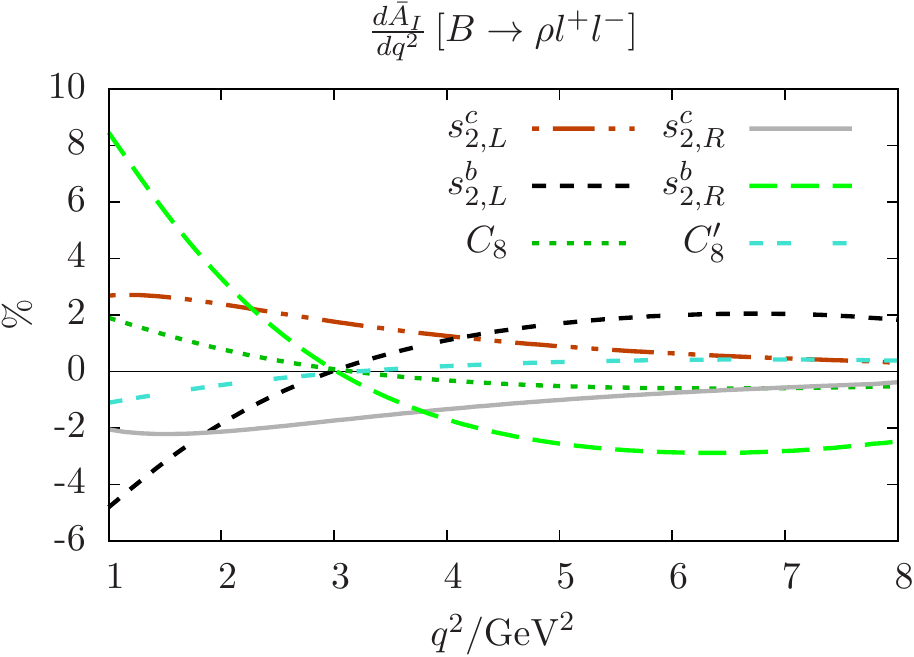}
\caption{\small  Breakdown of contributions of 
WA ($a_i$), QLSS ($s^f_{x,\chi}$) and ${\cal O}_8^{(')}$  to the isospin asymmetry 
$B \to \rho ll$  in the linear approximation \eqref{eq:dai}.
We have split the contributions as  detailed in Tab.~\ref{tab:breakdown} into different graphs in order to make them more readable.
Note that we use $a^u_i=0.1$, $\tilde a^d_i=0.1$ and $s^f_{x,\chi}=1$ to produce these figures, as in the tables.}
\label{fig:2rho}
\end{figure}

The extension of the SM basis was discussed throughout the main text and summarised 
in subsection \ref{sec:selection}. Possibly we should reemphasize, for the sake of clarity, that 
$\calO_{7,9,10}'$-operators of \VpA{}-chirality are of no interest to the isospin asymmetry 
as they do not violate isospin. Of course if they become extremely large then they would affect the rate. Yet it is already known that they cannot be too large e.g. \cite{Altmannshofer:2011gn,Altmannshofer:2012az,DescotesGenon:2012zf}.
The various contributions of the extended basis 
are detailed in Figs.~\ref{fig:2},\ref{fig:2rho} and tabulated in Tabs.~\ref{tab:2},\ref{tab:1}\&\ref{tab:1rho} (appendix \ref{app:tables}) for the $K$, $K^*$ and $\rho$ channels respectively.
One aspect that is immediately apparent from these graphs is that there are overwhelmingly many contributions that can give rise to a sizeable isospin asymmetry 
at low $q^2$. In fact there are so many that by the rules of probability one would  
expect cancellations in the generic case. Fortunately this is where the $q^2$-spectrum should help us, should there be new physics, one cannot expect to be unlucky over the entire 
$q^2$-range.

In subsection \ref{sec:non-relation} we have discussed that only for leading twist and 
SM chirality the $K$ and the $K^*_\parallel$ are related. As noted there this link breaks down in the presence of right-handed currents, which are only partially constrained, 
and thus in a generic scenario the link between the $K$ and the $K^*$ isospin asymmetry is lost. The reader can convince him or herself of this fact directly 
from the corresponding tables and figures referred to above.

\begin{table}[h]
\center
\begin{tabular}{c|ccc|c|ccc}
$B\to K^*\gamma$ & Min. & SM & Max. & & Min. & SM & Max. \\
\hline
$a_2^u$ & -0.39 & -0.068 & 0.25 & $a_2^d$ & -0.24 & -0.068 & 0.11 \\
$a_4^u$ & -0.38 & -0.068 & 0.25 & $a_4^d$ & -0.24 & -0.068 & 0.10 \\
$a_5^u$ & -0.41 & $-0.021+0.019 i$ & 0.37 & $a_5^d$ & -0.67 & -0.028 & 0.61 \\
$a_6^u$ & -0.62 & $0.021 -0.019 i$  & 0.57 & $a_6^d$ & -1.0 & 0.028 & 1.0 \\
$a_9^u$ & -0.049 & 0 & 0.049 & $a_9^d$ & -0.080 & 0 & 0.080 \\
$a_{10}^u$ & -0.048 & 0 & 0.048 & $a_{10}^d$ & -0.080 & 0 & 0.080
\end{tabular}
\caption{\small Constraints on WCs $a_i^q$ from $B\to K^*\gamma$ at
$2\sigma$,
assuming no accidental cancellations occur, along with SM values.
We assume $0<\bar a_I(K^*\gamma)<10\%$, and derive constraints from
Tab.~ \ref{tab:1}
assuming that only a single coefficient $a_i^q$ deviates from its SM value.
SM values are calculated from \eqref{eq:aSM} and table
\ref{tbl:wilson-coefficient-example}.
All constraints are for the \emph{real} part of these coefficients, the
imaginary part is not
constrained by the isospin asymmetry.}
\label{tbl:kstar-constraints}
\end{table}

\begin{table}[h]
\center
\begin{tabular}{c|cc|c|cc}
$B\to \rho\gamma$ & SM & Bound & & SM & Bound \\
\hline
$a_2^u$ & -0.068 & -4.1 & $\tilde a_2^d$ & -0.068 & -2.1 \\
$a_4^u$ & -0.068 & -4.0 & $\tilde a_4^d$ & -0.068 & -2.0 \\
$a_5^u$ & $-0.021+0.402i$ & $0.26 e^{-i 87^\circ}$ & $\tilde a_5^d$ & $-0.028+0.088 i$ & $0.42 e^{-i 72^\circ}$ \\
$a_6^u$ & $0.021-0.402i$ & $-0.40 e^{-i 87^\circ}$  & $\tilde a_6^d$ & $0.028-0.088 i$ & $-0.68 e^{-i 72^\circ}$ \\
$a_9^u$ & 0 & 0.56 & $\tilde a_9^d$ & 0 & 0.94 \\
$a_{10}^u$ & 0 & 0.56 & $\tilde a_{10}^d$ & 0 & 0.93
\end{tabular}
\center
\caption{\small Constraints on operators $a_i^q$ from $B\to \rho\gamma$ at
$2\sigma$,
assuming no accidental cancellations occur, along with SM values.
We assume $6\%<\bar a_I(\rho\gamma)<67\%$, and derive as described
below Tab.~ \ref{tbl:kstar-constraints}.
 The fact that the real part of $a_{5,6}^u$ is 
the same (on the level given digits)
as for $B \to K^*$ in tab \ref{tbl:kstar-constraints} is an numerical accident.
The large imaginary parts in $a_{5,6}^u$ and  $\tilde a_{5,6}^u$ imply that 
the constraints applies in the same direction in the complex plane. 
Note: in the published version (arXiv v2)
only real WC were assumed in which case 
the constraint on $a_{5}^u$ becomes $0.26/ \cos(-i 87^\circ) \simeq 4.9$ which matches the earlier result.
Since our calculated SM value is the lower bound of this range (within
uncertainties), we quote the
SM value of the coefficient and the other bound; the true value is
expected to lie in this range. Complementary constraints from $B \to (\pi/\rho)  (\pi/\rho)$ are presented in 
\cite{Bobeth:2014rda} (c.f. figure 1).}
\label{tbl:rho-constraints}
\end{table}

\subsection{Constraints on isospin sensitive four Fermi operators}
\label{sec:constraints}

We shall now briefly turn to the question to what extent these operators are already 
constrained. We identify  non-leptonic decays\footnote{$\Delta F=2$-constraints from neutral meson oscillations, usually rather severe, are presumably not very strong. 
More precisely if we are to compare SM/MFV type operator $\calO^{\rm MFV}_{\Delta B = 2} \approx  |\lambda_t|^2 G_F/16\pi^2(\bar b \Gamma_1 s_L)(\bar b \Gamma_2 s_L)$
(with $\Gamma_{1,2}$ specific Dirac structures) then integrating out 
either $b$-quarks or saturating light quarks with intermediate hadronic states one would 
expect to get a
$ G_F \times (m_b^2, \Lambda_{\rm QCD}^2) \approx  (10^{-4},10^{-6})$ 
suppression in each case.} as well as  $B \to \rho/K^* \gamma$ isospin asymmetries 
themselves as the main sources for constraints:
\begin{itemize}
\item \emph{$B \to \rho/K^* \gamma$ isospin asymmetries}: the experimental values 
are quoted in \eqref{eq:HFAG}, \eqref{eq:Delta_rho} respectively. 
These isospin asymmertries are  sensitive to $a^q_{2,4,5,6,9,10}$ of WA in particular.
Of course  one number 
such as  $\bar a_I(K^* \gamma)$ can  not seriously bound twelve numbers.
We might though give indicative constraints by imposing that each of the coefficients 
shall not be more than two standard deviations away from the central value,
which roughly  amounts to $ 0 < \bar a_I(K^* \gamma) < 10\%$ and $ 6 < \bar a_I(\rho \gamma) < 67\%$ ($  -80\%  < \Delta(\rho\gamma) < -12\%$). The results of this procedure  are collected 
in Tab.~\ref{tbl:kstar-constraints} and Tab.~\ref{tbl:rho-constraints} respectively.
\item 
\emph{Non-leptonic decays}: four Fermi operators do affect non-leptonic decays 
such as $B \to \rho/\pi K^{(*)}$, $B_s \to K^{(*)}\phi$ etc.  The disadvantage is that they are 
difficult to predict from a theoretical viewpoint. Especially in the absence of a first principle approach to final state rescattering. The uncertainty in strong phases obscures interference effects which affects all observables,
let alone CP-asymmetries.  The advantage though is that there is a plethora of channels 
which allows theorists to constrain certain weak topologies e.g. \cite{Fleischer:2010ib,Ciuchini:2012gd,Bhattacharya:2012ph} and permits them to cross-check their methods.  Interesting constraints on four Fermi operators, such as the so-called electroweak penguins present in the SM, have been obtained in Ref.~\cite{Hofer:2010ee,Hofer:2012vc} for instance in the framework of QCD factorisation.
We would like to add two remarks. First, these operators do partially overlap with ours and would indeed bring in additional constraints. Yet  only  global fits lead to solid constraints which 
is beyond the scope of this work. Second, from the plots  Ref.~\cite{Hofer:2010ee,Hofer:2012vc}  
one infers that it is rather unlikely that the NP contributions to the WC exceeds the SM values by a factor of five but could easily be out by a factor of two.  
In view of the multitude of channels this might very well be true for SM-operators.
It seems  more difficult to come to a quick judgement for non SM-operators (by which we mean operators with small WCs). Partial studies do exist: e.g. an interesting direction, in view of right-handed currents, is the investigation of polarisation in $B \to VV$ decays 
\cite{Bhattacharya:2012ph} which was carried out in \cite{Beneke:2006hg} in the framework of QCD factorisation. 
\end{itemize}

\subsection{$B \to K^*/\rho \gamma$ isospin asymmetry splitting - (quasi) SM null test}
\label{sec:dai}

\begin{figure}
\center
\includegraphics[width=0.5\textwidth]{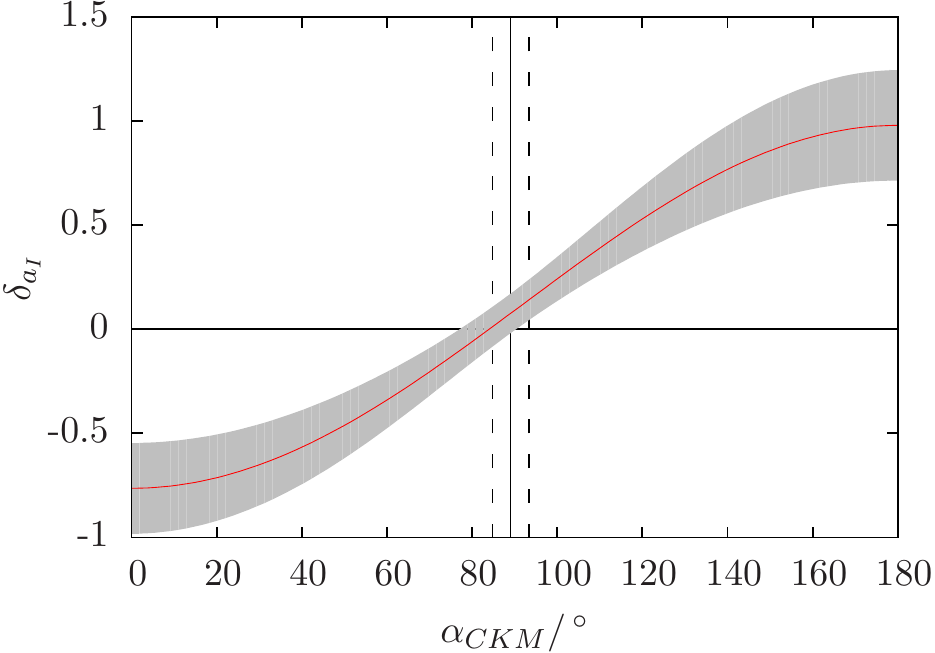}
\caption{\small Plot of the effective $\rho$ to $K^*$ isospin asymmetry difference $\delta_{a_I}$ \eqref{eq:delta-ai-def}.
Vertical lines indicate the current experimental value of $\alpha_{\rm CKM}$ \cite{Beringer:1900zz} and its uncertainty.
At the present small value of $\cos\alpha_{\rm CKM}$, $\delta_{a_I}$ is well determined theoretically.}
\label{fig:delta-ai}
\end{figure}

The closeness of $\alpha_{\rm CKM}$ to ninety degrees may be exploited to 
predict an oberservable with much smaller theoretical uncertainty,
albeit at the expense of larger current experimental uncertainty. 

Our prediction is essentially that $\bar a_I(\rho\gamma)$ and $\bar a_I(K^*\gamma)$
should be similar, up to form factor and hadronic parameter differences\footnote{An extension to $q^2 > 0$ is not straightforward as the isospin asymmetries of the $K^*$ and $\rho^*$ do differ qualitatively:
$C_9^{\rm eff}(q^2)$ contributes a small weak phase to the leading amplitude which partially alleviates the $\cos\alpha_{\rm CKM}$ suppression.}. A major source of uncertainty in determining $\bar a_I(K^*\gamma)$ and $\bar a_I(\rho\gamma)$ however is the renormalisation scale used to compute the WCs,
and because of this it is worthwhile to calculate a quantity in which the leading scale dependence and form factor differences cancel, namely
\begin{equation}
\delta_{a_I} \equiv 1 - \frac{\bar a_I(\rho\gamma)}{\bar a_I(K^*\gamma)}
R_{\rho K^*} 
=1 + \frac{\Delta(\rho\gamma)}{(2+\Delta(\rho\gamma))  \bar a_I(K^*\gamma)} R_{\rho K^*}  \;,
\label{eq:delta-ai-def}
\end{equation}
where 
\begin{equation}
\label{eq:RrhoKs}
R_{\rho K^*}  \equiv \sqrt{\frac{\bar \Gamma(B\to\rho\gamma)}{\bar \Gamma(B\to K^*\gamma)}}\left|\frac{V_{ts}}{V_{td}}\right|  \;,
\end{equation}
and a barred partial width $\bar \Gamma$ implies a CP-average, and omission of charges implies an isospin average\footnote{For the $\rho$-meson
this implies $\bar \Gamma(B\to\rho\gamma)=\frac{1}{2}\bar \Gamma(B^+\to\rho^+\gamma) + \bar \Gamma(B^0\to\rho^0\gamma)$
due to $\rho^0\sim (\bar uu-\bar dd)/\sqrt{2}$ as discussed previously.}.
The dominant contributions to the RHS of \eqref{eq:delta-ai-def} are
\begin{align}
\label{eq:crude}
a_I(V\gamma) &\approx \frac{C_6 + C_5/3}{C_7^{\rm eff}} \frac{f_V^\perp F^{\rm WA}(0) }{T_1^V(0)} + \dots &
\bar\Gamma(B\to V\gamma) &\approx \frac{3\alpha c_F}{8\pi} \left|\lambda_t C_7^{\rm eff}\right|^2 |T_1^V(0)|^2
\end{align}
where  the dots stand for $C_{3,4}$-contributions, which are small as the $K^*$ and $\rho$ cases are very similar,  quark masses and
$B^0\to\rho^0$ diagrams at $\calO(\alpha_s^2)$ where the different structure of the $\rho^0$ matters even for small $\cos\alpha_{\rm CKM}$. The function $f_V^\perp F^{\rm WA}(0)$
stands for final state emission where we have explicitly factored out the $f^\perp$ decay constant as compared with \eqref{eq:wa-if-split}. More precisely: 
$f_V^\perp F^{\rm WA}(q^2) = \sum_{i=2,4} \left(F^d_i(q^2) - F^u_i(q^2) \right)$ ($i =2,4$ are the operators proportional to $C_6 + C_5/3$)
in the notation of 
\eqref{eq:wa-if-split}, and $F^{\rm WA}$ is the same for the $K^*$ and the $\rho$ in our approximation up to small corrections from different Gegenbauer moments.
The correction factor $R_{\rho K^*}$ 
serves the purpose of eliminating the form factor ratio as $\bar a_I(\rho\gamma)/\bar a_I(K^*\gamma) \simeq  T_1^{B \to K^*}(0)/T_1^{B \to \rho}(0)$ which follows 
from Eq.~\eqref{eq:crude}.
Since the WA contribution $f^\perp F^{\rm WA}(0)$ is essentially proportional to $f^\perp$, 
it then follows that
\begin{equation}
\delta_{a_I}=1-f_\rho^\perp/f_{K^*}^\perp + \text{small corrections}\;,
\end{equation}
where the principal source of uncertainty, the scale dependence of $C_6 + C_5/3$, drops out.
Note that by `small corrections' we mean small as compared to $1$. 
The quantity 
$\delta_{a_I}$ is  particularly sensitive to  corrections to the isospin asymmetry and we therefore  include terms quadratic in WA amplitudes present in \eqref{eq:dai-def} but neglected in \eqref{eq:dai} and elsewhere.
Comparing our prediction with a naive combination of PDG \cite{Beringer:1900zz} results for these quantities gives:
\begin{align}
\left[\delta_{a_I}\right]_{\rm exp} &= -4.0(3.5)  \;, &
\left[\delta_{a_I}\right]_{\rm LZ} &= 0.10(11) \;.
\label{eq:delta-ai-exp}
\end{align}
The theoretical uncertainty should be compared with $1$ as it is a ratio, as above and thus is at $11\%$,  is under rather good control, as compared to roughly fifty percent in the individual asymmetries. 
See appendix \ref{app:delta-ai-calc} for the experimental input used as well as brief comments on the uncertainty. Let us briefly add that the uncertainty due to the difference in $B$-meson lifetimes and $|V_{ts}/V_{td}|$ is negligible. 
Experimental and theoretical values agree within uncertainties, though the central value is very different and an improved
experimental determination is desirable since the theoretical errors are under good control.
The experimental uncertainty due to all four branching fractions involved in $\delta_{a_I}$ is rather similar,
and thus \textit{all} of them need to be reduced to significantly improve the overall uncertainty. It should also be added that if the asymmetries are measured in the same experiment (some) systematic uncertainties can be expected to cancel.

\begin{table}[h]
\center
\begin{tabular}{c|rr}
$x$ & $\delta_{a_I}^{a_1^d + x}$  & $\delta_{a_I}^{a_8^d + x}$ \\
\hline
-0.3 & $  1.16(15)$ & $  1.71(20)$\\
-0.2 & $  0.82(11)$ & $  1.21(13)$\\
-0.1 & $  0.47(9)$ & $  0.67(9)$\\
0.1 & $ -0.29(14)$ & $ -0.51(15)$\\
0.2 & $ -0.68(18)$ & $ -1.14(21)$\\
0.3 & $ -1.08(23)$ & $ -1.78(28)$\\
\end{tabular}
\caption{\small
Effect of varying $a^d_i$ from their SM values on the isospin splitting $\delta_{a_I}$.
We fix all $a^q_i$ to their SM values and then alter a single one by the specified amount.
More precisely $\delta_{a_I}[{a_8^d + x}]  \leftrightarrow \delta_{a_I}[{ a^{d,\rm SM}_8+x}]$ above.
The resulting variation of $\delta_{a_I}$ can be large and is primarily the result of the $\rho^0$
coupling to a different combination of $a_i$ as discussed in section \ref{sec:Heffrho},
and is therefore an example of UV isospin violation (but not MFV violation).
The uncertainties quoted in this table do not include uncertainty from varying the renormalisation
scale: this would require a computation of the scale dependence of the Hamiltonian \eqref{eq:h-effective-wa-8}
and is thus beyond the scope of this work.
The renormalisation scale is taken to be the usual central value, $\mu=4.7\GeV$.
}
\label{tab:dai-violation}
\end{table}

The behaviour of $\delta_{a_I}$ and its uncertainty as a function of $\alpha_{\rm CKM}$ are shown in Fig.~\ref{fig:delta-ai}.
We provide a fit for this plot,
\begin{equation}
\left[\delta_{a_I}\right]_{\rm LZ} = 0.10 - 0.87\cos\alpha_{\rm CKM}  \;,
\end{equation}
where the $1\sigma$ bounds are given by $\left[\delta_{a_I}\right]_{\rm LZ} \pm \left[\sigma\right]_{\rm LZ}$ where $[\sigma]_{\rm LZ} = 0.10 + 0.14\cos^2\alpha_{\rm CKM}$ is a good fit for 
the error band shown in Fig.~\ref{fig:delta-ai}.

It is clear that the structure of the SM is responsible for the smallness of 
$\delta_{a_I}$ \eqref{eq:delta-ai-def}.  In general the quantity $\delta_{a_I}$ is
 thus highly sensitive to new physics. 
Two examples are:
\begin{itemize} 
\item \emph{Non-MFV isospin violation}: E.g. $(a_i^q|_{b \to s}) / ( a_i^q|_{b \to d}) 
\neq \lambda^{bs}_t /  \lambda^{bd}_t$ C.f.  Eq.~\eqref{eq:OWA} and therafter for 
the definition of $a_i^q$.
\item \emph{UV isospin violation}: Four fermi operators of the type \eqref{eq:OWA} in unequal 
proportion of $q =u,d$ quarks. Such a difference is  sensitive to the structure 
of the $K^{*,0}$ and $\rho^0$ parton content.  
\end{itemize}
We provide some example values for the case of UV isospin violation in Tab.~\ref{tab:dai-violation}.

\section{Conclusion}
\label{sec:discussion}

Isospin violating effects considered in this paper are of the UV-type, i.e. $H^{\rm eff}$ is asymmetric under $u \leftrightarrow d$, as well as IR-isospin violation which manifests itself in photon emission from the spectator quark.  
We have found that CP-averaged isospin asymmetries in 
the SM are small, below $1.5\%$, for $B \to (K,K^*\rho) ll$ at lepton  pair momentum $1\GeV^2  \leq q^2   \leq 4m_c^2$ as can be inferred from Figs.~\ref{fig:1},\ref{fig:2} 
as well as the actual breakdown of the various operator contributions.
In fact in the SM the $(K,K^*)ll$ and $\rho ll$ are somewhat accidentally small. 
In the former case  the large Wilson coefficient $C_2^u$ is  suppressed by a small CKM-prefactor  $|\lambda_u/\lambda_t| \simeq \lambda^2 \approx 0.04$ and in the latter case 
it is the smallness of $\cos \alpha_{\rm CKM}$ which suppresses the tree-level WC $C_2^u$.
The latter point is also the reason why the non CP-averged isospin asymmetry for 
$B  \to \rho ll$ deviates from the CP-averged one c.f. section \ref{sec:CP-averaging}.
Not performing the CP-average, which is possible for the $\rho$ and $K^*$, is certainly 
an interesting option for the former per se c.f. Fig.\ref{fig:2}(bottom) and for the latter in presence of new weak phases.

Isospin asymmetries of $B \to K^*/\rho \gamma$ are a bit higher, around $5\%$ each, 
due to the photon pole enhancement and are measured with reasonable accuracy (\ref{eq:HFAG},\ref{eq:Delta_rho}).  Whereas the $K^*$ experimental result is in perfect agreement with our prediction the $\rho$-asymmetry is off by two standard deviations
and calls for further experimental data. In both cases we use these results to give indicative constraints on the WA WCs, c.f. Tabs.~\ref{tbl:kstar-constraints},\ref{tbl:rho-constraints}, by demanding that no coefficient is more than two standard deviations away from the experimental results.
The smallness of $\cos \alpha_{\rm CKM}$ implies that the $K^*/\rho \gamma$ SM isospin asymmetries are
structurally very similar, resulting in the almost identical numerical result, which prompted us define a (quasi) null test of the SM $\delta_{a_I}$ in section \ref{sec:dai}.

We have not systematically investigated the isospin asymmetry in the high $q^2$-region in this work.
Nevertheless we have argued that it has to be small as it is a) no longer artificially enhanced as at low $q^2$ by the photon pole, through which isospin effects propagate, b) 
its contribution is further suppressed relative to the form factor contributions 
as the latter feel the closeness of the t-channel pole at $q^2 = m_{B_s^*}^2$. On grounds 
of these arguments, modulo magic 
cancellation at low $q^2$, one expects the isospin asymmetry to decrease. 
We should add, that 
the authors of reference \cite{Beylich:2011aq} had come to the same conclusion 
using arguments of a high $q^2$-OPE.

We have introduced the most general basis of dimension six operators 
for WA \eqref{eq:OWA} and QLSS \eqref{eq:OQLSS} and have detailed 
various contributions in Tabs.~\ref{tab:2},\ref{tab:1},\ref{tab:1rho} in appendix \ref{app:tables} as well as Figs.\ref{fig:2},\ref{fig:2rho} in section \ref{sec:isospinBSM} 
respectively.  
Generic selection rules for the $B \to Kll$, valid for any scalar $\to$ scalar $ll$ transition, 
were discussed in section \ref{sec:Ksel}.  Selection rules  for WA, which are more stringent, and  were worked out in section \ref{sec:WAsel} for vectors and pseudoscalars. After applying all selection rules $24$ and $10$ operators remain for 
a vector and pseudoscalar meson final state, which 
compares with $7$ operators in the SM for the $K^*$ and the $K$-meson.
 In view of the large number of operators that can contribute,
as detailed in Fig.~\ref{fig:1} and Tabs.~\ref{tab:2},\ref{tab:1} respectively, one might 
even wonder whether by the laws of probability cancellation of new physics is the 
rule rather than the exception. One would hope that a refined experimental analysis in 
$q^2$ would reveal the deviation in one bin or another.
In this paper we have not attempted to constrain the four Fermi operators through non-leptonic decays but have, for the time being\footnote{The tables \ref{tab:2},\ref{tab:1} and \ref{tab:1rho} can be obtained from the authors on request.}, contented ourselves with a few generic remarks in section \ref{sec:constraints}. 
In the future data from isospin asymmetries in $B \to P,V \gamma/ll$ could be combined 
with data in non-leptonic decays $B \to PP,PV,VV$ to constrain four Fermi operators 
of $b s(d) qq$-type more effectively.

On the theoretical side the SM isospin prediction would benefit from an evaluation 
at ${\cal O}(\alpha_s)$ of the WA contribution. This computation would also 
be beneficial to understand  $D^0 \to V ll/\gamma$ decays \cite{LZ12}. 
For the BSM analysis a computation of QLSS within LCSR would be desirable  for 
the reasons mentioned at the beginning of the section \ref{sec:QLSS}.

 We explained why the $K$ and the $K^*_\parallel$ contribution are linked 
 at leading twist and for left-handed currents only. Thus the relation 
 between the $Kll$- and $K^*ll$-asymmetry is therefore already only approximate in the SM and lost entirely should there be sizeable \VpA{} structures.
 In view of the experimental results we therefore conclude: 
 whereas it is very plausible 
that the $K$ isospin asymmetry is larger or very different from the $K^*$ isospin asymmetry, it remains mysterious at this moment why it would be sizeable 
at high $q^2$ at all. In fact, in view of this and the smallness of the prediction in the low $q^2$-region, c.f. Fig.~\ref{fig:1} (top,right), the measured deviation of the integrated isospin asymmetry in $B \to Kll$ of the LHCb collaboration \cite{Aaij:2012cq} away from zero by four standard deviations is somewhat puzzling. 
More statistics, especially in the neutral channel, is therefore eagerly awaited.  \\

\section*{Acknowledgements}

This work has benefited from discussion and or correspondence with Christoph Bobeth, Greig Cowan, Ulrik Egede, Thorsten Feldmann, Uli Haisch, Gudrun Hiller, Franz Muheim, Leonardo Vernazza and especially Steve Playfer.
Part of the computations in this paper were performed by the help of FeynCalc 
\cite{Mertig:1990an}.
RZ gratefully acknowledges the support of an advanced STFC fellowship.

\appendix
\numberwithin{equation}{section}

\newpage
 
\section{${\cal B}(B^0 \to K^{*0}\gamma)/{\cal B}(B_s \to \phi \gamma)$} 
\label{app:update}

The LHCb collaboration has recently measured \cite{Aaij:2012ita} the ratio of branching fractions
of $B^0 \to K^{*0}\gamma$ to $B_s \to \phi \gamma$ to be
\begin{equation}
\label{eq:Rbr}
R_{K^*\phi} \equiv \frac{\mathcal B (B^0\to K^{*0}\gamma)}{\mathcal B (B^0_s\to\phi\gamma)}  = 1.23(6)_{\rm stat} (4)_{\rm syst} (10)_{f_s/f_d} \;,
\end{equation}
where the uncertainties are statistical, systematic and due to  $s,d$-fragmentation. 

In the SM the difference to unity of \eqref{eq:Rbr} is mainly due to the ratio of form factors. Generically a difference can arise from WA and this where it connects to the rest of this work.  We shall give an update of the form factor ratio below, discuss an example of how 
an enhancement of one of the operators in \eqref{eq:OWA} can lead to sizeable deviations from the SM-value.

\subsection{Form factor ratio update}
We present a phenominological update of the form factor ratio, 
\begin{equation}
\label{eq:r}
r_{K^*\phi} = \frac{T_1^{B \to K^* \gamma}(0)}{T_1^{B_s \to \phi \gamma}(0)} = 0.89(10)\% \;,
\end{equation}
using the results in \cite{Ball:2004rg} with the same hadronic input as in \cite{DLZ12}. 
The uncertainty consists of an estimate of violation of semi-global quark hadron duality 
as well as a parametric error.
 The first uncertainty is obtained by varying the continuum thresholds $s_0^{B \to K^*,B_s \to \phi}$ separately 
and adding them in quadrature. It leads to an $\Delta_{s_0} \approx 4\%$ uncertainty. 
We fix $s_0^{B \to K^*},s_0^{B_s \to \phi} = 35(1),36(1)\GeV^2$ which 
is consistent with   $s_0^{B \to K^*}-s_0^{B_s \to \phi} \approx m_{B_s}^2 - m_{B_d}^2$. 
The second uncertainty is obtained by varying all other parameters and  adding  them in quadrature which leads to a $\Delta_{\rm para}  \approx 6\%$-uncertainty. 
Possibly we should add that we vary the $K^*$ and $\phi$ decay constants separately 
but vary $f^\perp$ and $f^\parallel$ in a correlated way as the ratio is known from 
lattice QCD to a high precision.  This leads to either a $(\Delta_{\rm para}^2+ \Delta_{s_0}^2)^{1/2} = 7\%$ or $\Delta_{\rm para} + \Delta_{s_0} = 10\%$ error depending
on whether the two uncertainties are added in quadrature or linearly. We chose to quote 
the more conservative error in Eq.~\eqref{eq:r} above.

It would seem worthwhile to compare the central value with previous determinations. 
Taking the ratio of the individual form factor predictions in  \cite{Ball:2004rg} we get 
$r_{K^*\phi} = 0.95 + 0.93 (a_1(K^*)-0.1)$. In this formula $a^\perp_1(K^*) = a_1^\parallel(K^*)$ was assumed which is still a reasonable rule in view of current determinations 
$a^\perp_1(K^*) = 0.04(3)$ and  $a_1^\parallel (K^*)=0.06(4)$ as used in \cite{Ball:2004rg}. Taking the average value of the two Gegenbauer moments one gets, $r_{K^*\phi} = 0.91$, a value rather close to  \eqref{eq:r}\footnote{In \cite{Ball:2006eu} $r_{K^*\phi} = 0.99(13)$ was quoted based on some input from lattice QCD on the ratio of $f_{B_s}/f_{B_d}$ for which there is no reason if the $f_{B_{d,s}}$ are taken from sum rules to the same order which is a consistent procedure.}.

One might further wonder why $r_{K^*\phi}$ is about $18\%$ lower than a naive estimate 
$f_{K^*}^\perp/f_\phi^\perp$. 
We identify four main effects: 1) $-6\%$ due to 
$m_\phi \neq m_{K^*}$ 2) $-3\%$ $s_0^{B \to K^*}\neq s_0^{B_s \to \phi}$, 3) $-5\%$ due to 
$a_1(K^*) \neq a_1(\phi)=0$ and 4) $-2\%$ due to $a_2(K^*) \neq a_2(\phi)$ which adds up to $16\%$ and consists of the bulk effect.

\subsection{Prediction of ${\cal B}(B^0 \to K^{*0}\gamma)/{\cal B}(B_s \to \phi \gamma)$ \& BSM-effect of WA}
The theoretical prediction in the SM is proportional to the form factor ratio 
\begin{equation}
\label{eq:Rschema}
R_{K^*\phi} = |r_{K^*\phi}|^2 c_{K^*\phi}(1+\delta_{WA})
\end{equation}
times a phase space factor (whose uncertainty is almost entirely from to the uncertainty in $\tau_{B_s}$)
\begin{equation}
c_{K^*\gamma} = \frac{\tau_{B^0}}{\tau_{B_s}}\left(\frac{m_{B^0}}{m_{B_s}}\right)^3 \left(\frac{1-m_{K^{*0}}^2/m_{B^0}^2}{1-m_\phi^2/m_{B_s}^2}\right)^3 = 1.01(2)\,;
\end{equation}
and a small correction for WA:  $\delta_{WA} = -0.02(2)$. Thus essentially 
$R_{K^*\phi}|_{\rm SM} \approx |r_{K^*\phi}|^2$
Finally we shall requote experimental ratio \eqref{eq:Rbr} besides 
our prediction assembling all three quantities in \eqref{eq:Rschema}:
\begin{align}
R_{K^*\phi}|_{\rm LHCb}  = 1.23(6)_{\rm stat} (4)_{\rm syst} (10)_{f_s/f_d} \;, \qquad \quad  
R_{K^*\phi}|_{\rm LZ} = 0.78(18) \;.
\end{align}
The theory uncertainty is almost entirely due to the form factor ratio uncertainty \eqref{eq:r}
which is after all not small. Thus new physics would need to manifest itself rather prominently\footnote{Similar remarks would apply to ratio of the kind 
${\cal B}(B^0 \to \rho^{*0}\gamma)/{\cal B}(B^0 \to \omega\gamma)$, as discussed in a previous footnote.} in order to surface above the form factor uncertainty.  
The possibility of which we shall illustrate just below.

The WA processes in $B_s\to\phi$ decay couples to a unique set of operators, and so we can modify the WCs in such
a way as to shift this amplitude without affecting any other process considered in this work.
By way of example, making the shift $a^{u,d,s}_7\to a^{u,d,s}_7 - 0.5$ leads to 
$\delta_{WA} \to 0.5(3)$, and therefore $R_{K^*\phi}\to1.2(3)$, without affecting any other FCNC process we are considering.
This result cannot be derived by a simple rescaling of the results in Tab.~\ref{tab:1} because the effect
of Gegenbauer moments in the $B_s\to\phi$ WA amplitudes are significant.

An extension of this analysis to the differential branching fractions 
$(B^0)B_s \to K^{*0}(\phi) \mu^+\mu^-$, which have recently been measured 
by the LHCb collaboration \cite{Aaij:2013aln}, would be interesting and is deferred 
to later work.

\section{Distribution amplitudes}
\label{app:DA}

\subsection{Light meson DA}
\label{app:light-DA}

We shall briefly summarise and define the DA used throughout this paper. For further 
references we refer the reader to the classic review \cite{Chernyak:1983ej}, the LCSR review \cite{Colangelo:2000dp} and the thorough  paper on higher twist DA  \cite{Ball:1998sk}. 
The $2$-particle DA  for the pseudoscalar at twist-2 ($\phi_K$) and -3 ($\phi_{p,\sigma}$) (e.g. \cite{Ball:2004ye}) is given by
\begin{equation}
\begin{split}
\label{eq:KDA}
\Braket{K(p)|\bar s(x)_a [x,z] q(z)_b|0} =&
\int_0^1 du e^{i(up\cdot x+\bar up\cdot z)} \Bigg[i\frac{f_K}{4 N_c}[\slashed p\gamma_5]_{ba}\phi_K(u) \\
- i\frac{\mu_K^2}{4 N_c} [\gamma_5]_{ba} \phi_p(u)
& - i\frac{\mu_K^2}{24 N_c} p_\mu (x-z)_\nu [\sigma^{\mu\nu}\gamma_5]_{ba} \phi_\sigma(u) \Bigg] + \text{higher twist} \;,
\end{split}
\end{equation}
where $a,b$ are Dirac indices, $\bar u\equiv 1-u$, $\mu_K^2 \equiv f_K m_K^2/(m_s+m_q)$ and the $[x,z]$, here and hereafter,  represent a QCD Wilson line to make the matrix element gauge invariant.
The asymptotic forms\footnote{By asymptotic we mean, as usual, for $\mu_F \to \infty$. All DA depend on the factorisation scale $\mu_F$ of the LC-OPE which we do not indicate explicitly.} of the DA functions are
\begin{align}
\phi_K(u) &= \phi_\sigma(u) = 6\bar uu & \phi_p(u) &= 1
\end{align}
From the appendix of \cite{Ball:2004ye} we see that upon neglecting quark masses and 3-particle DAs,
equations of motion constrain $\phi_{p,\sigma}(u)$ to their asymptotic forms.
$\phi_K(u)$ is expanded in Gegenbauer moments as usual.

The $2$-particle DA  for the vector meson at twist-2 ($\phi_{\parallel,\perp}$) and -3 ($g^{(v,a)}_{\perp}$)  (e.g. \cite{Ball:2004rg}) is
\begin{equation}
\begin{split}
 \Braket{K^*(p,\eta)|\bar s(x)_a [x,z] q(z)_b|0}
= \int_0^1 du\,e^{i(up\cdot x+\bar up\cdot z)}\Bigg\{\frac{f^\perp_{K^*}}{4 N_c}\Bigg[(\slashed\eta\slashed p)_{ba}\phi_\perp(u) \\
 - \frac{i}{2}(1)_{ba}(\eta\cdot(x-z)) m_{K^*}^2 h^{(s)}_\parallel(u)
 -i ( \sigma_{\mu \nu})_{ba} p^\mu  (x- z)^\nu   \frac{\eta\cdot (x-z)}{(p\cdot (x-z))^2} m_{K^*}^2 h^{(t)}_\parallel(u) \Bigg]\\
 + \frac{m_{K^*}f_{K^*}}{4 N_c}\Bigg[(\slashed p)_{ba} \frac{\eta\cdot (x-z)}{p\cdot (x-z)}  \phi_\parallel(u)
 + \left(\slashed\eta - \slashed p \frac{\eta\cdot(x-z)}{p\cdot(x-z)}\right)_{ba} g^{(v)}_\perp(u) \\
 + \frac{1}{4}\epsilon_{\mu\nu\rho\sigma}\eta^\nu p^\rho (x-z)^\sigma \left(\gamma^\mu\gamma_5\right)_{ba} g^{(a)}_\perp(u)\Bigg]\Bigg\} + \text{higher twist} \;.
\end{split}
\label{eq:vector-da}
\end{equation}
The asymptotic DAs are
\begin{gather}
\phi_\perp(u) = \phi_\parallel(u) = g^{(a)}_\perp(u) = h_\parallel^{(s)}(u) = 6\bar uu  \nonumber \\
\begin{align}
g^{(v)}_\perp(u) &= \frac{3}{4}\left(1 + (u-\bar u)^2\right) &
h^{(t)}_\parallel(u) &= 3(u-\bar u)^2 \;.
\end{align}
\end{gather}
In fact, these functions overparametrise the $K^*$ state and are related by QCD equations of motion \cite{Ball:1998sk}.
In the limit of three massless quark flavours, the relevant constraint for QLSS reads\cite{Kagan:2001zk}\footnote{This may be obtained from the equation in \cite[eq.8]{Kagan:2001zk}
assuming that $\phi_\parallel(u)=\phi_\parallel(\bar u)$ and likewise for $g^{(v)}_\perp$ and $g^{(a)}_\perp$ which is valid up to small isospin violating terms.
It may more properly be derived directly from the equation of motion in \cite{Ball:1998sk}} 
\begin{equation}
\label{eq:WW}
\int_0^u dv \left[\phi_\parallel(v) - g^{(v)}_\perp(v)\right] = \bar u\left(g^{(v)}_\perp(u) - \frac{g^{(a)'}_\perp(u)}{4}\right) - \frac{g_\perp^{(a)}(u)}{4}\;,
\end{equation}
which is used to eliminate the integral from \eqref{eq:charm-loop-perp}.
We also require the identity
\begin{equation}
\label{eq:WW2}
u h_\parallel^{(t)}(u) + \frac{u}{2} h_\parallel^{(s)'}(u) = 2\int_0^u \left(h_\parallel^{(t)}(v) - \phi_\perp(v)\right) dv
\end{equation}
in order to show gauge invariance in in $O^{\rm WA}_2$ results in appendix \ref{app:wa-formulae}.
To this end we note that
Eqs.(\ref{eq:WW},\ref{eq:WW2}) follow from equations (4.15/16) and (3.21/22) in \cite{Ball:1998sk}.

\subsection{Photon DA}
\label{app:photon-DA}
The leading twist $2$ photon DA \cite{Ball:2002ps} is:
\begin{alignat}{2}
& \Braket{\gamma(q,\epsilon)|\bar q_a(x) [x,z] q_b(z)|0} =  i e  \int_0^1 d^4 y 
\epsilon^*_\mu e^{i q \cdot y} \matel{0}{T \bar q_a(x) [x,z] q_b(z) j_{\rm em}^\mu(y)}{0} =  
\nonumber \\[0.1cm]
&     \qquad     \frac{iQ_q\Braket{\bar qq}}{4 N_c}\int_0^1 du e^{i(uq\cdot x + \bar uq\cdot z)}\big(
 \phi_\gamma(u) \sigma^{\alpha \beta} \epsilon_\alpha q_\beta+ (x-z) \!\cdot\! \epsilon \big)_{ba} + \text{higher twist}\;.
\label{eq:photon-da}
\end{alignat}
The first and second term on the last line correspond to the LHS of equation (2.7) in   \cite{Ball:2002ps} and second term on the RHS  of the same equation.
The reason Eq.~\eqref{eq:photon-da} is not gauge invariant is that  $[x,z]$ does not contain the QED (quantum electrodynamical) Wilson line as we expand in the external field to first order. 
Furthermore we have assumed the Lorentz gauge $\partial \cdot  A =0$ through
$A_\mu \to \epsilon_\mu e^{i q \cdot x}\;$\footnote{By working with a plane wave the Lorentz gauge is a natural choice. Note still adhering to the plane wave picture 
the axial gauge  $n \cdot A=0$ with $A_\mu \to ( \epsilon_\mu - (n \cdot \epsilon)/(n \cdot q) q_\mu) e^{i q \cdot x}$ is an alternative. It would amount to replacing the polarisation vector accordingly in the formulae above.}.  
Note that the perturbative photon contribution has to be included separately.
The asymptotic photon DA is given by
\begin{equation}
\phi_\gamma(u) = 6\chi\bar uu \;,
\end{equation}
where $\chi$ is the magnetic susceptibility of the quark condensate, calculated to be $\chi=-3.15(10)\GeV^{-2}$ at $\mu=1\GeV$ in \cite{Ball:2002ps} (the sign is adjusted to our convention of the covariant derivative).

\subsection{$B$-meson DA}
\label{app:B-DA}

The $B$-meson DA used in the QLSS diagrams is given in 
\cite{Beneke:2000wa,Beneke:2000ry}, 
\begin{equation}
\label{eq:fuller-BDA}
\begin{split}
\matel{0}{\bar q_a(x) [x,z] b_b(0) }{B(p_B)}
&= \frac{-i f_B m_B}{4 N_c}\int_0^\infty dl_+ e^{-il\cdot x} \\
\times &\left.\left[
\frac{1+\slashed v}{2}\left\{\phi_+(l_+)\slashed n_+ + \phi_-(l_+)\left(\slashed n_- - l_+\gamma^\nu_\perp\frac{\partial}{\partial l^\nu_\perp}\right)\right\}\gamma_5
\right]_{ba}\right|_{l=\frac{l_+ n_+}{2}}
\end{split}
\end{equation}
where $p_B=m_B v$ and $n_+$ and $n_-$ are light-like vectors
\begin{align}
n_+^2 &= n_-^2 = 0 \;, &
n_+\cdot n_- &= 2 \;,
\end{align}
for which $n_\pm = (1,0,0,\pm1)$ is a possible parametrisation.
This allows an arbitrary vector $x$ to be written as
\begin{equation}
\label{eq:lc-dec}
x^\mu = \frac{x_+ n_+^\mu + x_- n_-^\mu}{2} + x_\perp^\mu\;
\end{equation}
and the scalar product of two such vectors reads:
\begin{equation}
\label{eq:lc-SP}
x \cdot y = \frac{1}{2}(x_+ y_- + x_- y_+) + x_\perp 
\cdot y_\perp \;.
\end{equation}
The kinematics required for $B\to K^{(*)}(\gamma^* \to ll)$ are
\begin{align}
\label{eq:lc-dynamics}
p_+ &= m_B - \frac{q^2}{m_B} \;,  & p_- &= 0 \;, &
q_+ &= \frac{q^2}{m_B} \;,  & q_- &= m_B\;,
\end{align}
with $p_\perp = q_\perp = 0$.

Furthermore we take $\phi_+$ and $\phi_-$ to be the model functions defined in \cite{Grozin:1996pq}
\begin{align}
\phi_+(\omega) &= \frac{\omega}{\omega_0^2}e^{-\omega/\omega_0}  \;,&
\phi_-(\omega) &= \frac{1}{\omega_0}e^{-\omega/\omega_0} \;,
\end{align}
with $\omega_0=2\Lambda_{\text{HQET}}/3 \simeq 0.4 \GeV$.
Our results contain the moment functions
\begin{equation}
\lambda_\pm^{-1}(q^2) = \int_0^\infty dl_+ \frac{\phi_\pm(\omega)}{l_+ - q^2/m_B - i\epsilon}
\end{equation}
which evaluate to
\begin{align}
\lambda_+^{-1}(q^2) &= \frac{1}{\omega_0}\left[1 + ye^{-y}(i\pi - \mathrm{Ei}(y))\right] \;,  &
\lambda_-^{-1}(q^2) &= \frac{e^{-y}}{\omega_0}(i\pi - \mathrm{Ei}(y)) \;,
\end{align}
where $y=q^2/\omega_0 m_B$ and the function $\mathrm{Ei}$ is the exponential integral.

\subsubsection{On corrections for QLSS within QCDF}
\label{app:O(q2/mB2)}
We would like to discuss the origin of the ${\cal O}(q^2/m_B^2)$-corrections due to neglecting the $l_-$-direction and the $l_\perp$-derivative alluded 
to in section \ref{sec:quark-loop-spectator}. Further comments can be found in that section.  

\begin{itemize}
\item \emph{Neglecting the $l_-$-direction:}
We wish to stress a particular feature of the $B$-meson DA:
it takes the form of a function of a light-cone coordinate.
 In the $B$-meson rest frame
all components of the spectator quark momentum are expected to be of comparable magnitude $\calO(\Lambda_{\mathrm{QCD}})$ and thus the special r\^ole of $l_+$ originates from 
the dynamics. 
It turns out that to leading order in $1/m_B$ the short distance part of the matrix element is only sensitive to the $l_+$
component of the light quark momentum\footnote{This is the component in the light-like direction which is parallel to the final state light meson.},
and hence a light-cone DA is what is required \cite{Beneke:2000ry}.
The next-leading order diagram  in Fig.~ \ref{fig:quark-loop-spectator} with photon emission next to the $B$-meson is sensitive to $(q-l)^2 = q^2 + l^2 - q_+ l_- - q_- l_+$ ($q_\perp =0$ in the notation of \eqref{eq:lc-SP}) with $q_- = m_B$ and 
$q_+ = m_B (q^2/m_B^2)$. 
Thus at $q^2=0$ the process 
depends only on the $l_+$ direction but becomes increasingly sensitive to $l_-$ direction as $q^2$ rises.  At $q^2 = 4 m_c^2$ this amounts to about a $30\%$-effect ($q_+/q_- \simeq 0.3$).

\item \emph{Neglecting the $l_\perp$ derivative:}
The derivative w.r.t. $l_\perp$ will be $1/m_B$ suppressed, as compared to the other term 
coupling to $\phi_-(l_+)$, except for the case where 
the photon is emitted from the light quark originating from the $B$-meson (same diagram as discussed in the previous point). 
The effect on the corresponding light quark propagator  $S_F$ is
\begin{equation}
l_+ \gamma_\perp^\rho \frac{\partial}{\partial l_\perp^\rho}\gamma_5\gamma^\mu S_F(q-l)
= i l_+ \gamma_5 \gamma_\perp^\rho \gamma^\mu \left(\frac{\gamma^\rho_\perp}{(q-l)^2} + 2\frac{\slashed q-\slashed l}{(q-l)^4} l^\rho_\perp\right) = {\cal O}(\Lambda_{\mathrm{QCD}}/m_B) \;.
\end{equation}
This comes about as follows: 
the second term vanishes by setting $l_\perp=0$ after taking the derivative.
To analyse the first term we contract $\gamma^\rho_\perp \slashed\epsilon \gamma_{\perp,\rho}$ with a polarisation vector $\epsilon_\mu$, using 
the light-cone decomposition \eqref{eq:lc-dec}, 
\begin{equation}
l_+ \gamma^\rho_\perp \slashed \epsilon \gamma_{\perp,\rho}
= -l_+\left[\slashed n_+\epsilon_+ + \slashed n_-\epsilon_-\right] + l_+ \underbrace{\gamma^\rho_\perp \slashed\epsilon_\perp \gamma_{\perp,\rho}}_{=0} \;,
\end{equation}
where the first part must be compared with the other structure coupling to $\phi_-(l_+)$:
\begin{equation}
i \slashed n_- \slashed \epsilon (\slashed q-\slashed l) =
\frac{i}{4} \slashed n_-\left[\epsilon_+ q_- + \slashed \epsilon_\perp \slashed n_+ (q_+ + l_+)\right]
\end{equation}
We therefore see that the derivative term is subleading in the $\epsilon_+$ coefficient by $l_+/q_-={\cal O}(\Lambda_{\mathrm{QCD}}/m_B)$.
The $\epsilon_-$ coefficient must be related by gauge invariance; see appendix \ref{app:QL-GI} for further discussion. At last we would like to mention that it would be interesting to study whether there is any significant change for $0 < q^2 < 4 m_c^2$.
\end{itemize}

\section{Helicity projectors}
\label{app:decay-width-tensors}

As hinted in the main text the basis ${\cal T}_{1,2,3}$, or more precisely ${\cal T}_{2,3}$, is not ideal for addressing
physical quantities. In the decay rate this emerges in two ways.  First a particular direction 
namely zero helicity has to be ${\cal O}(m_V)$ \eqref{eq:X2X3general} and furthermore the directions ${\cal T}_{2,3}$ are not orthogonal to each other.  
In the main text we have given the transformation to the helicity basis $h_{0,\pm}$ in Eq.~\eqref{eq:Btrafo}. We shall give the Lorentz structures for the latter and discuss 
a few more details.
As in \eqref{eq:amplitude-definition} we define:

\begin{align}
_{\rm out}\left\langle V(p,\eta) l^+(l_1) l^-(l_2)\mid B(p+q)\right\rangle_{\rm in} &= \frac{G_F}{\sqrt 2}\lambda_t \frac{\alpha}{q^2\pi}
 \left(\calT^{V\mu}\bar u(l_1)\gamma_\mu v(l_2) + \calT^{A\mu}\bar u(l_1)\gamma_\mu\gamma_5 v(l_2)\right) \;, \nonumber 
\end{align}
where
\begin{equation}
\calT^{(V,A)\mu} = \sum_{i = \pm,0}  \h^{(V,A)} P_i^\mu  \;,
\end{equation}
and the helicity basis tensors are given by 
\begin{align}
P^\mu_\pm &= 
\frac{1}{\sqrt{2}}\left[2\epsilon^{\mu\nu\rho\sigma}\eta_\nu p_\rho q_\sigma \mp  \frac{i}{\sqrt{\lambda_V}}
\left(\lambda_{V} m_B^2 \eta^\mu -
2(\eta\cdot q)\left((1-\hat  m_V^2-\hat  q^2)p^\mu - 2\hat  m_V^2 q^\mu\right)  \right)\right]
\;,  \nonumber \\ 
P_0^\mu &= \frac{4 i \hat m_V}{ \sqrt{2\hat   q^2 \lambda_{V}}}(\eta\cdot q) \left[2 \hat q^2 p^\mu - (1- \hat m_V^2-\hat  q^2) q^\mu \right] \;. \nonumber
\end{align}
Writing $\vec{P}_h = (P_0,P_+,P_-)$ and $\vec{P} = (P_1,P_2,P_3)$, suppressing a Lorentz index for the time being, the transformation 
follows from the transformation \eqref{eq:Btrafo} through $\vec{P}_h = (B^T)^{-1} \vec{P}$.
The Lorentz structures have the following properties: $q \cdot P_{\pm,0} = 0$
and $p \cdot P_{\pm} = 0$. It seems worthwhile to mention that the
 $m_V$-factor in $P_0^\mu$ cancels against the $1/m_V^2$ originating from the polarisation sum:
$\sum_{\text{pol.}} = \eta_\mu \eta_\nu  =  (p_\mu p_\nu/m_V^2-g_{\mu\nu})$. This 
assures finiteness of the rate as $m_V$ approaches zero, provided that $\h_0 = {\cal O}(m_V^0)$. 

\section{Gauge invariance}
\label{app:gauge-invariance}

\subsection{WA contact terms and gauge invariance}
\label{app:wa-gauge-invariance}

Before discussing the problem in more detail let us state a few general facts, some of which have already been stated in the main text.  
\begin{itemize}
\item[(i)] GI \eqref{eq:WIem} is satisfied in $Q_b$ and $Q_q$ terms separately.
\item[(ii)] When there is ISR as well as FSR then one needs to approximate ISR and FSR 
consistently in order for (i) to be true.
\item[(iii)] In the neutral case (i.e. $Q_b = Q_q$), (ii) can be circumvented, at the cost of (i), 
as ISR and FSR are separately gauge invariant.
\item[(iv)] Statement (ii) can be circumvented in the case when the four quark operator is
of the current-current type (${\cal O}^{{\rm WA}}_{5-8}$ \eqref{eq:OWA}).
In this case FSR corresponds to a contact term, up  
to corrections ${\cal O}(m_{q,s})$,  which is easily computed using the weak WI.
\end{itemize}

In previous computations \cite{Ali:1995uy,Khodjamirian:1995uc}, 
as discussed in \cite{Khodjamirian:2001ga} in more clarity, (iv) applied as in the SM
only  ${\cal O}^{{\rm WA}}_{5-8}$-type \eqref{eq:aSM} are significant
in the absence of CKM suppression;
then (ii) does not apply as there is either only ISR or FSR c.f.Tab.~\ref{tab:OWA}.
Inspecting Tab.~\ref{tab:OWA} we see that ${\cal O}^{{\rm WA}}_4$ for the $K$ is the only problematic case which we shall discuss in some more detail in subsection \ref{app:wa-contact-terms} below.  Furthermore statement (iv) is explained in section \ref{app:vector-current}. 
The modification of the issue of contact terms  for $q^2 =0$, which implies substituting the quark condensate terms $\vev{qq}/q^2$ for the the photon DA, 
is outlined in appendix \ref{app:qq-photonDA}.

\subsubsection{Eliminating parasitic cuts - spurious momentum $k$}
\label{app:wa-contact-terms}

\begin{figure}
\center
\includegraphics[width=0.44\textwidth]{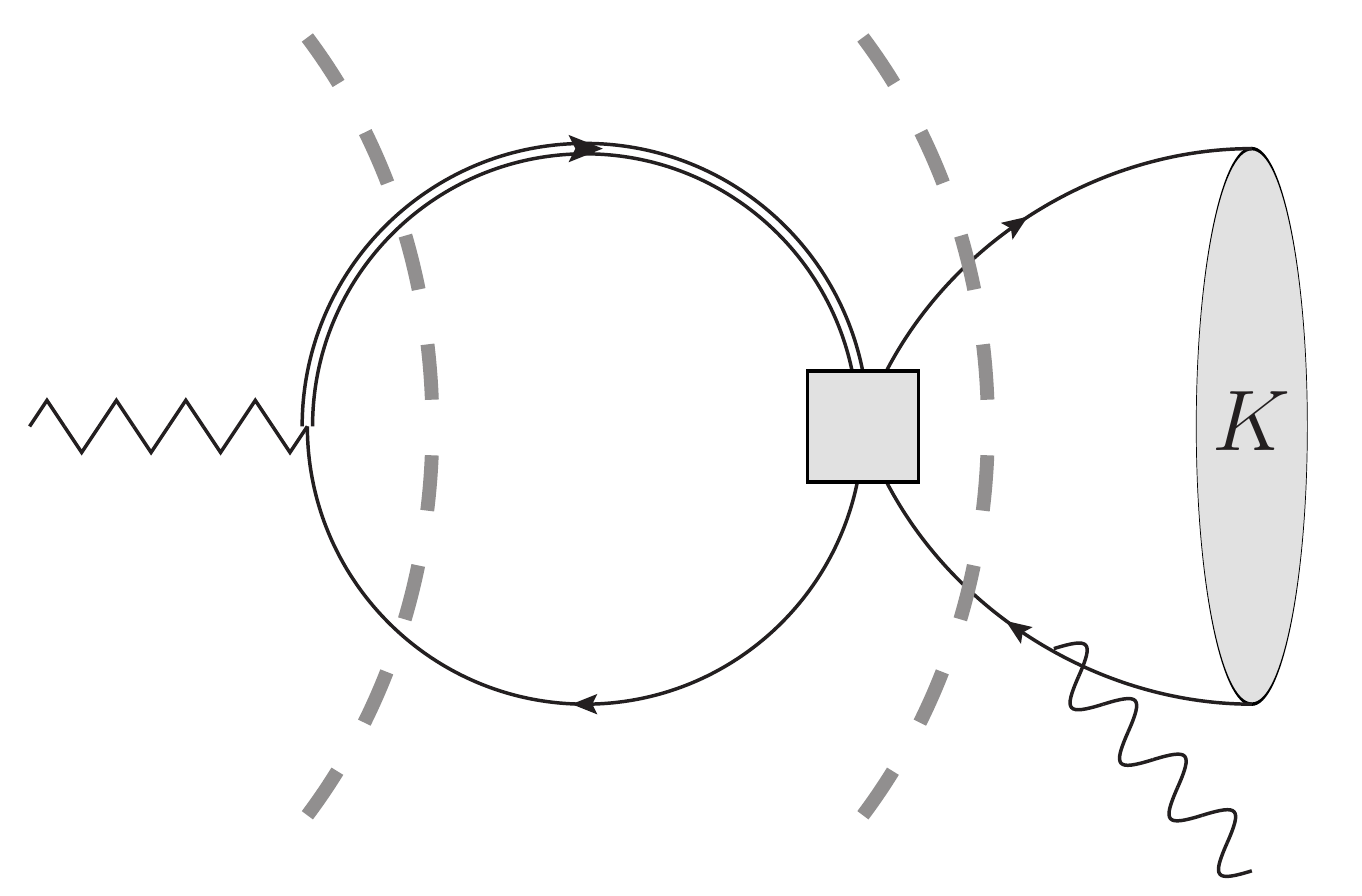}
\includegraphics[width=0.51\textwidth]{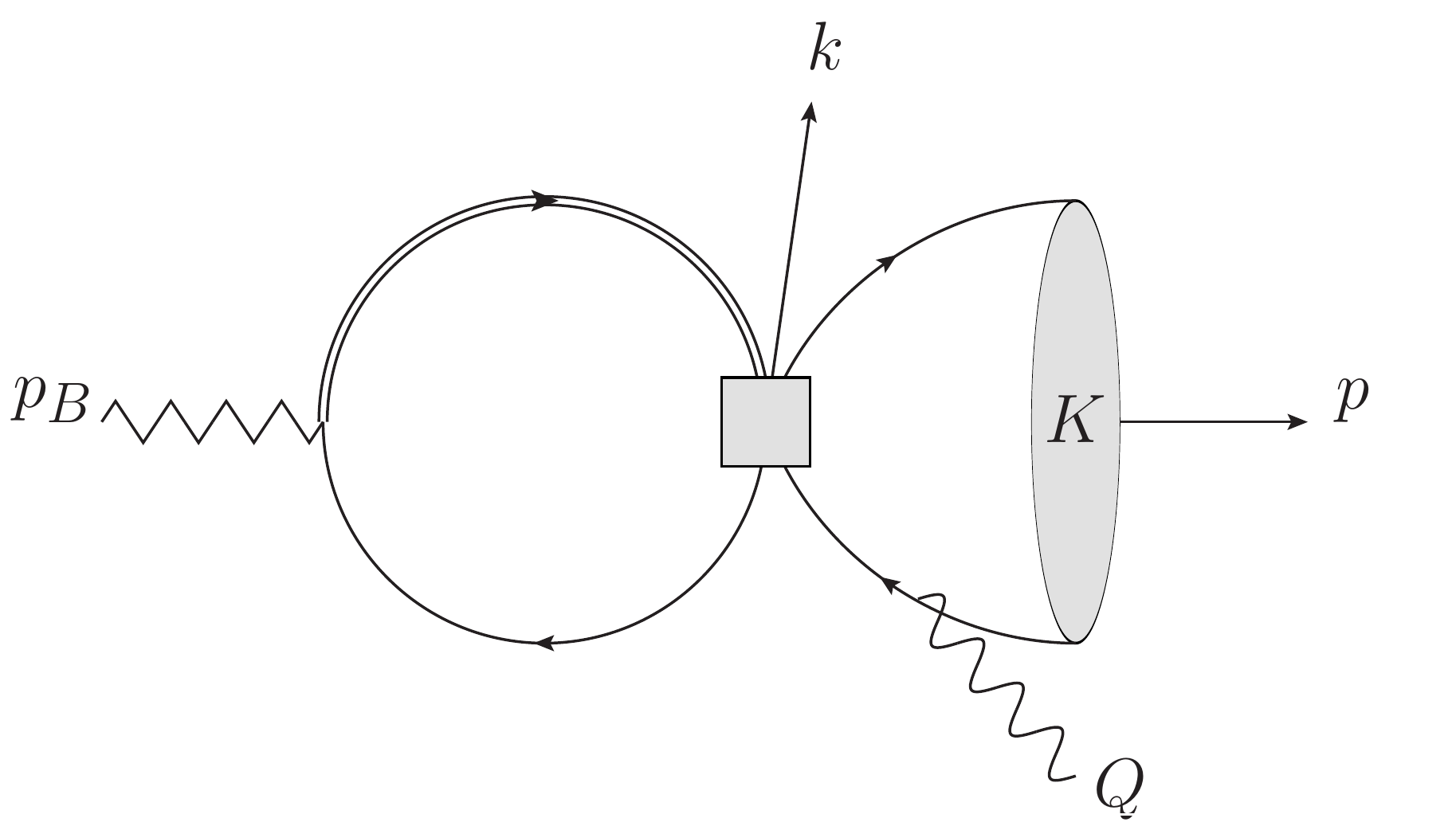}
\caption{\small
(left) Momentum assignments for $B\to K\gamma^*$ including an additional momentum at the weak vertex.  (right) Dashed lines denote possible cuts with momentum $(p+q)^2$ flowing through them, which contribute to a na\"ive sum rule.
The right hand cut is a parasite because the lines it cuts do not have the quantum numbers of the $B$-meson
so should not contribute to a dispersion relation for the $B$-meson current current $J_B$.}
\label{fig:wa-k-extended}
\end{figure}

It is at this point we must point out that factorising the WA matrix elements conceals a problem in constructing sum rules: the problem of parasitic cuts.
As shown in Fig.~\ref{fig:wa-k-extended}(left) a na\"ively constructed sum rule will receive contributions from cuts
which do not have the same quantum numbers as the $B$-meson.
A solution to this problem was introduced in \cite{Khodjamirian:2000mi} where spurious  momentum $k$ is introduced at the weak vertex,
which gives the second cut in Fig.~\ref{fig:wa-k-extended}(left) momentum $(p_B-k)^2$ which is 
distinct from  $p_B^2$ of the first cut. How the effect of the momentum $k$ 
is eliminated from the final result is to be discussed shortly below. 
We use the momentum assignments shown in Fig.~\ref{fig:wa-k-extended}(right) and reuse the modified basis tensors from \cite{DLZ12},
which are given by
\begin{equation}
p_T^\rho  =  i  \left[Q^\rho - \frac{q^2}{Q \cdot (p_B+p)} (p_B + p)^\rho\right] \;, \quad 
\bar p_{\bar T}^\rho  =  i \left[k^\rho - \frac{k \cdot Q}{Q \cdot (p_B+p)} (p_B + p)^\rho\right]
\;,
\label{eq:ps-extended-basis}
\end{equation}
where $Q \equiv q - k$. Amongst the possible six invariants four are fixed as  \begin{align}
p^2 &= m_{V,P}^2 =  0 &
k^2 &= 0 &
Q^2 &= q^2
\end{align}
and the remaining two invariants  $p_B^2$ and  $P^2 = (p_B -k)^2$ correspond  
to the two cuts shown in Fig.~\ref{fig:wa-k-extended}(left).
$p_B^2$ is the dispersion variable and the second cut variable $P^2 = (p_B -k)^2$, is  
the only trace  of the spurious momentum. This is eliminated by setting $P^2 = m_B^2$ as  $p_B^2 \simeq m_B^2$ by virtue of 
Eq.~\eqref{eq:final_SR}. 
After projection onto this extended basis, the coefficient of $p_T^\rho$ corresponds to the coefficient of $P_T^\rho$ which appears in the decay rate.

\subsubsection{Vector \& axial $4$-quark operators and the weak WI }
\label{app:vector-current}

In this subsection we dwell in more detail on how point (iv) at the beginning of this appendix unfolds. According to statement (ii)  ISR and FSR then ought to be treated 
consistently, as for instance outlined in the previous subsection, in order 
to maintain GI. 
The problem in distinguishing the two cuts shown in Fig.\ref{fig:wa-k-extended}(right)
applies only to the FSR diagram\footnote{For ISR radiation the second cut corresponds to the kaon as the momentum flowing into $4$ quark operator is the kaon final state momentum.}. We shall see just below that in the case where 
the current with kaon quantum numbers is of the vector and axial type 
the diagram is a pure contact term by virtue of the weak WI and only produces a 
gauge variant part. 
This means that it does not carry any non-trivial dynamics and that  
its sole purpose is to render the matrix element gauge invariant.  
Furthermore we note that adding a spurious momentum does not have any impact on the diagram. 
In two subsequent paragraphs  we are going to show this through the weak WI 
and infer the same result by sketching an explicit
computation. 

\paragraph{Weak WI}
\label{app:weakWI}
Consider the FSR for the operator ${\cal O}_6^{{\rm WA}}$ at leading order ${\cal O}(\alpha_s^0)$, 
\begin{eqnarray}
 \matel{K^* \gamma}{\bar s\gamma^\mu q}{0} \matel{0}{\bar q\gamma_\mu\gamma_5 b}{B}
&=& -e  f_B (p_B)_\mu \epsilon_\nu \int_x  e^{-ip_B \cdot x} \matel{K^*}{T \bar s\gamma^\mu q(x) J^\nu_{\text{em}}(0)}{0}  \nonumber  \\[0.1cm]
&=&  - e  f_B \epsilon_\nu  \int_x  e^{-ip_B \cdot x} \matel{K^*}{ i \partial_\mu  \left\{ T \bar s\gamma^\mu q(x) J^\nu_{\text{em}}(0)  \right\} }{0}  \nonumber  \\[0.1cm]
&=&  ie(Q_q-Q_s)  f_B \epsilon_\nu \matel{K^*(p,\eta)}{\bar s\gamma^\nu q(0)}{0}  \nonumber  \\[0.1cm]  &=&   ie (Q_q-Q_s) f_B f_{K^*}m_{K^*}(\eta\cdot\epsilon)   
\label{eq:wa-by-wti}
\end{eqnarray}
where we have used $\matel{0}{\bar q\gamma_\mu\gamma_5 b(0)}{B(p+q)} = if_B(p+q)_\mu$ in the first equality, 
\begin{equation}
\label{eq:weakWI}
\text{weak WI:} \quad  \partial_\mu  \bar s\gamma^\mu q(x) = {\cal O}(m_s-m_q) \to 0
\end{equation}
in the second equality 
and $[Q, {\cal O}] = Q_{\cal O} {\cal O}$ ($Q=\int d^3 x J_{\rm em}^0$)  in the third equality. The last 
step is due to the  
definition of the $K^*$ decay constant: $\matel{K^*(p,\eta)}{\bar s\gamma^\nu q(0)}{0}=m_{K^*}f_{K^*}\eta^\nu$.  As stated previously this contribution 
is a pure gauge variant term.
Adding a spurious momentum $p_B \to p_B -k$ would not change anything in 
the derivation. Similarly we find for the $K$:
\begin{equation}
\label{eq:seconds}
\matel{K \gamma }{\bar s\gamma^\mu\gamma_5 q}{0} \matel{0}{\bar q\gamma_\mu\gamma_5 b}{B} = e  (Q_q-Q_s)f_B f_K (p\cdot\epsilon)
\end{equation}
We would like to remark that the results in (\ref{eq:wa-by-wti},\ref{eq:seconds}) are correct to all orders in QCD (with $m_q = 0$).

In summary the weak WI \eqref{eq:weakWI} replaces the computation. 
In the next paragraph we shall outline the main points of  the explicit  LC-OPE computation which comes to the same conclusion.

\paragraph{Explicit computation}
\label{app:explicit}

We would like to mention one additional point:
the reader may wonder whether we could have simply used the $K^*$ DAs in \eqref{eq:vector-da},
worked out the result and not had to concern ourselves with arguments based on WIs.
It turns out that \eqref{eq:vector-da} is in fact insufficient for this purpose 
and the Wandzura--Wilczek type relation \eqref{eq:WW} has to be used.
We have checked that this leads to the same result up to $O(m_{K^*}^2)$ terms.  
The latter are of twist-$4$ and 
expected to be there as we have consistently neglected them throughout this work.

\subsubsection{Remarks on gauge invariance and contact terms at $q^2  =0$.}
\label{app:qq-photonDA}

 At $q^2 =0$  the $\vev{\bar q q}$-term 
from the light quark propagator, originating from the interpolating current $J_B$, 
have to replaced by the photon DA. This gives rise to a puzzle as the former are gauge variant whereas DA are usually GI.  The resolution is, as we shall see, that the photon DA 
is QED gauge variant.
 
Generally for $q^2 \neq 0$ and for ${\cal O}^{{\rm WA}}_{5-8}$ 
the term with FSR produces solely a gauge variant contact term proportional 
to $Q_q-Q_b$ c.f. \eqref{eq:wa-by-wti} as discussed in subsection \ref{app:weakWI}. GI is restored by 
a gauge variant term coming from ISR.  
As discussed in the main text section \ref{sec:WAq2eq0} for  $q^2 =0$ the 
the $Q_q \vev{\bar qq}$ is replaced by a photon DA term as depicted in 
Fig.~\ref{fig:wa-diagram-lcope}. This means that 
the ISR and FSR cancellation of gauge non-invariant terms  at the 
$(Q_q - Q_b) \vev{\bar q q}$-level implies that the matrix element used for the photon DA, 
which is usually gauge invariant,  is gauge variant!
This is indeed the case as  the QED Wilson line is absent in the matrix element \eqref{eq:photon-da}  as we expand in the external electromagnetic field.
This can be seen explicitly from the corresponding matrix element which is the sum of an 
 explicit gauge invariant plus a gauge variant term. 
We have checked that, by working in the Lorentz gauge $\partial \cdot  A =0$\footnote{Note, the Lorentz gauge does still allow for residual gauge transformations of the 
form $\epsilon_\mu \to \epsilon_\mu + q_\mu$ for example.}, which is consistent with $A_\mu \to \epsilon_\mu e^{i q \cdot x}$, that the
gauge variant term (corresponding to the second term on the RHS of \eqref{eq:photon-da})
conspires with the  gauge variant terms from the other diagram in Fig.~\ref{fig:wa-diagram-ope}(right) to produce a term proportional to $Q_q-Q_b$ which combines with the contact 
term  Eq.~\ref{eq:wa-by-wti} and leads to a gauge invariant result.

\subsection{QLSS and gauge invariance}
\label{app:QL-GI}

The issue of GI for spectator scattering, in section \ref{sec:quark-loop-spectator}, at $q^2 \neq 0$ is not straightforward. 
In principle we would expect the two diagrams, by which we mean photon emission form the spectator 
quark, in Fig.~ \ref{fig:quark-loop-spectator} to be GI. The computation used in  \cite{Feldmann:2002iw}\footnote{These authors do not discuss the QED GI.}, which we reproduced in this paper for non-SM operators, can only be expected to respect GI at leading order.  Yet, since GI mixes different orders a rigourous test cannot be expected. The recipe of  the pragmatist is then subtract the amount of next leading term 
that renders the leading term GI.
We shall discuss it in more detail below and see that 
a pole in $1/q^2$ supports our argumentation from another point of view. 

In full generality, the $B\to K^* l^+l^-$ decay may be parametrised:
\begin{equation}
\begin{split}
& \matel{K^*(\eta,p) \gamma^*(q,\mu)}{\mathcal H_{\mathrm{eff}}}{B(p+q)}  \\
& \equiv  U^\mu(q^2) = (\eta\cdot q)p^\mu U_p(q^2) + (\eta\cdot q) q^\mu U_q(q^2) + \eta^\mu (p\cdot q) U_\eta(q^2) + i\epsilon^{\mu\nu\rho\sigma} \eta_\nu p_\rho q_\sigma U_\epsilon(q^2) \;.
\end{split}
\end{equation}
QED GI requires the following WI to hold:
\begin{equation}
\label{eq:WIx}
0 = q_\mu U^\mu(q^2) = (\eta\cdot q)\left[(p\cdot q) U_p(q^2) + q^2 U_q(q^2) + (p\cdot q) U_\eta(q^2)\right] \;.
\end{equation}
In best of all worlds, where GI is obeyed exactly, we may choose to eliminate any function by virtue of the equation above.  For example we could solve for either
\begin{alignat}{2}
\label{eq:failure}
& \text{\cross}  \qquad  &U_q(q^2) \to&  -\frac{p\cdot q}{q^2} \left[U_p(q^2) + U_\eta(q^2)\right] \;, \\
\label{eq:procedure}
& \text{\tick}     &U_p(q^2) \to& -U_\eta(q^2) - \frac{q^2}{p\cdot q}U_q(q^2)\; .
\end{alignat}
For\eqref{eq:failure} to be well defined the following relation must hold:
\begin{equation}
\label{eq:lacondition}
U_p(0) + U_\eta(0) = 0 \;,
\end{equation} 
or $U_q(q^2) $ behaves as $1/q^2$, which is not acceptable.\footnote{More precisely integrability of the rate, which we expect, is incompatible with $1/q^2$-behaviour. Note this pole cannot be compensated by virtual correction as in 
$\gamma 
\to ll$ which leads to $1/q^2$ in $|h_\pm|^2$ 
\eqref{eq:vector-decay-rate}.}  
Since $U_p p_\mu$ is power suppressed with respect to $U_q q_\mu$, c.f. \eqref{eq:lc-dynamics}, we cannot expect \eqref{eq:lacondition} to hold,
however applying \eqref{eq:procedure} does work since in contrast to \eqref{eq:failure} it does not contain any $m_B^2/q^2$ enhancement.
Note for the $K$-meson the same discussion applies with $U_\eta|_{K} = 0$ from the start.
The term $U_\epsilon$ is of no relevance for the discussion here.

We will  illustrate this procedure in the case of $B\to K l^+ l^-$.
The result in this case is
\begin{equation}
U^\mu \propto \frac{(l_+ m_B - 2q^2)p^\mu + 2(p \cdot q) q^\mu}{l_+ m_B - q^2}
= 2\frac{(p\cdot q) q^\mu - q^2 p^\mu}{l_+ m_B - q^2} + \frac{l_+ m_B p^\mu}{l_+ m_B - q^2} \;.
\end{equation}
Our procedure \eqref{eq:procedure} 
($U_{\eta}|_{K}=0$) demands that $U_p = -(q^2/(p \cdot q)) \, 2 (p \cdot q) (l_+m_B -q^2)$, 
which amounts to dropping the second term 
on the RHS. This ensures   GI. We reemphasize  that if the $1/m_b$-expansion was implemented to all orders GI would have been automatic.

\section{Details of calculation}

\subsection{Input values}
\label{app:inputvalues}

Here we summarise the numerical input to our calculation for convenience of the reader.
We compute $\alpha_s$ using 2-loop running with 4 or 5 active flavours, with $M_Z=91.1876(21)\GeV$, $\alpha_s(M_Z)=0.1184(7)$ and $m_b(m_b)=4.18(3)\GeV$ in the $\overline{\rm MS}$ scheme \cite{Beringer:1900zz}.
We use  $b$ and $c$ quark masses adapted for the pole mass scheme  
$m_c = 1.4(1)$ and $m_b=4.7(1)\GeV$ in all other cases.
We use a lattice average $f_B=191(5)$ \cite{Bazavov:2011aa,Na:2012kp} when not calculating sum rules, i.e. in the computation of $F_{(2,4),i}$ \eqref{eq:wa-if-split}.
Inputs used to compute WCs are given below Table.~\ref{tbl:wilson-coefficient-example}.
We compute the CKM matrix elements using a Wolfenstein parametrisation expanding up to $\calO(\lambda^2)$ \cite{Wolfenstein:1983yz,Buras:1994ec} with
the parameters $\lambda=0.2254(7)$, $A=0.81(2)$, $\bar\rho=0.131(26)$ and $\bar\eta=0.345(14)$ \cite{Beringer:1900zz}.
For the $\alpha_{\rm CKM}$ dependence of $\delta_{a_I}$ \eqref{eq:delta-ai-def}, we fix the magnitude $|\lambda_u/\lambda_t|$ from the Wolfenstein parametrisation.
All hadronic inputs for the light mesons are as given in our previous paper \cite{DLZ12}.
The new input $\mu_K^2=f_K m_K^2/m_s$ is computed using $m_s(2\GeV)=95(5)\MeV$ \cite{Beringer:1900zz}.
The condensates $\vev{\bar qq}$ and $\vev{\bar qGq}$ are taken to be
$\vev{\bar qq}(1\GeV)=(-0.24(1)\GeV)^3$ and $\vev{\bar qGq}(1\GeV)=(0.8(1) \GeV)^2 \vev{\bar qq}$ \cite{DLZ12}.

\subsection{Error estimation}
\label{app:uncertainty}

We compute error estimates in the following way: the central value of a result is computed using
the central values of all inputs. To compute the error, we then generate a list of pseudo-random sample points
from the probability distributions of the input parameters, and compute the result for each sample point.

For a function $f(x)$, where $x$ represents all $N$ input parameters and is thus $N$-dimensional, the variance is estimated as
\begin{equation}
\sigma^2 = \frac{1}{n-1}\sum_{i=1}^n (f(x_i) - f(x_c))^2 \;,
\end{equation}
where $x_c$ is the central values of the input parameters, and not included in $x_i$, and
$n$ is the number of sample points used to compute an error estimate, excluding the central value $x_c$.
The points $x_i$ are generated from the $N$-dimensional probability distribution of input parameters;
note that $x_i$ is varied for all parameters simultaneously, so \textit{none} of its elements are equal to
the central value of any input parameter.
In effect, this is a primitive Monte Carlo integration over the input parameter distribution space.
All input parameters are assumed to be Gaussian distributed with standard deviation equal to their
quoted error, except for the renormalisation scale, to be discussed below.
We assign the functions $h_{0,\pm}$ and $h_T$ an error of 20\%, which arises from the uncertainty in the
form factors $T_i$, $A_{1,3}$, $V$ and $f_{+,T}$ and non form factor corrections.
We impose this at the level of the $h$ functions so that constraints such as \eqref{eq:X2X3general}
and $h_+\sim\calO(1/m_B)$ are maintained.

To compute the scale uncertainty only 3 points are sampled: $\mu$, $\mu/2$ and $2\mu$.
The renormalisation scale is set to $\mu$ to compute the central value of a result.
The error is computed as
\begin{equation}
\sigma^2_\mu = \frac{1}{2n-1}\sum_{i=1}^{n}\left[\left(f\left(\frac{\mu}{2}\right) - f(\mu)\right)^2 + \left(f(2\mu) - f(\mu)\right)^2\right] \;,
\end{equation}
although in practice this is implemented by generating a $2n$ pseudo-random numbers $y_i$ in $[0,1]$
and selecting $\mu/2$ or $2\mu$ depending on whether $y>0.5$.
This may then be incorporated into the same procedure as sampling all other input parameters.
We take the central renormalisation scale $\mu=m_b=4.7 \GeV$ for all processes except 
QLSS and $\calO_8$, which we take to be $\mu'=\sqrt{\Lambda_H\mu}$,
where $\Lambda_H=0.5(2)\GeV$ as in \cite{DLZ12}.

\subsection{Wilson coefficients}
\label{app:wilson-coefficients}

Although we specify our results in the BBL basis \cite{Buchalla:1995vs} our calculation is carried out in the CMM basis \cite{Chetyrkin:1996vx}
\begin{equation}
\mathcal H_{\mathrm{eff}} = \frac{G_F}{\sqrt 2}\left( \sum_{i=1}^2 
(\lambda_u C^{\rm CMM}_i \mathcal Q_i^u + \lambda_c C^{\rm CMM}_i \mathcal Q_i^c ) - \lambda_t \sum_{i=3}^{10} C^{\rm CMM}_i \mathcal Q_i \right) \;,
\end{equation}
where $\mathcal Q_{7-10}=\mathcal O_{7-10}$ and the four quark operators are given by
\begin{align}
\mathcal Q_1^q &= 4(\bar s_L\gamma_\mu T^a q_L)(\bar q\gamma^\mu T^a b_L) &
\mathcal Q_2^q &= 4(\bar s_L\gamma_\mu q_L)(\bar q\gamma^\mu b_L) \nonumber \\
\mathcal Q_3 &= 4(\bar s_L\gamma_\mu b_L) \sum_q (\bar q\gamma^\mu q) &
\mathcal Q_4 &= 4(\bar s_L\gamma_\mu T^a b_L) \sum_q (\bar q\gamma^\mu T^a q) \nonumber \\
\mathcal Q_5 &= 4(\bar s_L\gamma_\mu\gamma_\nu\gamma_\rho b_L) \sum_q (\bar q\gamma^\mu\gamma^\nu\gamma^\rho q) &
\mathcal Q_6 &= 4(\bar s_L\gamma_\mu\gamma_\nu\gamma_\rho T^a b_L) \sum_q (\bar q\gamma^\mu\gamma^\nu\gamma^\rho T^a q) \;,
\end{align}
with $2q_L = (1-\gamma_5)q$.
We transform into the BBL basis following the recipe in appendix A of \cite{Beneke:2001at};
our $C_i$ are equivalent to the $\bar C_i$ in \cite{Beneke:2001at} and are \textit{defined} by a linear transform from $C^{\rm CMM}_i$.
This linear transform reproduces BBL WCs at $O(\alpha_s^0)$.

We calculate the WCs to NNLL order.
The calculation is carried out as described in the appendix of \cite{Beneke:2001at} using
the full anomalous dimension matrix computed in \cite{Czakon:2006ss}, at fixed $N_f=5$,
and initial conditions are computed at $\mu=M_W$ as described therein using the NNLO expressions
in \cite{Bobeth:1999mk} for $C_{1-6}$ and $C_{9,10}$ and those in \cite{Misiak:2004ew} for
$C_{7,8}^{\mathrm{eff}}$.
An example of the result of this calculation is given in Tab.~ \ref{tbl:wilson-coefficient-example}.

\begin{table}
\center
\begin{tabular}{c|cc|cc|}
\cline{2-5}
& \multicolumn{2}{|c|}{$\mu=m_b$} & \multicolumn{2}{|c|}{$\mu=\sqrt{m_b\Lambda_H}$} \\
\cline{2-5}
& CMM & BBL & CMM & BBL \\
\hline
\multicolumn{1}{|c|}{$C_{1}$} & -0.2622 & -0.1311 & -0.5636 & -0.2818 \\
\multicolumn{1}{|c|}{$C_{2}$} & 1.0087 & 1.0524 & 1.0299 & 1.1238 \\
\multicolumn{1}{|c|}{$C_{3}$} & -0.0051 & 0.0110 & -0.0175 & 0.0194 \\
\multicolumn{1}{|c|}{$C_{4}$} & -0.0778 & -0.0316 & -0.1718 & -0.0524 \\
\multicolumn{1}{|c|}{$C_{5}$} & 0.0003 & 0.0087 & 0.0012 & 0.0132 \\
\multicolumn{1}{|c|}{$C_{6}$} & 0.0009 & -0.0371 & 0.0042 & -0.0775 \\
\multicolumn{1}{|c|}{$C_{7}^{\mathrm{eff}}$} & \multicolumn{2}{|c|}{-0.2975} & \multicolumn{2}{|c|}{-0.3351} \\
\multicolumn{1}{|c|}{$C_{8}^{\mathrm{eff}}$} & \multicolumn{2}{|c|}{-0.1569} & \multicolumn{2}{|c|}{-0.1828} \\
\multicolumn{1}{|c|}{$C_{9}$} & \multicolumn{2}{|c|}{4.0354} & \multicolumn{2}{|c|}{4.4207} \\
\multicolumn{1}{|c|}{$C_{10}$} & \multicolumn{2}{|c|}{-4.2496} & \multicolumn{2}{|c|}{-4.2496} \\
\hline
\end{tabular}
\caption{\small WCs at $\mu=m_b$ and $\mu=\sqrt{m_b\Lambda_H}$ at NNLL order
for $m_b=4.7 \mathrm{GeV}$, $M_W=80.4 \GeV$, $\sin^2\theta_W=0.23$,
$m_t=177 \mathrm{GeV}$, $\Lambda_H=0.5 \mathrm{GeV}$ and
$\Lambda^{(5)}_{\mathrm{QCD}}=214 \mathrm{MeV}$ in two different bases.
Three loop running for $\alpha_s$ is used.
The BBL basis we use is that defined in \cite{Beneke:2001at};
it is equivalent to the traditional basis defined in \cite{Buchalla:1995vs} at leading order.
We use the CMM basis \cite{Chetyrkin:1996vx} for loop calculations.
The BBL coefficients presented here are defined in terms of a linear transform
of the CMM coefficients as explained in the text.
}
\label{tbl:wilson-coefficient-example}
\end{table}

\subsection{Numerical evaluation of  $\delta_{a_I}$ from PDG values}
\label{app:delta-ai-calc}

In terms of experimentally measured quantities, $\delta_{a_I}$ is
\begin{equation}
\delta_{a_I} = 1 -
\left|\frac{V_{\rm ts}}{V_{\rm td}}\right|
\frac{2\frac{\tau_{B^+}}{\tau_{B^0}} \mathcal B(B^0\to\rho^0\gamma) - \mathcal B(B^+\to\rho^+\gamma)}{\frac{\tau_{B^+}}{\tau_{B^0}}\mathcal B(B^0\to K^{*0}\gamma) - \mathcal B(B^+\to K^{*+}\gamma)}
\sqrt{\frac{\frac{\tau_{B^+}}{\tau_{B^0}}\mathcal B(B^0\to K^{*0}\gamma) + \mathcal B(B^+\to K^{*+}\gamma)}{2\frac{\tau_{B^+}}{\tau_{B^0}} \mathcal B(B^0\to\rho^0\gamma) + \mathcal B(B^+\to\rho^+\gamma)}}
\end{equation}
We use the values \cite{Beringer:1900zz}
\begin{align}
\frac{\tau_{B^+}}{\tau_{B^0}} &= 1.079(7) &
\left|\frac{V_{\rm td}}{V_{\rm ts}}\right| &= 0.211(7) \nonumber \\
\mathcal B(B^+\to\rho^+\gamma) &= 9.8(2.5)\times10^{-7} &
\mathcal B(B^0\to\rho^0\gamma) &= 8.6(1.5)\times10^{-7} \\
\mathcal B(B^+\to K^{*+}\gamma) &= 4.21(18)\times10^{-5} &
\mathcal B(B^0\to K^{*0}\gamma) &= 4.33(15)\times10^{-5} \nonumber
\end{align}
and combine all errors in quadrature to get the result \eqref{eq:delta-ai-exp}.
The main error comes from the ratio of differences, that is to say the isospin asymetries themselves, with approximately equal parts from the numerator and the denominator.

\section{Results}
\label{app:results}

\subsection{Tabulated results for four Fermi operators}
\label{app:tables}

We provide numerical data corresponding to Figs.~ \ref{fig:1} and \ref{fig:2}
in Tabs.~ \ref{tab:1} and \ref{tab:2}, at $1 \GeV^2$ intervals in $q^2$.
Data for Figs.~ \ref{fig:1rho} and \ref{fig:2rho} are given in Tab.~ \ref{tab:1rho}.
We also provide data for $B\rightarrow (K^*,\rho)\gamma$, denoted by $q^2=0$ in Tabs.~ \ref{tab:1} and \ref{tab:1rho} respectively.

\begin{table}[t]
\center
\begin{tabular}{l|rrrrrrrr}
 & \multicolumn{8}{|c}{$q^2/\GeV^2$} \\
$B\rightarrow Kll$ & 1 & 2 & 3 & 4 & 5 & 6 & 7 & 8 \\
\hline
$a^u_{4} = 0.1$ &  0.35\% &  0.14\% & -0.03\% & -0.15\% & -0.23\% & -0.27\% & -0.28\% & -0.25\% \\
$a^u_{8}$ &  0.68\% &  0.60\% &  0.63\% &  0.64\% &  0.62\% &  0.58\% &  0.53\% &  0.47\% \\
\hline
$a^d_{4}$ & -0.10\% & -0.13\% & -0.18\% & -0.21\% & -0.22\% & -0.23\% & -0.22\% & -0.20\% \\
$a^d_{8}$ &  0.35\% &  0.31\% &  0.33\% &  0.33\% &  0.32\% &  0.30\% &  0.27\% &  0.24\% \\
\hline
$s^{SU(3)}_{1(R,L)}$ = 1 &  1.28\% &  0.68\% &  0.35\% &  0.18\% &  0.08\% &  0.04\% &  0.01\% & -0.01\% \\
$s^{c}_{1(R,L)}$ &  0.88\% &  0.60\% &  0.39\% &  0.25\% &  0.16\% &  0.11\% &  0.07\% &  0.04\% \\
$s^{b}_{1(R,L)}$ & -0.20\% & -0.34\% & -0.31\% & -0.25\% & -0.20\% & -0.15\% & -0.12\% & -0.09\% \\
$s^{c}_{2(R,L)}$ & -4.68\% & -3.94\% & -3.13\% & -2.46\% & -1.96\% & -1.57\% & -1.25\% & -0.91\% \\
$s^{b}_{2(R,L)}$ &  5.03\% &  1.75\% &  0.03\% & -1.04\% & -1.76\% & -2.25\% & -2.51\% & -2.44\% \\
\hline
$\CBBL_{1}$ Tab.~\ref{tbl:wilson-coefficient-example}  & -0.00\% & -0.00\% & -0.00\% & -0.00\% & -0.00\% & -0.00\% & -0.00\% & -0.00\% \\
$\CBBL_{2}$ & -0.84\% & -0.45\% & -0.22\% & -0.10\% & -0.03\% &  0.01\% &  0.03\% &  0.04\% \\
$\CBBL_{3}$ &  0.02\% &  0.04\% &  0.04\% &  0.04\% &  0.04\% &  0.04\% &  0.03\% &  0.03\% \\
$\CBBL_{4}$ & -0.11\% & -0.21\% & -0.28\% & -0.31\% & -0.31\% & -0.29\% & -0.27\% & -0.24\% \\
$\CBBL_{5}$ &  0.01\% & -0.00\% & -0.01\% & -0.02\% & -0.03\% & -0.03\% & -0.03\% & -0.03\% \\
$\CBBL_{6}$ &  0.20\% &  0.23\% &  0.30\% &  0.35\% &  0.39\% &  0.41\% &  0.39\% &  0.34\% \\
$\CBBL_8^{\rm eff}$ & -0.22\% & -0.09\% & -0.02\% &  0.02\% &  0.05\% &  0.08\% &  0.09\% &  0.09\% \\
\hline
SM total & -0.93\% & -0.48\% & -0.20\% & -0.01\% &  0.12\% &  0.20\% &  0.24\% &  0.24\% \\
\end{tabular}
\caption{\small Breakdown of contributions to $B\rightarrow Kll$ isospin asymmetry in SM
operator coefficients $\CBBL_i$, and in a generalised basis of four quark WA operators with coefficients
$a_i$ and QLSS contributions with coefficients $s^q_{x\chi}$.
We use $a^q_i=0.1$ and $s^q_{x,\chi}=1$ to produce these values.}
\label{tab:2}
\end{table}

\begin{table}
\center
\begin{tabular}{l|rrrrrrrrr}
 & \multicolumn{9}{|c}{$q^2/\GeV^2$} \\
$B\rightarrow K^*ll$ & 0 & 1 & 2 & 3 & 4 & 5 & 6 & 7 & 8 \\
\hline
$a^u_{2} = 0.1$ & -1.55\% & -0.22\% & -0.00\% &  0.06\% &  0.08\% &  0.08\% &  0.07\% &  0.06\% &  0.06\% \\
$a^u_{4}$ & -1.58\% & -0.33\% & -0.09\% & -0.00\% &  0.02\% &  0.03\% &  0.04\% &  0.03\% &  0.03\% \\
$a^u_{5}$ &  1.29\% & -0.07\% &  0.02\% &  0.00\% & -0.00\% &  0.00\% &  0.01\% &  0.02\% &  0.03\% \\
$a^u_{6}$ & -0.84\% & -0.53\% & -0.64\% & -0.67\% & -0.65\% & -0.60\% & -0.54\% & -0.47\% & -0.42\% \\
$a^u_{9}$ &  10.3\% & -0.20\% &  0.23\% &  0.03\% & -0.02\% &  0.02\% &  0.10\% &  0.17\% &  0.26\% \\
$a^u_{10}$ &  10.5\% &  0.43\% &  0.40\% &  0.13\% &  0.08\% &  0.14\% &  0.24\% &  0.34\% &  0.47\% \\
\hline
$a^d_{2} = 0.1$ & -2.85\% & -0.40\% & -0.00\% &  0.12\% &  0.15\% &  0.15\% &  0.14\% &  0.12\% &  0.11\% \\
$a^d_{4}$ & -2.91\% & -0.61\% & -0.17\% & -0.01\% &  0.05\% &  0.07\% &  0.07\% &  0.07\% &  0.06\% \\
$a^d_{5}$ &  0.78\% &  0.00\% &  0.02\% &  0.00\% & -0.00\% & -0.00\% & -0.00\% &  0.00\% &  0.01\% \\
$a^d_{6}$ & -0.50\% & -0.30\% & -0.34\% & -0.35\% & -0.33\% & -0.31\% & -0.27\% & -0.24\% & -0.21\% \\
$a^d_{9}$ &  6.23\% &  0.18\% &  0.20\% &  0.02\% & -0.04\% & -0.03\% &  0.00\% &  0.04\% &  0.08\% \\
$a^d_{10}$ &  6.29\% &  0.45\% &  0.24\% &  0.03\% & -0.03\% & -0.01\% &  0.04\% &  0.09\% &  0.16\% \\
\hline
$s^{SU(3)}_{1R} = 1$ &  0.00\% & -1.26\% & -0.75\% & -0.38\% & -0.18\% & -0.08\% & -0.03\% & -0.01\% &  0.00\% \\
$s^{c}_{1R}$ &  0.00\% & -0.90\% & -0.67\% & -0.43\% & -0.26\% & -0.16\% & -0.10\% & -0.06\% & -0.03\% \\
$s^{b}_{1R}$ &  0.01\% &  0.20\% &  0.38\% &  0.34\% &  0.26\% &  0.19\% &  0.14\% &  0.10\% &  0.08\% \\
$s^{SU(3)}_{1L}$ & -0.28\% &  0.81\% &  0.58\% &  0.36\% &  0.21\% &  0.12\% &  0.06\% &  0.03\% & -0.00\% \\
$s^{c}_{1L}$ & -0.40\% &  0.67\% &  0.57\% &  0.42\% &  0.30\% &  0.22\% &  0.16\% &  0.11\% &  0.07\% \\
$s^{b}_{1L}$ &  0.95\% & -0.00\% & -0.28\% & -0.33\% & -0.31\% & -0.26\% & -0.22\% & -0.18\% & -0.14\% \\
$s^{c}_{2R}$ &  1.59\% & -1.82\% & -1.96\% & -1.64\% & -1.27\% & -0.97\% & -0.73\% & -0.54\% & -0.36\% \\
$s^{b}_{2R}$ &  5.03\% &  8.33\% &  3.25\% &  0.12\% & -1.56\% & -2.38\% & -2.67\% & -2.62\% & -2.28\% \\
$s^{c}_{2L}$ &  0.02\% &  2.43\% &  2.21\% &  1.66\% &  1.19\% &  0.85\% &  0.61\% &  0.44\% &  0.29\% \\
$s^{b}_{2L}$ &  0.05\% & -4.60\% & -1.84\% & -0.05\% &  0.98\% &  1.56\% &  1.83\% &  1.88\% &  1.69\% \\
\hline
$\CBBL_{1}$ Tab.~\ref{tbl:wilson-coefficient-example}  & -0.01\% & -0.00\% & -0.00\% & -0.00\% & -0.00\% & -0.00\% & -0.00\% & -0.00\% & -0.00\% \\
$\CBBL_{2}$ &  0.11\% & -0.71\% & -0.44\% & -0.24\% & -0.12\% & -0.06\% & -0.02\% &  0.01\% &  0.03\% \\
$\CBBL_{3}$ &  0.09\% &  0.01\% &  0.04\% &  0.04\% &  0.05\% &  0.04\% &  0.04\% &  0.03\% &  0.03\% \\
$\CBBL_{4}$ & -0.98\% & -0.08\% & -0.25\% & -0.30\% & -0.31\% & -0.30\% & -0.28\% & -0.26\% & -0.24\% \\
$\CBBL_{5}$ & -0.51\% & -0.09\% & -0.02\% &  0.01\% &  0.02\% &  0.02\% &  0.02\% &  0.02\% &  0.01\% \\
$\CBBL_{6}$ &  6.41\% &  1.40\% &  0.40\% &  0.03\% & -0.11\% & -0.17\% & -0.18\% & -0.18\% & -0.17\% \\
$\CBBL_8^{\rm eff}$ & -0.19\% & -0.34\% & -0.14\% & -0.02\% &  0.05\% &  0.09\% &  0.10\% &  0.10\% &  0.09\% \\
\hline
SM total &  4.92\% &  0.18\% & -0.42\% & -0.48\% & -0.44\% & -0.38\% & -0.33\% & -0.28\% & -0.24\% \\
\end{tabular}
\caption{\small Breakdown of contributions to $B\rightarrow K^*ll$ isospin asymmetry in SM
operator coefficients $\CBBL_i$, and in a generalised basis of four quark WA operators with coefficients
$a_i$ and QLSS contributions with coefficients $s^q_{x\chi}$.
We use $a^q_i=0.1$ and $s^q_{x,\chi}=1$ to produce these values.
The $q^2=0$ value corresponds to the process $B\rightarrow K^*\gamma$ and is computed slightly differently
to $B\rightarrow K^*ll$ as described in section \ref{sec:WAq2eq0}. The value for $s^f_{1R}$ and $s^f_{2L}$ are zero at $q^2 = 0$ as a consequence of $h_+(0) = 0$ in our approximation.}
\label{tab:1}
\end{table}

\begin{table}
\center
\begin{tabular}{l|rrrrrrrrr}
 & \multicolumn{9}{|c}{$q^2/\GeV^2$} \\
$B\rightarrow \rho ll$ & 0 & 1 & 2 & 3 & 4 & 5 & 6 & 7 & 8 \\
\hline
$a^u_{2} = 0.1$ & -1.55\% & -0.23\% & -0.01\% &  0.05\% &  0.07\% &  0.07\% &  0.07\% &  0.06\% &  0.06\% \\
$a^u_{4}$ & -1.59\% & -0.33\% & -0.09\% & -0.01\% &  0.02\% &  0.04\% &  0.04\% &  0.04\% &  0.03\% \\
$a^u_{5}$ &  1.25\% & -0.06\% &  0.02\% &  0.00\% & -0.00\% &  0.00\% &  0.01\% &  0.02\% &  0.03\% \\
$a^u_{6}$ & -0.81\% & -0.62\% & -0.72\% & -0.74\% & -0.71\% & -0.66\% & -0.59\% & -0.52\% & -0.46\% \\
$a^u_{9}$ &  11.0\% & -0.20\% &  0.26\% &  0.04\% & -0.02\% &  0.03\% &  0.11\% &  0.19\% &  0.28\% \\
$a^u_{10}$ &  11.1\% &  0.45\% &  0.43\% &  0.14\% &  0.07\% &  0.12\% &  0.23\% &  0.33\% &  0.47\% \\
\hline
$\tilde{a}^d_{2} = 0.1$ & -3.10\% & -0.46\% & -0.03\% &  0.11\% &  0.14\% &  0.15\% &  0.14\% &  0.12\% &  0.11\% \\
$\tilde{a}^d_{4}$ & -3.17\% & -0.65\% & -0.18\% & -0.01\% &  0.05\% &  0.07\% &  0.07\% &  0.07\% &  0.07\% \\
$\tilde{a}^d_{5}$ &  0.76\% &  0.00\% &  0.02\% &  0.00\% & -0.00\% & -0.00\% & -0.00\% &  0.00\% &  0.01\% \\
$\tilde{a}^d_{6}$ & -0.48\% & -0.34\% & -0.38\% & -0.38\% & -0.37\% & -0.34\% & -0.30\% & -0.26\% & -0.23\% \\
$\tilde{a}^d_{9}$ &  6.62\% &  0.19\% &  0.22\% &  0.03\% & -0.04\% & -0.03\% &  0.00\% &  0.04\% &  0.09\% \\
$\tilde{a}^d_{10}$ &  6.68\% &  0.48\% &  0.27\% &  0.04\% & -0.03\% & -0.01\% &  0.03\% &  0.09\% &  0.16\% \\
\hline
$s^{SU(3)}_{1R} = 1$  &  0.00\% & -1.39\% & -0.81\% & -0.41\% & -0.19\% & -0.08\% & -0.03\% & -0.01\% &  0.00\% \\
$s^{c}_{1R}$ &  0.00\% & -1.01\% & -0.74\% & -0.47\% & -0.28\% & -0.17\% & -0.10\% & -0.06\% & -0.03\% \\
$s^{b}_{1R}$ &  0.01\% &  0.22\% &  0.42\% &  0.37\% &  0.29\% &  0.21\% &  0.15\% &  0.11\% &  0.08\% \\
$s^{SU(3)}_{1L}$ & -0.40\% &  0.94\% &  0.64\% &  0.39\% &  0.22\% &  0.12\% &  0.06\% &  0.03\% & -0.00\% \\
$s^{c}_{1L}$ & -0.44\% &  0.77\% &  0.64\% &  0.45\% &  0.32\% &  0.23\% &  0.16\% &  0.12\% &  0.07\% \\
$s^{b}_{1L}$ &  1.12\% &  0.00\% & -0.31\% & -0.36\% & -0.33\% & -0.28\% & -0.23\% & -0.19\% & -0.15\% \\
$s^{c}_{2R}$ &  1.76\% & -2.05\% & -2.13\% & -1.74\% & -1.34\% & -1.01\% & -0.76\% & -0.57\% & -0.38\% \\
$s^{b}_{2R}$ &  4.02\% &  8.46\% &  3.29\% &  0.06\% & -1.70\% & -2.55\% & -2.86\% & -2.82\% & -2.47\% \\
$s^{c}_{2L}$ &  0.02\% &  2.69\% &  2.38\% &  1.76\% &  1.26\% &  0.89\% &  0.64\% &  0.46\% &  0.30\% \\
$s^{b}_{2L}$ &  0.05\% & -4.81\% & -1.85\% &  0.03\% &  1.11\% &  1.70\% &  1.99\% &  2.04\% &  1.83\% \\
\hline
$\CBBL_{1}$  Tab.~\ref{tbl:wilson-coefficient-example}  &  0.01\% &  0.02\% &  0.01\% &  0.01\% &  0.00\% &  0.00\% & -0.00\% & -0.00\% & -0.00\% \\
$\CBBL_{2}$ &  0.01\% & -1.46\% & -1.00\% & -0.65\% & -0.40\% & -0.23\% & -0.11\% & -0.02\% &  0.09\% \\
$\CBBL_{3}$ &  0.08\% &  0.01\% &  0.04\% &  0.05\% &  0.05\% &  0.05\% &  0.04\% &  0.04\% &  0.03\% \\
$\CBBL_{4}$ & -0.93\% & -0.09\% & -0.28\% & -0.34\% & -0.35\% & -0.33\% & -0.31\% & -0.28\% & -0.26\% \\
$\CBBL_{5}$ & -0.54\% & -0.10\% & -0.02\% &  0.01\% &  0.02\% &  0.02\% &  0.02\% &  0.02\% &  0.01\% \\
$\CBBL_{6}$ &  6.74\% &  1.51\% &  0.46\% &  0.06\% & -0.10\% & -0.16\% & -0.18\% & -0.18\% & -0.18\% \\
$\CBBL_8^{\rm eff}$ & -0.14\% & -0.35\% & -0.14\% & -0.01\% &  0.06\% &  0.09\% &  0.11\% &  0.11\% &  0.09\% \\
\hline
SM total &  5.22\% & -0.45\% & -0.93\% & -0.87\% & -0.72\% & -0.57\% & -0.43\% & -0.32\% & -0.21\% \\
\end{tabular}
\caption{\small Breakdown of contributions to $B\rightarrow \rho ll$ isospin asymmetry in SM
operator coefficients $\CBBL_i$, and in a generalised basis of four quark WA operators with coefficients
$a_i$ and QLSS contributions with coefficients $s^q_{x\chi}$.
We use $a^u_i=0.1$, $\tilde a^d_i=0.1$ and $s^q_{x,\chi}=1$ to produce these values.
The modified four quark coefficients $\tilde a_i$ are explained in section \ref{sec:rho-meson}.
The $q^2=0$ value corresponds to the process $B\rightarrow \rho\gamma$ and is computed slightly differently
to $B\rightarrow \rho ll$ as described in section \ref{sec:WAq2eq0}.
The value for $s^f_{1R}$ and $s^f_{2L}$ are zero at $q^2 = 0$ as a consequence of $h_+(0) = 0$ in our approximation.}
\label{tab:1rho}
\end{table}

\subsection{Effective coefficients in $B^0\to \rho^0$ decay}
\label{app:rho0-effective-coefficients}

Here we collect the formulae for $\tilde a_i^d$ omitted from section \ref{sec:rho-meson}.

\begin{align}
\begin{split}
\tilde a_2^d =& a_2^d
+ \frac{1}{6}(-a_2^u - a_3^u + 4a_6^u - 4a_7^u + 3a_{10}^u) + \frac{2}{9}(-a_2^{8u} - a_3^{8u} + 4a_6^{8u} - 4a_7^{8u} + 3a_{10}^{8u}) \\
& + \frac{1}{12}(-a_2^d - a_3^d + 4a_6^d - 4a_7^d + 3a_{10}^d) + \frac{1}{9}(-a_2^{8d} - a_3^{8d} + 4a_6^{8d} - 4a_7^{8d} + 3a_{10}^{8d})
\end{split} \\
\begin{split}
\tilde a_4^d =& a_4^d
+ \frac{1}{6}(-a_1^u - a_4^u + 4a_5^u - 4a_8^u + 3a_9^u) + \frac{2}{9}(-a_1^{8u} - a_4^{8u} + 4a_5^{8u} - 4a_8^{8u} + 3a_9^{8u}) \\
& + \frac{1}{12}(-a_1^d - a_4^d + 4a_5^d - 4a_8^d + 3a_9^d) + \frac{1}{9}(-a_1^{8d} - a_4^{8d} + 4a_5^{8d} - 4a_8^{8d} + 3a_9^{8d})
\end{split} \displaybreak[0] \\
\begin{split}
\tilde a_5^d =& a_5^d + \frac{1}{12}\left(2(a_5^d - a_5^u) + 2(a_8^d-a_8^u) + (a_4^d-a_4^u) - (a_1^d-a_1^u)\right) \\
& + \frac{1}{9}\left(2(a_5^{8d}-a_5^{8u}) + 2(a_8^{8d}-a_8^{8u}) + (a_4^{8d}-a_4^{8u}) - (a_1^{8d}-a_1^{8u})\right)
\end{split} \\
\begin{split}
\tilde a_6^d =& a_6^d + \frac{1}{12}\left(2(a_6^d - a_6^u) + 2(a_7^d-a_7^u) + (a_2^d-a_2^u) - (a_3^d-a_3^u)\right) \\
& + \frac{1}{9}\left(2(a_6^{8d}-a_6^{8u}) + 2(a_7^{8d}-a_7^{8u}) + (a_2^{8d}-a_2^{8u}) - (a_3^{8d}-a_3^{8u})\right)
\end{split} \displaybreak[0] \\
\tilde a_9^d =& a_9^d + \frac{1}{6}(a_9^d-a_9^u) + \frac{1}{12}((a_1^d-a_1^u)+(a_4^d-a_4^u)) + \frac{1}{9}((a_1^{8d}-a_1^{8u})+(a_4^{8d}-a_4^{8u})) \\
\tilde a_{10}^d =& a_{10}^d + \frac{1}{6}(a_{10}^d-a_{10}^u) + \frac{1}{12}((a_2^d-a_2^u)+(a_3^d-a_3^u)) + \frac{1}{9}((a_2^{8d}-a_2^{8u})+(a_3^{8d}-a_3^{8u}))
\end{align}

\subsubsection{Effective coefficients in $B^0\to \rho^0$ decay in the SM}
\label{app:rho0-effective-coefficientsSM}

The effective coefficients in $B^0\to \rho^0$ decay in the SM for  $B^0\to\rho^0$ 
are
\begin{align}
\tilde a_2^d &= \tilde a_4^d = 2\left(\frac{C_5}{N_c} + C_6\right) \nonumber \\
\tilde a_5^d &= -\tilde a_6^d = \left(\frac{C_3}{N_c} + C_4\right) + \frac{\lambda_u}{\lambda_t}\left(C_1 + \frac{C_2}{N_c}\right) \\
\tilde a_9^d &= \tilde a_{10}^d = 0 \;, \nonumber
\end{align}
where we have used the formulae of the previous section and Eqs.~(\ref{eq:aSM},\ref{eq:a8SM}). Note, we recognise the well known colour suppressed tree-level combination $C_1 + C_2/N_c$ in the formula above.

\subsection{Weak annihilation formulae}
\label{app:wa-formulae}

We list the functions defined on the RHS of \eqref{eq:wa-if-split}.
Any function not listed is zero and there are many as can be inferred from Tab.~\ref{tab:OWA}.
The functions $\rho_{C_b}$ an $\rho_{C_d}$ are derived from the dispersion representation of the Passarino-Veltman functions
\begin{align}
C_b &= C_0(p_B^2,p_B^2-m_B^2,q^2,0,m_b^2,0) &
C_d &= C_0(p_B^2,p_B^2-m_B^2,q^2,m_b^2,0,m_b^2)
\end{align}
which are given in appendix H of \cite{DLZ12} (note that $C_b=C_a|_{u=1}$ and $C_d=C_c|_{u=1}$).

The functions in \eqref{eq:wa-if-split} which apply at $|q^2|>1\GeV^2$ are given in subsection \ref{app:q2neq0},
and the funtions in \eqref{eq:wa-if-split-0} which apply at $q^2=0$ are given in subsection \ref{app:q2eq0}.

\subsubsection{WA formulae $|q^2| > 1 \GeV^2$}
\label{app:q2neq0}

Deinfing, as before, $d  \equiv -\frac{\sqrt{2} m_B m_V}{\sqrt{q^2} E}$. we get:

\begin{dmath}f_{2,A}^q(q^2,u)=2 \pi ^2 \phi _{\perp}(u) \left(\frac{Q_q}{(u-1) m_B^2-u q^2}-\frac{Q_b}{u m_B^2-u q^2+q^2}\right)\end{dmath}
\begin{dmath}
d \cdot f^q_{2,0}(q^2,u) = \frac{32\pi^2 m_{K^*}^2 m_B^2}{(m_B^2-q^2)^2} h_\parallel^{(s)'}(u)\left[\frac{\bar uQ_b}{u m_B^2 + \bar u q^2} - \frac{uQ_q}{\bar u m_B^2 + uq^2}\right]
\end{dmath}
\begin{dmath}f_{4,V}^q(q^2,u)=-2 \pi ^2 \phi _{\perp}(u) \left(\frac{Q_b}{u m_B^2-u q^2+q^2}+\frac{Q_q}{-u m_B^2+m_B^2+u q^2}\right)\end{dmath}
\begin{dmath}\rho_{5,V}^q(q^2,s)=\frac{3}{2} m_b f_{K^*} m_{K^*} \left(s \left(s-q^2\right)^3\right)^{-1} \left(\left(m_b^2-s\right) \left(Q_b-Q_q\right) \left(s^2-\left(q^2\right)^2\right)-s Q_b \left(2 m_b^2 q^2-s q^2+s^2\right) \log \left(\frac{s m_b^2}{m_b^2 q^2-s q^2+s^2}\right)+s Q_q q^2 \left(2 m_b^2+q^2-s\right) \log \left(\frac{s \left(m_b^2+q^2-s\right)}{m_b^2 q^2}\right)\right)\end{dmath}
\begin{dmath}V_{5,V}^q(q^2)=\frac{2 \pi ^2 f_{K^*} m_{K^*} \left(m_b^2 Q_q-Q_b q^2\right)}{m_b^2 q^2}\end{dmath}
\begin{dmath}\rho_{6,A}^q(q^2,s)=\frac{3}{2} m_b f_{K^*} m_{K^*} \left(s^2 \left(s-q^2\right)^3 \left(q^2-m_B^2\right)\right){}^{-1} \left(\left(m_b^2-s\right) \left(s-q^2\right) \left(m_b^2 \left(Q_b-Q_q\right) \left(-s q^2+\left(q^2\right)^2+2 s^2\right)-s \left(s-q^2\right) \left(s \left(Q_b-Q_q\right)-2 Q_b q^2\right)\right)+s^2 Q_q q^2 \left(-2 m_b^2 \left(s-q^2\right)+2 m_b^4+\left(s-q^2\right)^2\right) \log \left(\frac{s \left(m_b^2+q^2-s\right)}{m_b^2 q^2}\right)+s^2 Q_b \left(-2 s m_b^2 \left(s-q^2\right)-2 m_b^4 q^2+s \left(s-q^2\right)^2\right) \log \left(\frac{s m_b^2}{m_b^2 q^2-s q^2+s^2}\right)\right)\end{dmath}
\begin{dmath}V_{6,A}^q(q^2)=-\frac{2 \pi ^2 f_{K^*} m_{K^*} \left(-m_b^2 q^2 \left(Q_b-3 Q_q\right)+m_b^4 Q_q+Q_b \left(q^2\right)^2\right)}{m_b^2 q^2 \left(m_B^2-q^2\right)}\end{dmath}
\begin{dmath}d \cdot  \rho_{6,0}^q(q^2,s)=3 m_b m_B^2 f_{K^*} m_{K^*} \left(s^2 \left(s-q^2\right)^3 \left(m_B^2-q^2\right)\right){}^{-1} \left(2 s^2 m_b^4 Q_b \log \left(\frac{s m_b^2}{m_b^2 q^2-s q^2+s^2}\right)-2 s^2 m_b^2 Q_q \left(m_b^2+q^2-s\right) \log \left(\frac{s \left(m_b^2+q^2-s\right)}{m_b^2 q^2}\right)+\left(m_b^2-s\right) \left(Q_b-Q_q\right) \left(s-q^2\right) \left(m_b^2 \left(q^2-3 s\right)+s \left(s-q^2\right)\right)\right)\end{dmath}
\begin{dmath}d \cdot  V_{6,0}^q(q^2)=\frac{8 \pi ^2 m_B^2 f_{K^*} m_{K^*} Q_q}{q^2 \left(m_B^2-q^2\right)}\end{dmath}
\begin{dmath}\rho_{9,V}^q(q^2,s)=\frac{3}{2} \left(s \left(s-q^2\right)^3\right)^{-1} f_{K^*}^{\perp} \left(-2 s m_b^4 Q_b q^2 \log \left(\frac{s m_b^2}{m_b^2 q^2-s q^2+s^2}\right)+\left(m_b^2-s\right) \left(s-q^2\right) \left(m_b^2 \left(Q_b-Q_q\right) \left(q^2+s\right)-s \left(Q_b+Q_q\right) \left(s-q^2\right)\right)+2 s m_b^4 Q_q q^2 \log \left(\frac{s \left(m_b^2+q^2-s\right)}{m_b^2 q^2}\right)\right)\end{dmath}
\begin{dmath}V_{9,V}^q(q^2)=\frac{4 \pi ^2 f_{K^*}^{\perp} \left(m_b^2 Q_q-Q_b q^2\right)}{m_b q^2}\end{dmath}
\begin{dmath}\rho_{10,A}^q(q^2,s)=-\frac{3}{2} f_{K^*}^{\perp} \left(s \left(s-q^2\right)^2 \left(q^2-m_B^2\right)\right){}^{-1} \left(-2 s m_b^4 Q_b q^2 \log \left(\frac{s m_b^2}{m_b^2 q^2-s q^2+s^2}\right)+\left(m_b^2-s\right) \left(s-q^2\right) \left(m_b^2 \left(Q_b-Q_q\right) \left(q^2+s\right)-s \left(Q_b+Q_q\right) \left(s-q^2\right)\right)+2 s m_b^4 Q_q q^2 \log \left(\frac{s \left(m_b^2+q^2-s\right)}{m_b^2 q^2}\right)\right)\end{dmath}
\begin{dmath}V_{10,A}^q(q^2)=\frac{4 \pi ^2 \left(m_b^2-q^2\right) f_{K^*}^{\perp} \left(m_b^2 Q_q-Q_b q^2\right)}{m_b q^2 \left(m_B^2-q^2\right)}\end{dmath}
\begin{dmath}d \cdot  \rho_{10,0}^q(q^2,s)=-12 m_B^2 m_{K^*}^2 f_{K^*}^{\perp} \left(\left(q^2-s\right)^2 \left(m_B^2-q^2\right){}^3\right){}^{-1} \left(m_b^2 Q_b \left(m_b^2 \left(q^2+s\right)+2 s \left(s-q^2\right)\right) \log \left(\frac{s m_b^2}{m_b^2 q^2-s q^2+s^2}\right)-\left(m_b^2-s\right) \left(s-q^2\right) \left(2 m_b^2 \left(Q_b-Q_q\right)+\left(Q_b+Q_q\right) \left(s-q^2\right)\right)+m_b^2 Q_q \left(q^2+s\right) \left(m_b^2+q^2-s\right) \left(-\log \left(\frac{s \left(m_b^2+q^2-s\right)}{m_b^2 q^2}\right)\right)\right)\end{dmath}
\begin{dmath}d \cdot  V_{10,0}^q(q^2)=\frac{16 \pi ^2 m_B^2 m_{K^*}^2 \left(m_b^2-q^2\right) f_{K^*}^{\perp} \left(m_b^2 Q_q-Q_b q^2\right)}{m_b q^2 \left(q^2-m_B^2\right){}^3}\end{dmath}
\begin{dmath}\rho_{4,T}^q(q^2,s)=-\frac{3}{2} \mu _K^2 \left(m_B+m_K\right) \left(s m_B^2 \left(2 m_B^2 q^2+m_B^4-4 s q^2+\left(q^2\right)^2\right)\right){}^{-1} \left(2 s Q_b m_B^2 \left(m_b^4+s \left(m_B^2-s\right)\right) \rho _{C_d}(s)+2 s m_B^2 Q_q \rho _{C_b}(s) \left(m_b^2 \left(m_B^2+q^2-2 s\right)+m_b^4+s \left(s-m_B^2\right)\right)+\left(m_b^2-s\right) \left(Q_b-Q_q\right) \left(m_b^2 \left(m_B^2+q^2-4 s\right)+s \left(-3 m_B^2-q^2+4 s\right)\right)\right)\end{dmath}
\begin{dmath}V_{4,T}^q(q^2)=-\frac{4 \pi ^2 m_b \mu _K^2 Q_q \left(m_B+m_K\right)}{m_B^2 q^2}\end{dmath}
\begin{dmath}
f^{q}_{4,T}(q^2,u) = \frac{2\pi^2}{m_B^2}(m_B+m_K)\left[\phi_P(u)\left(\frac{\bar u Q_b}{u m_B^2 + \bar u q^2} - \frac{u Q_q}{\bar u m_B^2 + u q^2}\right) +
\frac{\phi_\sigma(u)}{6}\left(Q_b\frac{u (1+2\bar u)m_B^2 + 2\bar u^2q^2}{u(u m_B^2 + \bar u q^2)^2}
- Q_q\frac{\bar u (1+2u)m_B^2 + 2u^2q^2}{\bar u(\bar u m_B^2 + u q^2)^2}\right)\right]
\end{dmath}
\begin{dmath}\rho_{8,T}^q(q^2,s)=\frac{3}{2} m_b f_K \left(m_B+m_K\right) \left(s^2 \left(s-q^2\right)^3\right)^{-1} \left(2 s^2 m_b^4 Q_b \log \left(\frac{s m_b^2}{m_b^2 q^2-s q^2+s^2}\right)-2 s^2 m_b^2 Q_q \left(m_b^2+q^2-s\right) \log \left(\frac{s \left(m_b^2+q^2-s\right)}{m_b^2 q^2}\right)+\left(m_b^2-s\right) \left(Q_b-Q_q\right) \left(s-q^2\right) \left(m_b^2 \left(q^2-3 s\right)+s \left(s-q^2\right)\right)\right)\end{dmath}
\begin{dmath}V_{8,T}^q(q^2)=\frac{4 \pi ^2 f_K Q_q \left(m_B+m_K\right)}{q^2}\end{dmath}

\subsubsection{WA formulae $q^2 = 0$}
\label{app:q2eq0}

\begin{align}
\widetilde{\rho}_{5,V}^{q,\gamma}(s) &=\frac{2 \pi ^2 f_{K^*} m_{K^*} Q_q \phi _{\gamma }\left(\frac{m_b^2}{s}\right)}{s}&
V_{5,V}^{q,\gamma} &=-\frac{2 \pi ^2 Q_b f_{K^*} m_{K^*}}{m_b^2}
\end{align}
\begin{align}
\widetilde{\rho}_{6,A}^{q,\gamma}(s) &=-\frac{2 \pi ^2 f_{K^*} m_{K^*} Q_q \left(s \phi _{\gamma }\left(\frac{m_b^2}{s}\right)-2\right)}{s m_B^2}&
V_{6,A}^{q,\gamma} &=\frac{2 \pi ^2 f_{K^*} m_{K^*} \left(Q_b-2 Q_q\right)}{m_B^2}
\end{align}
\begin{align}
\widetilde{\rho}_{9,V}^{q,\gamma}(s) &=\frac{4 \pi ^2 m_b Q_q f_{K^*}^{\perp} \phi _{\gamma }\left(\frac{m_b^2}{s}\right)}{s}&
V_{9,V}^{q,\gamma} &=-\frac{4 \pi ^2 Q_b f_{K^*}^{\perp}}{m_b}
\end{align}
\begin{align}
\widetilde{\rho}_{10,A}^{q,\gamma}(s) &=\frac{4 \pi ^2 m_b Q_q f_{K^*}^{\perp} \phi _{\gamma }\left(\frac{m_b^2}{s}\right)}{m_B^2}&
V_{10,A}^{q,\gamma} &=-\frac{4 \pi ^2 m_b Q_b f_{K^*}^{\perp}}{m_B^2}
\end{align}

\bibliographystyle{utphys}
\bibliography{references}

\end{document}